\documentclass[10pt,journal,compsoc]{IEEEtran}
\usepackage[T1]{fontenc} % optional
\usepackage[english]{babel}
\usepackage[ruled, vlined]{algorithm2e}
\usepackage{amsthm}
\usepackage{bm}
\usepackage{algpseudocode}
\usepackage{amssymb}
\usepackage{colortbl}
\usepackage{arydshln}
\usepackage{blkarray, bigstrut}
\usepackage{bm}
\usepackage{booktabs}
\usepackage{braket}
\usepackage{cite}
\usepackage{cuted}
\usepackage{cancel}
\usepackage{enumitem}
\usepackage{graphicx}
\usepackage{glossaries}
\usepackage{indentfirst}
\usepackage[mathscr]{euscript}
\usepackage[switch]{lineno}
\usepackage{mathrsfs}
\usepackage{mathtools}
\usepackage{multirow}
\usepackage{multicol}
\usepackage{makecell}
\usepackage{lipsum}
\usepackage{xfrac}
\usepackage{pifont}
\usepackage{pgfplots}
\pgfplotsset{compat=1.16}
\usepackage{filecontents}
\usepackage{adjustbox}
\usepackage{rotating}
\usepackage{float}
\usepackage{stfloats}
\usepackage[skip=0pt,font=footnotesize]{caption} % affects figures and tables fontsize and vertical spacing
\usepackage[aboveskip=0pt, belowskip=0pt]{subcaption}
\usepackage{url}
\usepackage{tikz}
\usetikzlibrary{positioning}
\usepackage{verbatimbox}
\usepgfplotslibrary{groupplots}
\usetikzlibrary{patterns}
\usetikzlibrary{spy}
\usetikzlibrary{shapes, backgrounds,calc, decorations}
\usetikzlibrary{matrix}
\usepackage{fancybox}
\usepackage[normalem]{ulem}
\usepackage[compact]{titlesec}
\usepackage{thmtools}
\usepackage{balance}
\usepackage{totpages}
\usepackage[nodisplayskipstretch]{setspace}

\usepackage{microtype}

\makeatletter
\renewcommand\subsubsection{\@startsection{subsubsection}{3}{\z@}%
                                     {-0.75ex\@plus -0.75ex \@minus -.2ex}%
                                     {-0em}%
                                     {\normalfont\normalsize\itshape}}
\makeatother

\tikzstyle{vertex} = [circle, draw, inner sep = 0pt, minimum size = 10pt]

\newcommand*\circled[1]{\tikz[baseline=(char.base)]{
            \node[shape=circle,draw,inner sep=2pt] (char) {#1};}}

\definecolor{blue1}{HTML}{b3cde0}
\definecolor{blue2}{HTML}{6497b1}
\definecolor{blue3}{HTML}{005b96}
\definecolor{blue4}{HTML}{03396c}
\definecolor{blue5}{HTML}{011f4b}

\definecolor{dcolor1}{HTML}{253494}
\definecolor{dcolor2}{HTML}{636363}
\definecolor{dcolor3}{HTML}{fcbba1}
\definecolor{dcolor4}{HTML}{fb6a4a}
\definecolor{dcolor5}{HTML}{cb181d}
\definecolor{dcolor6}{HTML}{67000d}
\definecolor{dcolor7}{HTML}{9ecae1}
\definecolor{dcolor8}{HTML}{4292c6}
\definecolor{dcolor9}{HTML}{08519c}

\definecolor{layer1}{HTML}{003f5c}
\definecolor{layer2}{HTML}{444e86}
\definecolor{layer3}{HTML}{955196}
\definecolor{layer4}{HTML}{dd5182}
\definecolor{layer5}{HTML}{ff6e54}
\definecolor{layer6}{HTML}{ffa600}

\definecolorset{rgb}{}{}{%
AliceBlue,.94,.972,1;%
AntiqueWhite,.98,.92,.844;%
Aqua,0,1,1;%
Aquamarine,.498,1,.83;%
Azure,.94,1,1;%
Beige,.96,.96,.864;%
Bisque,1,.894,.77;%
Black,0,0,0;%
BlanchedAlmond,1,.92,.804;%
Blue,0,0,1;%
BlueViolet,.54,.17,.888;%
Brown,.648,.165,.165;%
BurlyWood,.87,.72,.53;%
CadetBlue,.372,.62,.628;%
Chartreuse,.498,1,0;%
Chocolate,.824,.41,.116;%
Coral,1,.498,.312;%
CornflowerBlue,.392,.585,.93;%
Cornsilk,1,.972,.864;%
Crimson,.864,.08,.235;%
Cyan,0,1,1;%
DarkBlue,0,0,.545;%
DarkCyan,0,.545,.545;%
DarkGoldenrod,.72,.525,.044;%
DarkGray,.664,.664,.664;%
DarkGreen,0,.392,0;%
DarkGrey,.664,.664,.664;%
DarkKhaki,.74,.716,.42;%
DarkMagenta,.545,0,.545;%
DarkOliveGreen,.332,.42,.185;%
DarkOrange,1,.55,0;%
DarkOrchid,.6,.196,.8;%
DarkRed,.545,0,0;%
DarkSalmon,.912,.59,.48;%
DarkSeaGreen,.56,.736,.56;%
DarkSlateBlue,.284,.24,.545;%
DarkSlateGray,.185,.31,.31;%
DarkSlateGrey,.185,.31,.31;%
DarkTurquoise,0,.808,.82;%
DarkViolet,.58,0,.828;%
DeepPink,1,.08,.576;%
DeepSkyBlue,0,.75,1;%
DimGray,.41,.41,.41;%
DimGrey,.41,.41,.41;%
DodgerBlue,.116,.565,1;%
FireBrick,.698,.132,.132;%
FloralWhite,1,.98,.94;%
ForestGreen,.132,.545,.132;%
Fuchsia,1,0,1;%
Gainsboro,.864,.864,.864;%
GhostWhite,.972,.972,1;%
Gold,1,.844,0;%
Goldenrod,.855,.648,.125;%
Gray,.5,.5,.5;%
Grey,.5,.5,.5;%
Green,0,.5,0;%
GreenYellow,.68,1,.185;%
Honeydew,.94,1,.94;%
HotPink,1,.41,.705;%
IndianRed,.804,.36,.36;%
Indigo,.294,0,.51;%
Ivory,1,1,.94;%
Khaki,.94,.9,.55;%
Lavender,.9,.9,.98;%
LavenderBlush,1,.94,.96;%
LawnGreen,.488,.99,0;%
LemonChiffon,1,.98,.804;%
LightBlue,.68,.848,.9;%
LightCoral,.94,.5,.5;%
LightCyan,.88,1,1;%
LightGoldenrodYellow,.98,.98,.824;%
LightGray,.828,.828,.828;%
LightGreen,.565,.932,.565;%
LightGrey,.828,.828,.828;%
LightPink,1,.712,.756;%
LightSalmon,1,.628,.48;%
LightSeaGreen,.125,.698,.668;%
LightSkyBlue,.53,.808,.98;%
LightSlateGray,.468,.532,.6;%
LightSlateGrey,.468,.532,.6;%
LightSteelBlue,.69,.77,.87;%
LightYellow,1,1,.88;%
Lime,0,1,0;%
LimeGreen,.196,.804,.196;%
Linen,.98,.94,.9;%
Magenta,1,0,1;%
Maroon,.5,0,0;%
MediumAquamarine,.4,.804,.668;%
MediumBlue,0,0,.804;%
MediumOrchid,.73,.332,.828;%
MediumPurple,.576,.44,.86;%
MediumSeaGreen,.235,.7,.444;%
MediumSlateBlue,.484,.408,.932;%
MediumSpringGreen,0,.98,.604;%
MediumTurquoise,.284,.82,.8;%
MediumVioletRed,.78,.084,.52;%
MidnightBlue,.098,.098,.44;%
MintCream,.96,1,.98;%
MistyRose,1,.894,.884;%
Moccasin,1,.894,.71;%
NavajoWhite,1,.87,.68;%
Navy,0,0,.5;%
OldLace,.992,.96,.9;%
Olive,.5,.5,0;%
OliveDrab,.42,.556,.136;%
Orange,1,.648,0;%
OrangeRed,1,.27,0;%
Orchid,.855,.44,.84;%
PaleGoldenrod,.932,.91,.668;%
PaleGreen,.596,.985,.596;%
PaleTurquoise,.688,.932,.932;%
PaleVioletRed,.86,.44,.576;%
PapayaWhip,1,.936,.835;%
PeachPuff,1,.855,.725;%
Peru,.804,.52,.248;%
Pink,1,.752,.796;%
Plum,.868,.628,.868;%
PowderBlue,.69,.88,.9;%
Purple,.5,0,.5;%
Red,1,0,0;%
RosyBrown,.736,.56,.56;%
RoyalBlue,.255,.41,.884;%
SaddleBrown,.545,.27,.075;%
Salmon,.98,.5,.448;%
SandyBrown,.956,.644,.376;%
SeaGreen,.18,.545,.34;%
Seashell,1,.96,.932;%
Sienna,.628,.32,.176;%
Silver,.752,.752,.752;%
SkyBlue,.53,.808,.92;%
SlateBlue,.415,.352,.804;%
SlateGray,.44,.5,.565;%
SlateGrey,.44,.5,.565;%
Snow,1,.98,.98;%
SpringGreen,0,1,.498;%
SteelBlue,.275,.51,.705;%
Tan,.824,.705,.55;%
Teal,0,.5,.5;%
Thistle,.848,.75,.848;%
Tomato,1,.39,.28;%
Turquoise,.25,.88,.815;%
Violet,.932,.51,.932;%
Wheat,.96,.87,.7;%
White,1,1,1;%
WhiteSmoke,.96,.96,.96;%
Yellow,1,1,0;%
YellowGreen,.604,.804,.196}

\definecolorset{rgb}{}{}{%
AntiqueWhite1,1,.936,.86;%
AntiqueWhite2,.932,.875,.8;%
AntiqueWhite3,.804,.752,.69;%
AntiqueWhite4,.545,.512,.47;%
Aquamarine1,.498,1,.83;%
Aquamarine2,.464,.932,.776;%
Aquamarine3,.4,.804,.668;%
Aquamarine4,.27,.545,.455;%
Azure1,.94,1,1;%
Azure2,.88,.932,.932;%
Azure3,.756,.804,.804;%
Azure4,.512,.545,.545;%
Bisque1,1,.894,.77;%
Bisque2,.932,.835,.716;%
Bisque3,.804,.716,.62;%
Bisque4,.545,.49,.42;%
Blue1,0,0,1;%
Blue2,0,0,.932;%
Blue3,0,0,.804;%
Blue4,0,0,.545;%
Brown1,1,.25,.25;%
Brown2,.932,.23,.23;%
Brown3,.804,.2,.2;%
Brown4,.545,.136,.136;%
Burlywood1,1,.828,.608;%
Burlywood2,.932,.772,.57;%
Burlywood3,.804,.668,.49;%
Burlywood4,.545,.45,.332;%
CadetBlue1,.596,.96,1;%
CadetBlue2,.556,.898,.932;%
CadetBlue3,.48,.772,.804;%
CadetBlue4,.325,.525,.545;%
Chartreuse1,.498,1,0;%
Chartreuse2,.464,.932,0;%
Chartreuse3,.4,.804,0;%
Chartreuse4,.27,.545,0;%
Chocolate1,1,.498,.14;%
Chocolate2,.932,.464,.13;%
Chocolate3,.804,.4,.112;%
Chocolate4,.545,.27,.075;%
Coral1,1,.448,.336;%
Coral2,.932,.415,.312;%
Coral3,.804,.356,.27;%
Coral4,.545,.244,.185;%
Cornsilk1,1,.972,.864;%
Cornsilk2,.932,.91,.804;%
Cornsilk3,.804,.785,.694;%
Cornsilk4,.545,.532,.47;%
Cyan1,0,1,1;%
Cyan2,0,.932,.932;%
Cyan3,0,.804,.804;%
Cyan4,0,.545,.545;%
DarkGoldenrod1,1,.725,.06;%
DarkGoldenrod2,.932,.68,.055;%
DarkGoldenrod3,.804,.585,.048;%
DarkGoldenrod4,.545,.396,.03;%
DarkOliveGreen1,.792,1,.44;%
DarkOliveGreen2,.736,.932,.408;%
DarkOliveGreen3,.635,.804,.352;%
DarkOliveGreen4,.43,.545,.24;%
DarkOrange1,1,.498,0;%
DarkOrange2,.932,.464,0;%
DarkOrange3,.804,.4,0;%
DarkOrange4,.545,.27,0;%
DarkOrchid1,.75,.244,1;%
DarkOrchid2,.698,.228,.932;%
DarkOrchid3,.604,.196,.804;%
DarkOrchid4,.408,.132,.545;%
DarkSeaGreen1,.756,1,.756;%
DarkSeaGreen2,.705,.932,.705;%
DarkSeaGreen3,.608,.804,.608;%
DarkSeaGreen4,.41,.545,.41;%
DarkSlateGray1,.592,1,1;%
DarkSlateGray2,.552,.932,.932;%
DarkSlateGray3,.475,.804,.804;%
DarkSlateGray4,.32,.545,.545;%
DeepPink1,1,.08,.576;%
DeepPink2,.932,.07,.536;%
DeepPink3,.804,.064,.464;%
DeepPink4,.545,.04,.312;%
DeepSkyBlue1,0,.75,1;%
DeepSkyBlue2,0,.698,.932;%
DeepSkyBlue3,0,.604,.804;%
DeepSkyBlue4,0,.408,.545;%
DodgerBlue1,.116,.565,1;%
DodgerBlue2,.11,.525,.932;%
DodgerBlue3,.094,.455,.804;%
DodgerBlue4,.064,.305,.545;%
Firebrick1,1,.19,.19;%
Firebrick2,.932,.172,.172;%
Firebrick3,.804,.15,.15;%
Firebrick4,.545,.1,.1;%
Gold1,1,.844,0;%
Gold2,.932,.79,0;%
Gold3,.804,.68,0;%
Gold4,.545,.46,0;%
Goldenrod1,1,.756,.145;%
Goldenrod2,.932,.705,.132;%
Goldenrod3,.804,.608,.112;%
Goldenrod4,.545,.41,.08;%
Green1,0,1,0;%
Green2,0,.932,0;%
Green3,0,.804,0;%
Green4,0,.545,0;%
Honeydew1,.94,1,.94;%
Honeydew2,.88,.932,.88;%
Honeydew3,.756,.804,.756;%
Honeydew4,.512,.545,.512;%
HotPink1,1,.43,.705;%
HotPink2,.932,.415,.655;%
HotPink3,.804,.376,.565;%
HotPink4,.545,.228,.385;%
IndianRed1,1,.415,.415;%
IndianRed2,.932,.39,.39;%
IndianRed3,.804,.332,.332;%
IndianRed4,.545,.228,.228;%
Ivory1,1,1,.94;%
Ivory2,.932,.932,.88;%
Ivory3,.804,.804,.756;%
Ivory4,.545,.545,.512;%
Khaki1,1,.965,.56;%
Khaki2,.932,.9,.52;%
Khaki3,.804,.776,.45;%
Khaki4,.545,.525,.305;%
LavenderBlush1,1,.94,.96;%
LavenderBlush2,.932,.88,.898;%
LavenderBlush3,.804,.756,.772;%
LavenderBlush4,.545,.512,.525;%
LemonChiffon1,1,.98,.804;%
LemonChiffon2,.932,.912,.75;%
LemonChiffon3,.804,.79,.648;%
LemonChiffon4,.545,.536,.44;%
LightBlue1,.75,.936,1;%
LightBlue2,.698,.875,.932;%
LightBlue3,.604,.752,.804;%
LightBlue4,.408,.512,.545;%
LightCyan1,.88,1,1;%
LightCyan2,.82,.932,.932;%
LightCyan3,.705,.804,.804;%
LightCyan4,.48,.545,.545;%
LightGoldenrod1,1,.925,.545;%
LightGoldenrod2,.932,.864,.51;%
LightGoldenrod3,.804,.745,.44;%
LightGoldenrod4,.545,.505,.298;%
LightPink1,1,.684,.725;%
LightPink2,.932,.635,.68;%
LightPink3,.804,.55,.585;%
LightPink4,.545,.372,.396;%
LightSalmon1,1,.628,.48;%
LightSalmon2,.932,.585,.448;%
LightSalmon3,.804,.505,.385;%
LightSalmon4,.545,.34,.26;%
LightSkyBlue1,.69,.888,1;%
LightSkyBlue2,.644,.828,.932;%
LightSkyBlue3,.552,.712,.804;%
LightSkyBlue4,.376,.484,.545;%
LightSteelBlue1,.792,.884,1;%
LightSteelBlue2,.736,.824,.932;%
LightSteelBlue3,.635,.71,.804;%
LightSteelBlue4,.43,.484,.545;%
LightYellow1,1,1,.88;%
LightYellow2,.932,.932,.82;%
LightYellow3,.804,.804,.705;%
LightYellow4,.545,.545,.48;%
Magenta1,1,0,1;%
Magenta2,.932,0,.932;%
Magenta3,.804,0,.804;%
Magenta4,.545,0,.545;%
Maroon1,1,.204,.7;%
Maroon2,.932,.19,.655;%
Maroon3,.804,.16,.565;%
Maroon4,.545,.11,.385;%
MediumOrchid1,.88,.4,1;%
MediumOrchid2,.82,.372,.932;%
MediumOrchid3,.705,.32,.804;%
MediumOrchid4,.48,.215,.545;%
MediumPurple1,.67,.51,1;%
MediumPurple2,.624,.475,.932;%
MediumPurple3,.536,.408,.804;%
MediumPurple4,.365,.28,.545;%
MistyRose1,1,.894,.884;%
MistyRose2,.932,.835,.824;%
MistyRose3,.804,.716,.71;%
MistyRose4,.545,.49,.484;%
NavajoWhite1,1,.87,.68;%
NavajoWhite2,.932,.81,.63;%
NavajoWhite3,.804,.7,.545;%
NavajoWhite4,.545,.475,.37;%
OliveDrab1,.752,1,.244;%
OliveDrab2,.7,.932,.228;%
OliveDrab3,.604,.804,.196;%
OliveDrab4,.41,.545,.132;%
Orange1,1,.648,0;%
Orange2,.932,.604,0;%
Orange3,.804,.52,0;%
Orange4,.545,.352,0;%
OrangeRed1,1,.27,0;%
OrangeRed2,.932,.25,0;%
OrangeRed3,.804,.215,0;%
OrangeRed4,.545,.145,0;%
Orchid1,1,.512,.98;%
Orchid2,.932,.48,.912;%
Orchid3,.804,.41,.79;%
Orchid4,.545,.28,.536;%
PaleGreen1,.604,1,.604;%
PaleGreen2,.565,.932,.565;%
PaleGreen3,.488,.804,.488;%
PaleGreen4,.33,.545,.33;%
PaleTurquoise1,.732,1,1;%
PaleTurquoise2,.684,.932,.932;%
PaleTurquoise3,.59,.804,.804;%
PaleTurquoise4,.4,.545,.545;%
PaleVioletRed1,1,.51,.67;%
PaleVioletRed2,.932,.475,.624;%
PaleVioletRed3,.804,.408,.536;%
PaleVioletRed4,.545,.28,.365;%
PeachPuff1,1,.855,.725;%
PeachPuff2,.932,.796,.68;%
PeachPuff3,.804,.688,.585;%
PeachPuff4,.545,.468,.396;%
Pink1,1,.71,.772;%
Pink2,.932,.664,.72;%
Pink3,.804,.57,.62;%
Pink4,.545,.39,.424;%
Plum1,1,.732,1;%
Plum2,.932,.684,.932;%
Plum3,.804,.59,.804;%
Plum4,.545,.4,.545;%
Purple1,.608,.19,1;%
Purple2,.57,.172,.932;%
Purple3,.49,.15,.804;%
Purple4,.332,.1,.545;%
Red1,1,0,0;%
Red2,.932,0,0;%
Red3,.804,0,0;%
Red4,.545,0,0;%
RosyBrown1,1,.756,.756;%
RosyBrown2,.932,.705,.705;%
RosyBrown3,.804,.608,.608;%
RosyBrown4,.545,.41,.41;%
RoyalBlue1,.284,.464,1;%
RoyalBlue2,.264,.43,.932;%
RoyalBlue3,.228,.372,.804;%
RoyalBlue4,.152,.25,.545;%
Salmon1,1,.55,.41;%
Salmon2,.932,.51,.385;%
Salmon3,.804,.44,.33;%
Salmon4,.545,.298,.224;%
SeaGreen1,.33,1,.624;%
SeaGreen2,.305,.932,.58;%
SeaGreen3,.264,.804,.5;%
SeaGreen4,.18,.545,.34;%
Seashell1,1,.96,.932;%
Seashell2,.932,.898,.87;%
Seashell3,.804,.772,.75;%
Seashell4,.545,.525,.51;%
Sienna1,1,.51,.28;%
Sienna2,.932,.475,.26;%
Sienna3,.804,.408,.224;%
Sienna4,.545,.28,.15;%
SkyBlue1,.53,.808,1;%
SkyBlue2,.494,.752,.932;%
SkyBlue3,.424,.65,.804;%
SkyBlue4,.29,.44,.545;%
SlateBlue1,.512,.435,1;%
SlateBlue2,.48,.404,.932;%
SlateBlue3,.41,.35,.804;%
SlateBlue4,.28,.235,.545;%
SlateGray1,.776,.888,1;%
SlateGray2,.725,.828,.932;%
SlateGray3,.624,.712,.804;%
SlateGray4,.424,.484,.545;%
Snow1,1,.98,.98;%
Snow2,.932,.912,.912;%
Snow3,.804,.79,.79;%
Snow4,.545,.536,.536;%
SpringGreen1,0,1,.498;%
SpringGreen2,0,.932,.464;%
SpringGreen3,0,.804,.4;%
SpringGreen4,0,.545,.27;%
SteelBlue1,.39,.72,1;%
SteelBlue2,.36,.675,.932;%
SteelBlue3,.31,.58,.804;%
SteelBlue4,.21,.392,.545;%
Tan1,1,.648,.31;%
Tan2,.932,.604,.288;%
Tan3,.804,.52,.248;%
Tan4,.545,.352,.17;%
Thistle1,1,.884,1;%
Thistle2,.932,.824,.932;%
Thistle3,.804,.71,.804;%
Thistle4,.545,.484,.545;%
Tomato1,1,.39,.28;%
Tomato2,.932,.36,.26;%
Tomato3,.804,.31,.224;%
Tomato4,.545,.21,.15;%
Turquoise1,0,.96,1;%
Turquoise2,0,.898,.932;%
Turquoise3,0,.772,.804;%
Turquoise4,0,.525,.545;%
VioletRed1,1,.244,.59;%
VioletRed2,.932,.228,.55;%
VioletRed3,.804,.196,.47;%
VioletRed4,.545,.132,.32;%
Wheat1,1,.905,.73;%
Wheat2,.932,.848,.684;%
Wheat3,.804,.73,.59;%
Wheat4,.545,.494,.4;%
Yellow1,1,1,0;%
Yellow2,.932,.932,0;%
Yellow3,.804,.804,0;%
Yellow4,.545,.545,0;%
Gray0,.333,.333,.333;%
Green0,0,1,0;%
Grey0,.745,.745,.745;%
Maroon0,.69,.19,.376;%
Purple0,.628,.125,.94}

\def\genbox#1#2#3#4#5#6{% #1=0/1, #2=color, #3=shape, #4=raise, #5=width, #6=width/2
    \leavevmode\raise#4bp\hbox to#5bp{\vrule height#5bp depth0bp width0bp
    \pdfliteral{q .5 w \csname #2COLOR\endcsname\space RG
                       \csname #3PDF\endcsname{#5}{#6} S Q
             \ifx1#1 q \csname #2COLOR\endcsname\space rg 
                       \csname #3PDF\endcsname{#5}{#6} f Q\fi}\hss}}

\titlespacing{\section}{0pt}{0.4ex}{0.4ex}
\titlespacing{\subsection}{0pt}{0.2ex}{0ex}
\titlespacing{\subsubsection}{0pt}{0.0ex}{0ex}

\renewcommand\arraystretch{0.8}

\DeclareMathSizes{10}{8.5}{7}{6}
\DeclareMathSizes{9.8}{8.5}{7}{6}
\DeclareMathSizes{9.6}{8.5}{7}{6}
\DeclareMathSizes{9.5}{8.5}{7}{6}
\DeclareMathSizes{9.4}{8.5}{7}{6}
\DeclareMathSizes{9.2}{8.5}{7}{6}
\DeclareMathSizes{9.1}{8.5}{7}{6}
\DeclareMathSizes{9}{8.5}{7}{6}

\makeatletter
\IEEEtriggercmd{\reset@font\normalfont\fontsize{7.6pt}{8pt}\selectfont}
\makeatother
\IEEEtriggeratref{1}

\makeatletter
\newcommand*\rel@kern[1]{\kern#1\dimexpr\macc@kerna}
\newcommand*\widebar[1]{%
  \begingroup
  \def\mathaccent##1##2{%
    \rel@kern{0.8}%
    \overline{\rel@kern{-0.8}\macc@nucleus\rel@kern{0.2}}%
    \rel@kern{-0.2}%
  }%
  \macc@depth\@ne
  \let\math@bgroup\@empty \let\math@egroup\macc@set@skewchar
  \mathsurround\z@ \frozen@everymath{\mathgroup\macc@group\relax}%
  \macc@set@skewchar\relax
  \let\mathaccentV\macc@nested@a
  \macc@nested@a\relax111{#1}%
  \endgroup
}
\makeatother

\everypar{\looseness=-1}

\DeclareMathSizes{10}{8}{7}{5}

\begin{document} 

\bstctlcite{IEEEexample:BSTcontrol}

\title{\huge{Radio Resource Management Design for RSMA: Optimization of Beamforming, User Admission, and Discrete/Continuous Rates with Imperfect SIC}}

\vspace{-3mm}

{
{

\author{
\IEEEauthorblockN{Luis F. Abanto-Leon\IEEEauthorrefmark{2}, 
Aravindh Krishnamoorthy\IEEEauthorrefmark{3}, 
Andres Garcia-Saavedra\IEEEauthorrefmark{4}, 
Gek Hong Sim\IEEEauthorrefmark{2}, \\
Robert Schober\IEEEauthorrefmark{3}, and 
Matthias Hollick\IEEEauthorrefmark{2}} \\ 
\IEEEauthorblockA{\IEEEauthorrefmark{2}Secure Mobile Networking Lab, Technische Universit\"{a}t Darmstadt, Germany} 
\IEEEauthorblockA{\IEEEauthorrefmark{3}Friedrich-Alexander-Universit\"{a}t Erlangen-N\"{u}rnberg, Germany~~}
\IEEEauthorblockA{\IEEEauthorrefmark{4}NEC Laboratories Europe, Germany}
}
}
}

% The paper headers
%\markboth{IEEE Transactions on Mobile Computing,~Vol.~33, No.~27, Feb~2024}%
%{Shell \MakeLowercase{\textit{et al.}}: Bare Demo of IEEEtran.cls for IEEE Communications Society Journals}

\IEEEtitleabstractindextext
{%
\begin{abstract}

		\noindent This paper investigates the radio resource management (RRM) design for multiuser rate-splitting multiple access (RSMA), accounting for various characteristics of practical wireless systems, such as the use of discrete rates, the inability to serve all users, and the imperfect successive interference cancellation (SIC). Specifically, failure to consider these characteristics in RRM design may lead to inefficient use of radio resources. Therefore, we formulate the RRM of RSMA as optimization problems to maximize respectively the weighted sum rate (WSR) and weighted energy efficiency (WEE), and jointly optimize the beamforming, user admission, discrete/continuous rates, accounting for imperfect SIC, which result in nonconvex mixed-integer nonlinear programs that are challenging to solve. Despite the difficulty of the optimization problems, we develop algorithms that can find high-quality solutions. We show via simulations that carefully accounting for the aforementioned characteristics, can lead to significant gains. Precisely, by considering that transmission rates are discrete, the transmit power can be utilized more intelligently, allocating just enough power to guarantee a given discrete rate. Additionally, we reveal that user admission plays a crucial role in RSMA, enabling additional gains compared to random admission by facilitating the servicing of selected users with mutually beneficial channel characteristics. Furthermore, provisioning for possibly imperfect SIC makes RSMA more robust and reliable. 

\end{abstract}
\vspace*{-3mm}
\begin{IEEEkeywords}
	Beamforming, user admission, discrete rates, rate splitting, imperfect SIC, spectral efficiency, energy efficiency.
\end{IEEEkeywords}
}

% make the title area
\maketitle
\IEEEdisplaynontitleabstractindextext
\vspace*{-0.8cm}

\section*{Nomenclature} \label{section_nomenclature}

\addcontentsline{toc}{section}{Acronyms}

\begin{IEEEdescription}[\IEEEusemathlabelsep\IEEEsetlabelwidth{$V_1,V_2,V_3$}]
\item[BnB] Branch-and-bound
%\item[DPC] Dirty paper coding
\item[EE] Energy efficiency
\item[IPM] Interior-point method
%\item[KKT] Karush-Kuhn-Tucker
\item[MCS] Modulation and coding scheme
\item[MINLP] Mixed-integer nonlinear program
\item[MISOCP] Mixed-integer second-order cone program
\item[NOMA] Non-orthogonal multiple access
\item[NOUM] Non-orthogonal unicast and multicast
\item[RRM] Radio resource management
\item[RSMA] Rate-splitting multiple access
\item[SCA] Successive convex approximation
\item[SDMA] Space-division multiple access
\item[SDR] Semidefinite relaxation
\item[SE] Spectral efficiency
\item[SIC] Successive interference cancellation
\item[SINR] Signal-to-interference-plus-noise ratio
\item[SOCP] Second-order cone program
\item[SR] Sum rate
\item[SSR] Sum secrecy rate
\item[WEE] Weighted energy efficiency
\item[WSR] Weighted sum rate
%\item[CQI] Channel quality indicator
%item[BLER] Block error rate
%\item[BS] Base station
%\item[UE] User equipment
%\item[LOS] Line-of-sight
%\item[NLOS] Non-line-of-sight
\end{IEEEdescription}

% Introduction
\section{Introduction} \label{section_introduction}

Rate-splitting multiple access (RSMA) has emerged as a promising technology capable of outperforming non-orthogonal multiple access (NOMA) and space-division multiple access (SDMA), owing to its superior ability to cope with multiuser interference \cite{clerckx2023:a-primer-rate-splitting-multiple-access-tutorial-myths-frequently-asked-questions, clerckx2021:is-noma-efficient-multi-antenna-networks-a-critical-look-next-generation-multiple-access-techniques, mao2018:rsma-downlink-communication-systems-bridging-generalizing-outperforming-sdma-noma}. RSMA is a power-domain non-orthogonal technology that relies on multi-antenna rate-splitting at the transmitter and successive interference cancellation (SIC) at the user side. Specifically, the transmitter partitions the message for each user into a common and a private portion. Then, it encodes the common portions of all users into a common stream and each of the private portions into an independent stream. Prior to over-the-air transmission, the common and private streams are precoded by the transmitter. Upon reception of the streams, each user employs SIC to decode and remove the common stream, which carries the common portions of other users and its own, before accessing the private streams to decode its private portion. With both the common and private portions available, the user can reassemble its message. By adjusting the partitioning of messages into common and private portions, RSMA flexibly controls the level of interference that each user can cancel, thus bridging smoothly between the two extreme strategies of fully decoding interference (as in NOMA) and fully treating it as noise (as in SDMA), leading to further performance gains \cite{3gpp.38.812, mao2018:rsma-downlink-communication-systems-bridging-generalizing-outperforming-sdma-noma, shi2008:rate-optimization-multiuser-mimo-systems-linear-processing}.

% A key element to ensure high RSMA performance is the radio resource management (RRM) design.
To date, several studies have demonstrated in various use cases \cite{mao2022:rate-splitting-multiple-access-fundamentals-survey-future-research-trends} the higher capabilities of RSMA compared to SDMA and NOMA, thus positioning RSMA as a formidable multiple access candidate with enormous potential to meet the stringent connectivity requirements of next-generation wireless communications systems \cite{liu2022:evolution-noma-toward-next-generation-multiple-access-6g}. Still, a key element that needs further advances to ensure high RSMA performance is the radio resource management (RRM) design. The RRM for RSMA systems has focused chiefly on the beamforming and power allocation design, which have been investigated for a plethora of use cases and different design goals, such as fairness, sum secrecy rate (SSR), sum rate (SR), weighted SR (WSR), and weighted energy efficiency (WEE)\footnote{In our work, EE (WEE) is defined as the ratio of SR (WSR) to total power consumption, with the goal being its maximization. An alternative but less commonly used definition of EE is expressed as the total energy expenditure, with the goal being its minimization \cite{reifert2022:energy-efficiency-rate-splitting-multiple-access-mixed-criticality}.} optimization, as summarized in the following.

The authors of \cite{xu2021:rate-splitting-multiple-access-multi-antenna-joint-radar-communications} studied the beamforming design for WSR maximization in joint radar and communications (JRC), whereas the authors of \cite{naser2022:interference-management-strategies-multiuser-multicell-mimo-vlc-systems} developed cooperative beamforming strategies to maximize the WSR in visible light communications (VLC). The authors of \cite{abanto2022:sequential-parametric-optimization-rate-splitting-precoding-nonorthogonal-unicast-multicast-transmissions, mao2018:rate-splitting-multi-antenna-non-orthogonal-unicast-multicast-transmission} investigated the beamforming design for WSR maximization in non-orthogonal unicast and multicast (NOUM) systems. In contrast, the security aspect was investigated in \cite{fu2021:secrecy-outage-constrained-robust-resource-allocation-design-mu-miso-rsma-systems}, where the beamforming and artificial noise were designed to maximize the secrecy rate fairness. Besides, the beamforming design for rate fairness maximization was investigated in \cite{pang2023:resource-allocation-rsma-based-coordinated-direct-relay-transmission} in the context of cooperative relaying networks. Driven by the same goal, the joint design of beamforming and the phase shifts of an intelligent reflecting surface (IRS) was researched in \cite{fu2021:resource-allocation-design-irs-aided-downlink-mu-miso-rsma-systems}. In addition, the beamforming design for WEE maximization was studied for unmanned aerial vehicle networks in \cite{rahmati2019:energy-efficiency-rsma-noma-cellular-connected-mmwave-uav-networks}, for VLC in \cite{xing2022:energy-efficiency-optimization-rate-splitting-multiple-access-based-indoor-visible-light-communication-networks}, for semantic communications in \cite{yang2023:energy-efficient-semantic-communication-wireless-networks-rate-splitting}, and for JRC in \cite{dizdar2022:energy-efficient-dual-functional-radar-communication-rate-splitting-multiple-access-low-resolution-dacs-rf-chain-selection}. The authors of \cite{mao2019:rate-splitting-multi-antenna-non-orthogonal-unicast-multicast-transmission-spectral-energy-efficiency, matthiesen2022:globally-optimal-spectrum-energy-efficient-beamforming-rate-splitting-multiple-access} investigated the beamforming design for maximization of respectively the WSR and WEE, whereas the authors of \cite{zhou2022:rate-splitting-multiple-access-multi-antenna-downlink-communication-systems-spectral-energy-efficiency-tradeoff} designed the beamforming for simultaneous WSR and WEE maximization. Power allocation was investigated for SR, SSR, and rate fairness maximization in \cite{hieu2021:optimal-power-allocation-rate-splitting-communications-deep-reinforcement-learning, yang2021:optimization-rate-allocation-power-control-rate-splitting-multiple-access, flores2022:robust-adaptive-power-allocation-techniques-rate-splitting-based-mu--mimo-systems}, \cite{cai2021:resource-allocation-secure-rate-splitting-multiple-access-adaptive-beamforming}, and \cite{hieu2023:joint-power-allocation-rate-control-rate-splitting-multiple-access-networks-covert-communications}, respectively. \emph{The body of work on beamforming and power allocation design of RSMA continues to grow and show promising results. However, the literature has so far ignored critical characteristics that are inherent to practical wireless systems, i.e., rate discretization, user admission, and imperfect SIC, which merit investigation.}

Concerning the first characteristic, most of the literature, assumed continuous rates modeled by Shannon's capacity formula, e.g., \cite{clerckx2023:a-primer-rate-splitting-multiple-access-tutorial-myths-frequently-asked-questions, clerckx2021:is-noma-efficient-multi-antenna-networks-a-critical-look-next-generation-multiple-access-techniques, mao2018:rsma-downlink-communication-systems-bridging-generalizing-outperforming-sdma-noma,
mao2022:rate-splitting-multiple-access-fundamentals-survey-future-research-trends,
xu2021:rate-splitting-multiple-access-multi-antenna-joint-radar-communications,
naser2022:interference-management-strategies-multiuser-multicell-mimo-vlc-systems,
fu2021:resource-allocation-design-irs-aided-downlink-mu-miso-rsma-systems,
fu2021:secrecy-outage-constrained-robust-resource-allocation-design-mu-miso-rsma-systems,
rahmati2019:energy-efficiency-rsma-noma-cellular-connected-mmwave-uav-networks,
xing2022:energy-efficiency-optimization-rate-splitting-multiple-access-based-indoor-visible-light-communication-networks,
yang2023:energy-efficient-semantic-communication-wireless-networks-rate-splitting,
dizdar2022:energy-efficient-dual-functional-radar-communication-rate-splitting-multiple-access-low-resolution-dacs-rf-chain-selection,
mao2019:rate-splitting-multi-antenna-non-orthogonal-unicast-multicast-transmission-spectral-energy-efficiency, matthiesen2022:globally-optimal-spectrum-energy-efficient-beamforming-rate-splitting-multiple-access,
zhou2022:rate-splitting-multiple-access-multi-antenna-downlink-communication-systems-spectral-energy-efficiency-tradeoff,
hieu2021:optimal-power-allocation-rate-splitting-communications-deep-reinforcement-learning, yang2021:optimization-rate-allocation-power-control-rate-splitting-multiple-access, flores2022:robust-adaptive-power-allocation-techniques-rate-splitting-based-mu--mimo-systems,
cai2021:resource-allocation-secure-rate-splitting-multiple-access-adaptive-beamforming,
hieu2023:joint-power-allocation-rate-control-rate-splitting-multiple-access-networks-covert-communications}. This assumption contrasts with the predominant use of discrete rates in practical wireless systems and raises questions as to whether RSMA's gains will hold when discrete rates are accounted for. Transmission rates in wireless systems are determined by the choice of a modulation and coding scheme (MCS), as specified by 3GPP \cite{3gpp2020:38.214}, leading to a finite set of discrete rates. Employing Shannon's capacity formula is mathematically more tractable than assuming discrete rates, hence its wide adoption. However, it renders continuous-valued rate upper bounds that are unachievable in practice. A naive solution to this problem is to project the continuous rates onto the discrete rate set, i.e., round them to the closest feasible discrete rate \cite{sim2017:finite-horizon-opportunistic-multicast-beamforming}. However, this may lead to performance degradation. Therefore, rate discretization must be properly accounted for in the RRM design to exploit the full potential of RSMA. \emph{To date, the incorporation of discrete rates into the RRM design of RSMA remains to be investigated.} An early study in \cite{dizdar2020:rsma-downlink-multiantenna-communications-physical-layer-design-link-level-simulations} showed that RSMA outperforms SDMA for discrete rates. The proposed design, however, did not account for predefined MCSs, as the authors assumed continuous rates and tailored the MCSs to achieve a SR close to the ensemble average obtained over multiple channel realizations. A few works investigated the impact of beamforming and discrete rates on the performance of SDMA. For instance, joint beamforming and discrete rate selection design was investigated in \cite{cheng2012:dynamic-rate-adaptation-multiuser-beamforming-mixed-integer-conic-programming,  cheng2015:joint-discrete-rate-adaptation-beamforming-mixed-integer-conic-programming} for SR maximization and in \cite{wai2016:discrete-sum-rate-maximization-miso-interference-broadcast-channels-convex-approximations-efficient-algorithms} for WSR maximization. However, these results are not applicable to RSMA, since RSMA is a more general framework that includes SDMA as a special case.

Concerning the second characteristic, access policies in wireless systems typically restrict the number of users served per time slot, e.g., due to the availability of a limited number of radio frequency (RF) chains. This characteristic is especially limiting for SDMA, which requires one RF chain per served user. In contrast, RSMA can support more users than RF chains are available since RSMA can exploit the multicast signal to aggregate information for several users. However, in this case, RSMA may degrade severely as the number of users increases since the multicast signal must be delivered to all users. This peculiarity becomes a limitation in RSMA and raises the need for selective user admission. \emph{The impact of user admission on performance has been studied for SDMA and NOMA, but has yet to be researched for RSMA.} For instance, the joint beamforming and user admission design for SDMA was investigated in \cite{ni2018:mixed-integer-semidefinite-relaxation-joint-admission-control-beamforming-soc-outer-approximation-provable-guarantees} to minimize the transmit power, and in \cite{bandi2020:joint-solution-scheduling-precoding-multiuser-miso} to maximize the SR and rate fairness. The authors of \cite{matskani2008:convex-approximation-techniques-joint-multiuser-downlink-beamforming-admission-control} developed joint beamforming and user admission strategies for SDMA to maximize the number of users served. With the same goal, the authors of \cite{wang2020:admission-control-power-allocation-noma-satellite-multibeam-network} designed the power and subchannel allocation for NOMA. Besides, the joint design of beamforming, user admission, and discrete rate selection for SR maximization of SDMA was investigated in \cite{abanto2022:radiorchestra-proactive-management-millimeter-wave-self-backhauled-small-cells-joint-optimization-beamforming-user-association-rate-selection-admission-control}. However, the solutions developed in the preceding studies are not valid for RSMA. Specifically, user admission for RSMA differs significantly from that for SDMA and NOMA, as RSMA delivers information to users via superimposed multicast and unicast precoders. Unicast precoders benefit from users with uncorrelated channels because interference is easier to mitigate. In contrast, the multicast precoder benefits from users with correlated channels as this facilitates transmitting shared information with less power. Therefore, given these conflicting objectives inherent to RSMA, including user admission in the RRM design is essential.

Finally, as the last key characteristic, it is important to account for imperfect SIC. Specifically, the performance of RSMA is highly dependent on the success of SIC. In practice, SIC is generally not perfect, which causes unmanaged self-interference that can compromise performance. \emph{Despite the importance of accounting for imperfect SIC in the RRM design of RSMA, SIC has been assumed to be perfect in most of the RSMA literature, with few exceptions.} For instance, the authors of \cite{ou2023:resource-allocation-mu-miso-rate-splitting-multiple-access-sic-errors-urllc-services} investigated the beamforming and subcarrier allocation design for SR maximization taking into account imperfect SIC. However, the proposed design assumed continuous rates and did not consider user admission. The impact of imperfect SIC on the SR of RSMA was also investigated in \cite{chopra2022:bounds-power-common-message-fractions-rsma-imperfect-sic}, where the authors derived bounds for power allocation but did not take user admission and discrete rates into account. NOMA can also be affected by imperfect SIC and, therefore, a number of works have investigated its impact. In particular, for NOMA, the impact of imperfect SIC and power allocation was studied in \cite{mouni2021:adaptive-user-pairing-noma-systems-imperfect-sic, mahady2019:sum-rate-maximization-noma-systems-imperfect-successive-interference-cancellation, wang2019:low-complexity-power-allocation-noma-systems-imperfect-sic-maximizing-weighted-sum-rate} for SR maximization, and in \cite{wang2017:energy-efficient-power-allocation-nonorthogonal-multiple-access-imperfect-successive-interference-cancellation} for EE maximization.

Motivated by the above discussion, the performance of RSMA systems in real-world deployments is anticipated to improve significantly if the characteristics above are taken into account. Thus, we propose to account for discrete rates, user admission, and imperfect SIC during RRM design. To the best of the authors' knowledge, RRM design for RSMA systems considering these characteristics has not been investigated yet. Due to the widespread adoption of Shannon's capacity formula for RRM design, we also investigate the integration of continuous rates, user admission, and imperfect SIC, which has not been studied before for RSMA systems. We adopt the maximization of the WSR and WEE as design objectives. This paper makes the following contributions:
\begin{itemize} [leftmargin = 0.3cm]

	\item We formulate two novel RRM problems to maximize respectively the WSR and WEE of RSMA by jointly optimizing the beamforming, user admission, and private and common discrete rates, while accounting for imperfect SIC. The resulting WSR and WEE problems, denoted by $ \mathcal{P}_\mathrm{DWSR}' $ and $ \mathcal{P}_\mathrm{DWEE}' $ in Section~\ref{subsection_problem_formulation_discrete_rates}, are nonconvex mixed-integer nonlinear programs (MINLPs) and are difficult to solve. In addition, we formulate problems $ \mathcal{Q}_\mathrm{CWSR}' $ and $ \mathcal{Q}_\mathrm{CWEE}' $ in Section~\ref{subsection_problem_formulation_continuous_rates}, which represent the continuous-rate counterparts of $ \mathcal{P}_\mathrm{DWSR}' $ and $ \mathcal{P}_\mathrm{DWEE}' $, and are also nonconvex MINLPs.

	\item In Section~\ref{subsection_proposed_algorithm_discrete_rates}, we propose an optimal mixed-integer second-order cone program (\texttt{OPT-MISOCP}) algorithm, which tackles the nonconvexities of $ \mathcal{P}_\mathrm{DWSR}' $ and $ \mathcal{P}_\mathrm{DWEE}' $ through a series of convex transformations. Specifically, instead of treating $ \mathcal{P}_\mathrm{DWSR}' $ and $ \mathcal{P}_\mathrm{DWEE}' $ as general nonconvex MINLPs, \texttt{OPT-MISOCP} approximates $ \mathcal{P}_\mathrm{DWSR}' $ and $ \mathcal{P}_\mathrm{DWEE}' $ as convex MINLPs $ \mathcal{P}_\mathrm{DWSR} $ and $ \mathcal{P}_\mathrm{DWEE} $, respectively, which can be solved in a globally optimal manner. However, circumventing the nonconvexities of $ \mathcal{P}_\mathrm{DWSR}' $ and $ \mathcal{P}_\mathrm{DWEE}' $ poses the risk of shrinking their feasible set, possibly resulting in a loss of optimality. Therefore, we derive an upper bound to evaluate the corresponding loss in performance and show via simulations that the globally optimal solutions for $ \mathcal{P}_\mathrm{DWSR} $ and $ \mathcal{P}_\mathrm{DWEE} $ result in near-optimal solutions for $ \mathcal{P}_\mathrm{DWSR}' $ and $ \mathcal{P}_\mathrm{DWEE}' $ with negligible degradation. In addition, the proposed \texttt{OPT-MISOCP} algorithm features customized cutting planes that reduce the runtime by a factor of $ 3 - 20 $ without impacting performance.

	\item In Section~\ref{subsection_proposed_algorithm_continuous_rates}, we solve $ \mathcal{Q}_\mathrm{CWSR}' $ and $ \mathcal{Q}_\mathrm{CWEE}' $ based on binary enumeration and convex transformations. In particular, we employ enumeration to list all possible user admission combinations, thus resulting in multiple subproblems. To solve each subproblem, we devise an optimal successive convex approximation with semidefinite relaxation (\texttt{OPT-SCA-SDR}) algorithm, which finds a Karush-Kuhn-Tucker (KKT) point. In addition, to comply with the finite set of discrete rates, the continuous rates obtained by \texttt{OPT-SCA-SDR} are projected, i.e., rounded to the closest feasible discrete rates.

	\item Our simulations show that RSMA designed for discrete rates achieves gains of up to $ 89.7\% $ (WSR) and $ 21.5\% $ (WEE) compared to projecting continuous rates. Also, user admission proves crucial for RSMA as it yields additional gains of up to $ 15.3\% $ (WSR) and $ 11.4\% $ (WEE) compared to random user admission when discrete rates are considered. Furthermore, accounting for imperfect SIC prevents severe performance degradation by mitigating the impact of self-interference. Overall, our simulation results reveal that accounting for characteristics of practical wireless systems in RRM of RSMA leads to improved exploitation of the radio resources, and therefore to higher spectral efficiency (SE) and energy efficiency (EE).

\end{itemize}

% Figure 1: System model
\begin{figure}[!t]
	% Figure 1a
 	\begin{subfigure}[b]{0.35\columnwidth}
		\begin{center}
			%\resizebox{!}{4cm}
			{%
			\includegraphics[width=1\textwidth]{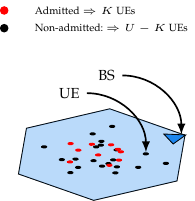}
			}
			\caption{System model}
			\label{figure_system_model_a}
		\end{center}
 	\end{subfigure}
 	\hspace{0.2cm}
 	\begin{subfigure}[b]{0.55\columnwidth}
		\begin{center}
			\resizebox{!}{4cm}
			{%
			\includegraphics[width=1\textwidth]{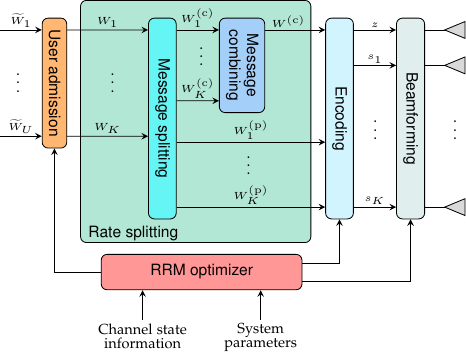}
			}
			\caption{RSMA with integrated  RRM optimizer}
			\label{figure_system_model_b}
		\end{center}
 	\end{subfigure}
	\caption{System model and RSMA with integrated RRM optimizer. In the system, $ K $ out of $ U $ UEs are admitted for downlink transmission. The messages for the admitted UEs are precoded via rate splitting and transmitted over the air.}
	\label{figure_system_model}
\end{figure}

The remainder of this paper is organized as follows. In Section~\ref{section_system_model}, we present the system model. In Section~\ref{section_problem_modeling_problem_formulation_proposed_algorithm_discrete_rates}, considering discrete rates, we formulate and solve the RRM as optimization problems for maximization of the WSR and WEE, respectively, while in Section~\ref{section_problem_modeling_problem_formulation_proposed_algorithm_continuous_rates}, we do the same for continuous rates. Simulation results are presented in Section~\ref{section_results}. Finally, we summarize our conclusions in Section~\ref{section_conclusions}.

\emph{Notation}: In this paper, $ \left| a \right| $ and the $ \left\| \mathbf{a} \right\|_2 $ represent the absolute value of scalar $ a $ and the $ \ell_2 $-norm of vector $ \mathbf{a} $, respectively. $ \mathbf{A}^\mathrm{T} $, $ \mathbf{A}^\mathrm{H} $, $ \mathrm{Rank} \left( \mathbf{A} \right) $, $ \mathrm{Tr} \left( \mathbf{A} \right) $, and $ \mathfrak{Re} \left\lbrace \mathbf{A} \right\rbrace $ and $ \mathfrak{Im} \left\lbrace \mathbf{A} \right\rbrace $ denote the transpose, Hermitian transpose, rank, trace, and real and imaginary part of matrix $ \mathbf{A} $, respectively. $ \mathbf{A} \succcurlyeq \mathbf{0} $ indicates that $ \mathbf{A} $ is a positive semidefinite matrix. $ \mathbb{C}^{N \times M} $ denotes the space of $ N \times M $ complex-valued matrices. $ \mathbf{I} $ is the identity matrix, $ j \triangleq \sqrt{-1} $ is the imaginary unit, and $ \mathbb{E} \left\lbrace \cdot \right\rbrace  $ denotes statistical expectation.

\section{System Model} \label{section_system_model}

In this section, we present the system model for the considered RRM optimization problems.

\subsection{System Architecture} \label{subsubsection_system_architecture}

We consider the downlink cellular system shown in Fig. \ref{figure_system_model_a}, where a base station (BS) is equipped with an antenna array with $ N_\mathrm{tx} $ elements, which can consume a maximum transmit power of $ P_\mathrm{tx}^\mathrm{max} $. There are $ U $ single-antenna user equipments (UEs), and the BS admits only $ K $ UEs, where $ K \leq U $. We index the UEs with the elements of set $ \mathcal{U} $, such that $ | \mathcal{U} | = U $. The UEs are distributed within a $ 120^{\circ} $ sector and are located at a maximum distance $ D_\mathrm{BS} $ from the BS. The BS estimates the channel state information (CSI) using uplink pilots exploiting channel reciprocity. The RRM optimizer at the BS uses the CSI and other system parameters, such as the maximum transmit power, to maximize the WSR or WEE.

\subsection{RSMA Principle} \label{subsubsection_rsma_principle}

RSMA allows for partial interference decoding, which facilitates the selective cancellation of interference components. Precisely, UEs focus on decoding their desired signals while coping only with specific interference components from other UEs. This strategy allows RSMA to achieve additional gains over NOMA, which attempts to decode interference entirely, and SDMA, which treats interference as noise and hence does attempt to decode it at all. Partial interference decoding in RSMA is accomplished by splitting the UEs' messages at the BS into common and private portions and sending them via multicast and unicast signals, respectively. As all UEs receive the multicast signal containing the common portions, each UE decodes information intended for other UEs. Precisely, by optimizing the partitioning of messages into common and private portions, RSMA dynamically adjusts the level of interference for each UE. This adaptive capability allows RSMA to seamlessly transition between fully treating interference as noise, as in SDMA, and fully decoding it, as in NOMA \cite{3gpp.38.812, mao2018:rsma-downlink-communication-systems-bridging-generalizing-outperforming-sdma-noma, shi2008:rate-optimization-multiuser-mimo-systems-linear-processing}. Although NOMA and RSMA both employ SIC, the former requires UEs to employ multiple SIC stages to sequentially decode interference from other UEs and thus recover their desired signals. Particularly, the number of SIC stages required increases with the number of UEs served. Furthermore, the order in which signals are decoded and removed can affect the performance and complexity of NOMA receivers. In contrast, regardless of the number of UEs, RSMA requires only one SIC stage as we consider single-layer RSMA \cite{clerckx2023:a-primer-rate-splitting-multiple-access-tutorial-myths-frequently-asked-questions}.

In the following, we explain the technical aspects of RSMA but exclude UE admission for ease of presentation. In Fig. \ref{figure_system_model_b}, every UE has a corresponding message denoted by $ \widetilde{W}_u $, $ u \in \mathcal{U} $, but only $ K $ UEs out of $ U $ are served. Thus, we assume that $ K $ UEs are pre-selected for downlink transmission, and denote this set of UEs by $ \mathcal{U}' $, such that $ | \mathcal{U}' | = K $, and by $ \mathsf{UE}_u $ the $ u $-th UE in $ \mathcal{U}' $. Now, every UE in $ \mathcal{U}' $ is served with a message denoted by $ W_u $, $ u \in \mathcal{U}' $, which is decomposed into two parts as $ W_u \triangleq \left( W^{(\mathrm{p})}_u, W^{(\mathrm{c})}_u \right) $, where $ W^{(\mathrm{p})}_u $ and $ W^{(\mathrm{c})}_u $ are respectively referred to as the private and common portions of $ W_u $. The private portion of $ \mathsf{UE}_u $ is encoded into a symbol $ s_u \in \mathbb{C} $ that is transmitted at a rate $ R^{(\mathrm{p})}_u $ in an unicast manner. On the other hand, the common portions $ W^{(\mathrm{c})}_u $ of all UEs are combined and encoded into a symbol $ s_0 \in \mathbb{C} $, which is transmitted at a rate $ R^\mathrm{(c)} $ in a multicast manner to all UEs. The symbols are assumed to be statistically independent, such that $ \mathbb{E} \left\lbrace \mathbf{s}^\mathrm{H} \mathbf{s} \right\rbrace = \mathbf{I} $ and $ \mathbf{s} = \left[ s_0, s_1, \dots, s_K \right]^\mathrm{T} \in \mathbb{C}^{(K + 1) \times 1} $. The rate portion of $ R^\mathrm{(c)} $ corresponding to $ \mathsf{UE}_u $ is denoted by $ C_u $, such that $ R^\mathrm{(c)} = \sum_u C_u $. As a result, $ \mathsf{UE}_u $ is served with an overall rate of $ R^{(\mathrm{p})}_u + C_u $. Each $ \mathsf{UE}_u $ recovers $ W^{(\mathrm{c})}_u $ first by decoding $ s_0 $ and then recovers $ W^{(\mathrm{p})}_u $ by decoding $ s_u $ with the assistance of SIC. With both portions $ W^{(\mathrm{c})}_u $ and $ W^{(\mathrm{p})}_u $ available at $ \mathsf{UE}_u $, the original message $ W_u $ can be reassembled. In addition, each $ \mathsf{UE}_u $ acquires the common portions $ W^{(\mathrm{c})}_{i \neq u} $, corresponding to the other UEs, which are used for interference decoding and cancellation. In particular, the size of the common portions $ W^{(\mathrm{c})}_u $ are adjusted according to the level of interference that can be canceled by the UEs \cite{clerckx2021:is-noma-efficient-multi-antenna-networks-a-critical-look-next-generation-multiple-access-techniques} and represents the amount of decodable interference, which is removed using SIC.

\subsection{Discrete and Continuous Rates} \label{subsubsection_discrete_continuous_rates}
Practical wireless systems, as defined, e.g., in cellular standards, support only a finite set of data rates \cite[p.~64]{3gpp2020:38.214}. These predefined rates are identified by their channel quality indicator (CQI) and correspond to specific MCSs. For each rate, a target SINR is required to ensure a given block error rate (BLER) \cite{leung2002:integrated-link-adaptation-power-control-improve-error-throughput-broadband-wireless-networks}. The rates and MCSs are typically standardized, e.g., by 3GPP. However, the target SINRs are specific to the equipment in use. We denote with $ J $ the total number of available MCSs supported by the system and with $ \mathcal{J} = \left\lbrace 1, \dots, J \right\rbrace $ the set that indexes them. Hence, for a given discrete rate $ R_j $, $ j \in \mathcal{J} $, there is a corresponding target SINR $ \Gamma_j $ that must be met to guarantee that rate. In addition, we assume that $ \mathcal{J} $ is an ordered set, such that $ R_{j+1} > R_j $ and $ \Gamma_{j+1} > \Gamma_j $. Thus, if an UE achieves an SINR of $ \widebar{\Gamma} $, the BS can allocate any discrete rate $ \widebar{R}_\mathrm{dis} \triangleq \left\lbrace R_j \mid \Gamma_j \leq \widebar{\Gamma}, j \in \mathcal{J} \right\rbrace  $ to the UE. On the contrary, when Shannon's capacity formula is used for rate allocation, the BS assigns continuous rate $ \widebar{R}_\mathrm{con} \triangleq \log_2 \left( 1 + \widebar{\Gamma} \right) $.

% Problem Formulation using Discrete Rates and Proposed Algorithm: Optimization of Beamforming, User Admission and Discrete Rates with Imperfect SIC
\section{Problem Formulation and Proposed Algorithm for Discrete-Rate RSMA} \label{section_problem_modeling_problem_formulation_proposed_algorithm_discrete_rates}

In this section, we formulate and solve the WSR and WEE maximization problems to optimize the beamforming, user admission, and discrete rates for imperfect SIC. For ease of presentation, we summarize the most important parameters and decision variables in Table \ref{table_parameters_variables}.
% Table 
\vspace{1mm}
\begin{table}[!h]
 	\centering
	\fontsize{7}{8}\selectfont
	\caption{Parameters and decision variables.}
	\begin{tabular}{|m{5.6cm} |c|}
		% Header
		\hline
		\centering {\bf Parameters and Decision Variables} & \bf Notation \\ 
		\hline
		% Rows
		Number of antennas at the BS 	& $ N_\mathrm{tx} $ 
		\\
		Number of UEs							& $ U $ 
		\\
		Number of admitted UEs					& $ K $ 
		\\
		Number of discrete rates							& $ J $ 
		\\
		Channel between the BS and $ \mathsf{UE}_u $		& $ \mathbf{h}_u $ 
		\\
		Noise power 										& $ \sigma^2 $ 
		\\
		Weight of $ \mathsf{UE}_u $							& $ \omega_u $ 
		\\
		Dynamic power consumption of the circuitry			& $ P_\mathrm{dyn} $ 
		\\				
		Static power consumption of the circuitry			& $ P_\mathrm{sta} $ 
		\\
		Conversion efficiency of the power amplifier		& $ \eta_\mathrm{eff} $ 
		\\  
		Common rate of $ \mathsf{UE}_u $	 				& $ C_u $ 
		\\
		Private precoder for $ \mathsf{UE}_u $	 			& $ \mathbf{w}_u $ 
		\\
		Common precoder for all admitted UEs		 		& $ \mathbf{m} $ 
		\\
		Binary variable for private rate selection 	 	& $ \alpha_{u,j} $ 
		\\
		Binary variable for common rate selection 		 	& $ \kappa_j $ 
		\\ 
		Binary variable for UE admission 					& $ \chi_u $ 
		\\ 
		Binary variable for private precoder design 		& $ \mu_u $ 
		\\ 	
		Binary variable for common precoder design 			& $ \psi $ \\ 
		\hline
	\end{tabular}
	\label{table_parameters_variables}
\end{table}

% Problem Formulation
\subsection{Problem Formulation} \label{subsection_problem_formulation_discrete_rates} 
\begin{figure*}[!b]
% ensure that we have normalsize text
\normalsize
\hrulefill
% Equation
% Problem P_prima
\begin{align*} 
	% Objective
	\begin{matrix} \mathcal{P}_\mathrm{DWSR}' \\ \mathcal{P}_\mathrm{DWEE}' \end{matrix}: & \max_{
			\substack{
						\mathbf{W}, \mathbf{m}, \mathbf{c}, \boldsymbol{\chi}, \boldsymbol{\mu}, \boldsymbol{\alpha}, \boldsymbol{\kappa}, \psi
			 }
	} 
	& & 
	\begin{cases}
		f_\mathrm{DWSR} \left( \mathbf{c}, \boldsymbol{\alpha} \right) \triangleq \sum_{u \in \mathcal{U}} \omega_u \big( \sum_{j \in \mathcal{J}} \alpha_{u,j} R_j + C_u \big) \nonumber 
		\\ 
		f_\mathrm{DWEE} \left( \mathbf{W}, \mathbf{m}, \mathbf{c}, \boldsymbol{\mu}, \boldsymbol{\alpha}, \psi \right) \triangleq \frac{ \sum_{u \in \mathcal{U}} \omega_u \left( \sum_{j \in \mathcal{J}} \alpha_{u,j} R_j + C_u \right) }{\frac{1}{\eta_\mathrm{eff}} \left( \sum_{u \in \mathcal{U}} \left\| \mathbf{w}_u \mu_u \right\|^2_2 + \left\| \mathbf{m} \psi \right\|^2_2 \right) + P_\mathrm{cir}} \nonumber
	\end{cases} & \begin{matrix} \phantom{-1} ~ (\text{\footnotesize{linear}}) \\ (\text{\footnotesize{nonconvex}}) \end{matrix}
	\\
	% Constraint C1
	& ~~~~~~~~~ \mathrm{s.t.} & \widebar{\mathrm{C}}_{1}: ~ & \chi_{u} \in \left\lbrace 0, 1\right\rbrace, \forall u \in \mathcal{U}, & (\text{\footnotesize{binary}}) \nonumber	
	\\
	% Constraint C2
	& & \widebar{\mathrm{C}}_{2}: ~ & \textstyle \sum_{u \in \mathcal{U}} \chi_u = K, & (\text{\footnotesize{linear}}) \nonumber
	\\
	% Constraint C3
	& & \widebar{\mathrm{C}}_{3}: ~ & \mu_u \in \left\lbrace 0, 1 \right\rbrace, \forall u \in \mathcal{U}, & (\text{\footnotesize{binary}}) \nonumber
	\\
	% Constraint C4
	& & \widebar{\mathrm{C}}_{4}: ~ & \mu_u \leq \chi_u, \forall u \in \mathcal{U}, & (\text{\footnotesize{linear}}) \nonumber
	\\
	% Constraint C5
	& & \widebar{\mathrm{C}}_{5}: ~ & \psi \in \left\lbrace 0, 1\right\rbrace, & (\text{\footnotesize{binary}}) \nonumber
	\\
	% Constraint C6
	& & \widebar{\mathrm{C}}_{6}: ~ & \textstyle \sum_{u \in \mathcal{U}} \left\| \mathbf{w}_u \mu_u \right\|^2_2 + \left\| \mathbf{m} \psi \right\|^2_2 \leq P_\mathrm{tx}^\mathrm{max}, & (\text{\footnotesize{nonconvex}})\nonumber
	\\
	% Constraint C7
	& & \widebar{\mathrm{C}}_{7}: ~ & \alpha_{u,j} \in \left\lbrace 0, 1\right\rbrace, \forall u \in \mathcal{U}, j \in \mathcal{J}, & (\text{\footnotesize{binary}}) \nonumber
	\\
	% Constraint C8
	& & \widebar{\mathrm{C}}_{8}: ~ & \textstyle \sum_{j \in \mathcal{J}} \alpha_{u,j} = \mu_u, \forall u \in \mathcal{U}, & (\text{\footnotesize{linear}}) \nonumber
	\\
	% Constraint C9
	& & \widebar{\mathrm{C}}_{9}: ~ & \mathsf{SINR}^{(\mathrm{p})}_u \geq \textstyle \sum_{j \in \mathcal{J}} \alpha_{u,j} \Gamma_j, \forall u \in \mathcal{U}, & (\text{\footnotesize{nonconvex}}) \nonumber 
	\\
	% Constraint C10
	& & \widebar{\mathrm{C}}_{10}: ~ & \kappa_j \in \left\lbrace 0, 1\right\rbrace, \forall j \in \mathcal{J}, & (\text{\footnotesize{binary}}) \nonumber
	\\
	% Constraint C11
	& & \widebar{\mathrm{C}}_{11}: ~ & \textstyle \sum_{j \in \mathcal{J}} \kappa_j = \psi, & (\text{\footnotesize{linear}}) \nonumber
	\\
	% Constraint C12
	& & \widebar{\mathrm{C}}_{12}: ~ & \mathsf{SINR}^{(\mathrm{c})}_u \geq \chi_u \textstyle \sum_{j \in \mathcal{J}} \kappa_j \Gamma_j , \forall u \in \mathcal{U}, & (\text{\footnotesize{nonconvex}}) \nonumber 	
	\\
	% Constraint C13
	& & \widebar{\mathrm{C}}_{13}: ~ & C_u \geq 0, \forall u \in \mathcal{U}, & (\text{\footnotesize{linear}}) \nonumber	
	\\ 
	% Constraint C14
	& & \widebar{\mathrm{C}}_{14}: ~ &  C_u \leq \chi_u \textstyle \sum_{j \in \mathcal{J}} \kappa_j R_j, \forall u \in \mathcal{U}, & (\text{\footnotesize{nonconvex}}) \nonumber	
	\\ 
	% Constraint C15
	& & \widebar{\mathrm{C}}_{15}: ~ & \textstyle \sum_{u \in \mathcal{U}} C_u = \textstyle \sum_{j \in \mathcal{J}} \kappa_j R_j, & (\text{\footnotesize{linear}}) \nonumber	
	\\ 
	% Constraint C16
	& & \widebar{\mathrm{C}}_{16}: ~ & \textstyle \sum_{j \in \mathcal{J}} \alpha_{u,j} R_j + C_u \geq R_\mathrm{min} \chi_u, \forall u \in \mathcal{U}, & (\text{\footnotesize{linear}}) \nonumber	
\end{align*}
\vspace*{4pt}
\end{figure*}
We consider two objectives, namely, WSR and WEE maximization, and define the corresponding optimization problems $ \mathcal{P}_\mathrm{DWSR}' $ and $ \mathcal{P}_\mathrm{DWEE}' $, shown at the bottom of next page. Specifically, $ \omega_u $ is the weight associated with $ \mathsf{UE}_u $, which can be set by the network operator, e.g., to improve rate fairness. Besides, we define $ \mathbf{W} = \left[ \mathbf{w}_1, \dots, \mathbf{w}_U \right] $, $ \mathbf{c} = \left[ C_1, \dots, C_U \right] $, $ \boldsymbol{\chi} = \left[ \chi_1, \dots, \chi_U \right] $, $ \boldsymbol{\mu} = \left[ \mu_1, \dots, \mu_U \right] $, $ \boldsymbol{\alpha} = \left[ \alpha_{1,1}, \dots, \alpha_{U,J} \right] $, and $ \boldsymbol{\kappa} = \left[ \kappa_1, \dots, \kappa_J \right] $. In addition, $ \eta_\mathrm{eff} $ represents the amplifier efficiency and $ P_\mathrm{cir} = N_\mathrm{tx} P_\mathrm{dyn} + P_\mathrm{sta} $ is the power consumed by the circuitry at the BS, where $ P_\mathrm{dyn} $ and $ P_\mathrm{sta} $ denote the dynamic and static parts, respectively \cite{zhou2022:rate-splitting-multiple-access-multi-antenna-downlink-communication-systems-spectral-energy-efficiency-tradeoff}. Next, we discuss the constraints of the optimization problems.

% User admission
\subsubsection{User admission} To indicate whether a given $ \mathsf{UE}_u $ is admitted, we introduce constraint $ \widebar{\mathrm{C}}_{1}: \chi_u \in \left\lbrace 0, 1\right\rbrace, \forall u \in \mathcal{U} $, i.e., $ \chi_u = 1 $ indicates that the BS serves $ \mathsf{UE}_u $, and $ \chi_u = 0 $ otherwise. Further, we have $ \widebar{\mathrm{C}}_2: \sum_{u \in \mathcal{U}} \chi_u = K $ as the number of admitted UEs is $ K $. An admitted UE can receive its message via the common signal only, the private signal only, or both. To indicate whether an admitted $ \mathsf{UE}_u $ is served via the private signal, we introduce $ \widebar{\mathrm{C}}_3: \mu_u \in \left\lbrace 0, 1 \right\rbrace, \forall u \in \mathcal{U} $, i.e., $ \mu_u = 1 $ indicates that $ \mathsf{UE}_u $ is served via the private signal, and $ \mu_u = 0 $ otherwise. We also include $ \widebar{\mathrm{C}}_4: \mu_u \leq \chi_u, \forall u \in \mathcal{U} $, to ensure that non-admitted UEs are not served by a private signal. Naturally, non-admitted UEs are also not served by the common signal but this is handled by constraint $ \widebar{\mathrm{C}}_{12} $, discussed in Section~\ref{subsubsection_rate_selection_common_signal}. Lastly, we introduce $ \widebar{\mathrm{C}}_5: \psi \in \left\lbrace 0, 1\right\rbrace $ to denote whether the common signal is used.

% Beamforming
\subsubsection{Beamforming} 
The BS employs a private precoder $ \mathbf{w}_u \mu_u \in \mathbb{C}^{N_\mathrm{tx} \times 1} $ to precode symbol $ s_u  $, and a common precoder $ \mathbf{m} \psi \in \mathbb{C}^{N_\mathrm{tx} \times 1} $ to precode symbol $ s_0 $. The private precoder is $ \mathbf{0} $ when $ \mathsf{UE}_u $ is not admitted. Thus, the downlink signal of the BS is given by $ \mathbf{x} = \sum_{u \in \mathcal{U}} \mathbf{w}_u \mu_u s_u +  \mathbf{m} \psi s_0  $. To account for the transmit power limitation of the BS, the precoders must satisfy $ \widebar{\mathrm{C}}_6: \sum_{u \in \mathcal{U}} \left\| \mathbf{w}_u \mu_u \right\|^2_2 + \left\| \mathbf{m} \psi \right\|^2_2 \leq P_\mathrm{tx}^\mathrm{max} $.

% Imperfect SIC
\subsubsection{Imperfect SIC} \label{subsubsection_imperfect_SIC} The signal received by $ \mathsf{UE}_u $ is expressed as $ y_u = \mathbf{h}^\mathrm{H}_u \mathbf{x} + \eta_u $, which is equivalent to 
% Equation
\begin{align} \nonumber
		y_u = \underbrace{\mathbf{h}^\mathrm{H}_u \mathbf{m} \psi s_0}
		_{\substack{\text{common signal} \\ y^{(\mathrm{c})}_u}} + \underbrace{\mathbf{h}^\mathrm{H}_u \mathbf{w}_u \mu_u s_u}
		_{\substack{\text{private signal}\\ y^{(\mathrm{p})}_u}} + \underbrace{ \textstyle \sum\nolimits_{i \neq u}{\mathbf{h}^\mathrm{H}_u \mathbf{w}_i \mu_i s_i }}_{\substack{\text{interference} \\ y^{(\mathrm{int})}_u}} + \underbrace{\eta_u}_{\text{noise}}, 
\end{align}
where $ y^{(\mathrm{c})}_u $ is the received common signal, $ y^{(\mathrm{p})}_u $ is the received private signal, and $ y^{(\mathrm{int})}_u $ is the interference at $ \mathsf{UE}_u $. Further, $ \eta_u \sim \mathcal{CN} \left( 0, \sigma^2 \right) $ denotes additive white Gaussian noise, and $\mathbf{h}_u \in {\mathbb{C}}^{N_\mathrm{tx} \times 1}$ represents the channel between the BS and $ \mathsf{UE}_u $. An admitted $ \mathsf{UE}_u $ utilizes SIC in order to recover its message from $ y_u $. Specifically, $ \mathsf{UE}_u $ decodes first the common symbol $ s_0 $ by treating signals $ y^{(\mathrm{p})}_u $ and $ y^{(\mathrm{int})}_u $ as noise. Next, $ \mathsf{UE}_u $ reconstructs the received common signal $ y^{(\mathrm{c})}_u $ and subtracts it from $ y_u $, yielding $ {y}_u^\mathrm{SIC} = y^{(\mathrm{p})}_u + y^{(\mathrm{int})}_u + \eta_u $, based on which it decodes its private symbol $ s_u $. However, in practice, the removal of $ y^{(\mathrm{c})}_u $ is not perfect, which can be caused by, e.g., hardware impairments \cite{mouni2021:adaptive-user-pairing-noma-systems-imperfect-sic, chopra2022:bounds-power-common-message-fractions-rsma-imperfect-sic}. Therefore, the signal after imperfect SIC can be expressed as $ {y}_u^\mathrm{iSIC} = \Delta_\mathrm{SIC} y^{(\mathrm{c})}_u + y^{(\mathrm{p})}_u + y^{(\mathrm{int})}_u + \eta_u $, where $ 0 \leq \Delta_\mathrm{SIC} \leq 1 $ is the percentage of the common signal that is not canceled, i.e., $ \Delta_\mathrm{SIC} = 0 $ implies perfect SIC. As a result, the SINRs of the common and private signals at $ \mathsf{UE}_u $ are $ \mathsf{SINR}^{(\mathrm{c})}_u = \frac{\left| \mathbf{h}^\mathrm{H}_u \mathbf{m} \psi \right|^2 } { \sum_{i \in \mathcal{U}} \left| \mathbf{h}^\mathrm{H}_u  \mathbf{w}_i \mu_i \right|^2 + {\sigma}^2 } $ and $ \mathsf{SINR}^{(\mathrm{p})}_u = \frac{ \left| \mathbf{h}^\mathrm{H}_u \mathbf{w}_u \mu_u \right|^2 }{ \left| \Delta_\mathrm{SIC} \mathbf{h}^\mathrm{H}_u \mathbf{m} \psi \right|^2  + \sum_{i \neq u} \left| \mathbf{h}^\mathrm{H}_u  {\mathbf{w}_i} \mu_i \right|^2 + \sigma^2} $, respectively. The exact value of $ \Delta_\mathrm{SIC} $ is usually not known by the BS. Therefore, it must be set properly to avoid performance degradation, and thus guarantee the target SINRs that enable the allocated rates. Typical values for $ \Delta_\mathrm{SIC} $ are in the range of $ 4\% $ and $ 10\% $ \cite{mouni2021:adaptive-user-pairing-noma-systems-imperfect-sic}.

% Private rates assignment
\subsubsection{Rate selection for the private signals} An $ \mathsf{UE}_u $ receiving a private signal at a rate $ R_j $, can only decode the message if $ \mathsf{SINR}^{(\mathrm{p})}_u \geq \Gamma_j $, where $ \Gamma_j $ is the target SINR that guarantees $ R_j $ (for numerical values, see Table \ref{table_rates_sinr} in Section~\ref{section_results}). To depict the assignment of private rates, we introduce constraint $ \widebar{\mathrm{C}}_7: \alpha_{u,j} \in \left\lbrace 0, 1 \right\rbrace, \forall u \in \mathcal{U}, j \in \mathcal{J} $, where $ \alpha_{u,j} = 1 $ indicates that $ \mathsf{UE}_u $ is served by a private signal transmitted at rate $ R_j $. In addition, we include $ \widebar{\mathrm{C}}_8:  \sum_{j \in \mathcal{J}} \alpha_{u,j} = \mu_u, \forall u \in \mathcal{U} $, to ensure that a rate is allocated to $ \mathsf{UE}_u $, if it is served by the private signal. Further, to associate the discrete rates and their corresponding target SINRs, we include $ \widebar{\mathrm{C}}_9: \mathsf{SINR}^{(\mathrm{p})}_u \geq \sum_{j \in \mathcal{J}} \alpha_{u,j} \Gamma_j, \forall u \in \mathcal{U} $, which ensures for $ \mathsf{UE}_u $ a private rate of $ \sum_{j \in \mathcal{J}} \alpha_{u,j} R_j $ if $ \mu_u = 1 $. Note that $ \mu_u = 0 $ does not indicate that $ \mathsf{UE}_u $ is not admitted since $ \mathsf{UE}_u $ can also be served by the common signal if $ C_u > 0 $.

% Common rate assignment
\subsubsection{Rate selection for the common signal} \label{subsubsection_rate_selection_common_signal}
An admitted $ \mathsf{UE}_u $ can only decode the common message transmitted at rate $ R_j $, if $ \mathsf{SINR}^{(\mathrm{c})}_u \geq \Gamma_j $. To this end, we introduce constraint $ \widebar{\mathrm{C}}_{10}: \kappa_j \in \left\lbrace 0, 1 \right\rbrace $, $ j \in \mathcal{J} $, where $ \kappa_j = 1 $ indicates that rate $ R_j $ is selected. We include $ \widebar{\mathrm{C}}_{11}: \sum_{j \in \mathcal{J}} \kappa_j = \psi $ to allow for the possibility that the common rate is zero (see constraint $ \widebar{\mathrm{C}}_{5} $). To unify user admission and the allocation of the common rate, we add $ \widebar{\mathrm{C}}_{12}: \mathsf{SINR}^{(\mathrm{c})}_u \geq \chi_u \sum_{j \in \mathcal{J}} \kappa_j \Gamma_j , \forall u \in \mathcal{U} $, which results in the common rate $ \sum_{j \in \mathcal{J}} \kappa_j R_j $ for all admitted UEs. Although the rate portions $ C_u $ are not continuous, they have very fine granularity because rate splitting is capable of dividing the messages $ W_u $ into portions of any size. Thus, we treat $ C_u $ as continuous-valued by adding $ \widebar{\mathrm{C}}_{13}: C_u \geq 0, \forall u \in \mathcal{U} $. To keep consistency with user admission, we include $ \widebar{\mathrm{C}}_{14}: C_u \leq \chi_u \sum_{j \in \mathcal{J}} \kappa_j R_j, \forall u \in \mathcal{U} $, to enforce $ C_u = 0 $ for non-admitted UEs or when the common rate is zero (see constraints $ \widebar{\mathrm{C}}_{5}, \widebar{\mathrm{C}}_{11} $). Moreover, we add $ \widebar{\mathrm{C}}_{15}: \sum_u C_u = \sum_{j \in \mathcal{J}} \kappa_j R_j $ to guarantee that the sum of all common portions $ C_u $ is equal to the overall common rate. Finally, we enforce a minimum rate $ R_\mathrm{min} $ per admitted UE by including constraint $ \widebar{\mathrm{C}}_{16}: \sum_{j \in \mathcal{J}} \alpha_{u,j} R_j + C_u \geq R_\mathrm{min} \chi_u $.

% Remark
\vspace{1mm}
\noindent \textit{{\textsc{Remark 1:}} Problems $ \mathcal{P}_\mathrm{DWSR}' $ and $ \mathcal{P}_\mathrm{DWEE}' $ are nonconvex MINLPs, which are challenging to solve. Specifically, the nonconvexity is due to constraints $ \widebar{\mathrm{C}}_{6} $, $ \widebar{\mathrm{C}}_{9} $, $ \widebar{\mathrm{C}}_{12} $, $ \widebar{\mathrm{C}}_{14} $ and the objective function $ f_\mathrm{DWEE} \left( \mathbf{W}, \mathbf{m}, \mathbf{c}, \boldsymbol{\mu}, \boldsymbol{\alpha}, \psi \right) $, which contain quotients of quadratic functions and multiplicative couplings. Further, a simple strategy to obtain the SDMA versions of $ \mathcal{P}_\mathrm{DWSR}' $ and $ \mathcal{P}_\mathrm{DWEE}' $ is to set $ \psi = 0 $ because RSMA includes SDMA as a special case.}

% Constraints
\begin{figure*}[!t]
% ensure that we have normalsize text
\normalsize
% Constraints D1-D6
\begin{equation} \nonumber
	\widebar{\mathrm{C}}_{12}, \widebar{\mathrm{C}}_{14} \Leftrightarrow
		\begin{cases}
			   	\widebar{\mathrm{D}}_{1}: \pi_{u,j} \in \left\lbrace 0, 1 \right\rbrace , \forall u \in \mathcal{U}, j \in \mathcal{J}, & (\text{\footnotesize{binary}})
			   	\\	
			   	\widebar{\mathrm{D}}_{2}: \pi_{u,j} \leq \chi_u, \forall u \in \mathcal{U}, j \in \mathcal{J}, ~~ \widebar{\mathrm{D}}_{3}: \pi_{u,j} \leq \kappa_j, \forall u \in \mathcal{U}, j \in \mathcal{J}, & (\text{\footnotesize{linear}})
			  	\\
			  	\widebar{\mathrm{D}}_{4}: \pi_{u,j} \geq \chi_u + \kappa_j - 1, \forall u \in \mathcal{U}, j \in \mathcal{J}, & (\text{\footnotesize{linear}})
			  	\\	
			   	\widebar{\mathrm{D}}_{5}: \frac{\left| \mathbf{h}^\mathrm{H}_u \mathbf{m} \psi \right|^2 } { \sum_{i \in \mathcal{U}} \left| \mathbf{h}_u^\mathrm{H} \mathbf{w}_i \mu_i \right|^2 + {\sigma}^2} \geq \sum_{j \in \mathcal{J}} \pi_{u,j} \Gamma_j, \forall u \in \mathcal{U}, & (\text{\footnotesize{nonconvex}})
			  	\\	
			   	\widebar{\mathrm{D}}_{6}: C_u \leq \textstyle \sum_{j \in \mathcal{J}} \pi_{u,j} R_j, \forall u \in \mathcal{U}. & (\text{\footnotesize{linear}})
		\end{cases}
\end{equation}
\hrulefill
% Proposition 2
\begin{equation} \nonumber
	\widebar{\mathrm{C}}_{6}, \widebar{\mathrm{C}}_{9}, \widebar{\mathrm{D}}_{5} \Leftrightarrow
		\begin{cases}
				  	\widebar{\mathrm{E}}_{1}: \left\| \mathbf{w}_u \right\|_2^2  \leq \mu_u P_\mathrm{tx}^\mathrm{max}, \forall u \in \mathcal{U}, ~~ \widebar{\mathrm{E}}_{2}: \left\| \mathbf{m} \right\|_2^2  \leq \psi P_\mathrm{tx}^\mathrm{max}, ~~ \widebar{\mathrm{E}}_{3}: \sum_{u \in \mathcal{U}} \left\| \mathbf{w}_u \right\|_2^2 + \left\| \mathbf{m} \right\|_2^2  \leq P_\mathrm{tx}^\mathrm{max},  & (\text{\footnotesize{convex}})	
				   	\\
				   	\widebar{\mathrm{E}}_{4}: \frac{\left| \mathbf{h}_u^\mathrm{H} {\mathbf{w}_u} \right|^2 }{ \Delta_\mathrm{SIC}^2 \left| \mathbf{h}^\mathrm{H}_u \mathbf{m} \right|^2 +  \sum_{ i \neq u, i \in \mathcal{U} } \left| \mathbf{h}_u^\mathrm{H} {\mathbf{w}_i} \right|^2 + \sigma^2 } \geq  \sum_{j \in \mathcal{J}} \alpha_{u,j} \Gamma_j, \forall u \in \mathcal{U},  & (\text{\footnotesize{nonconvex}}) 
				   	\\
				   	\widebar{\mathrm{E}}_{5}: \frac{\left| \mathbf{h}_u^\mathrm{H} \mathbf{m} \right|^2 }
				   			{  \sum_{i \in \mathcal{U}} \left| \mathbf{h}_u^\mathrm{H} {\mathbf{w}_i} \right|^2 + {\sigma}^2 } \geq  \sum_{j \in \mathcal{J}} \pi_{u,j} \Gamma_j , \forall u \in \mathcal{U}.  & (\text{\footnotesize{nonconvex}}) 	
			\end{cases}
\end{equation}
\hrulefill
% Proposition 3
\begin{equation} \nonumber
	\widebar{\mathrm{E}}_{4}, \widebar{\mathrm{E}}_{4} \Leftrightarrow
		\begin{cases}
				   	\widebar{\mathrm{F}}_{1}: \frac{\left| \mathbf{h}_u^\mathrm{H} {\mathbf{w}_u} \right|^2 }{ \Delta_\mathrm{SIC}^2 \left| \mathbf{h}^\mathrm{H}_u \mathbf{m} \right|^2 +  \sum_{ i \neq u, i \in \mathcal{U} } \left| \mathbf{h}_u^\mathrm{H} {\mathbf{w}_i} \right|^2 + \sigma^2 } \geq \alpha_{u,j} \Gamma_j, \forall u \in \mathcal{U}, j \in \mathcal{J},  & (\text{\footnotesize{nonconvex}})
				   	\\
				   	\widebar{\mathrm{F}}_{2}: \frac{\left| \mathbf{h}_u^\mathrm{H} \mathbf{m} \right|^2 }
				   			{ \sum_{i \in \mathcal{U}} \left| \mathbf{h}_u^\mathrm{H} {\mathbf{w}_i} \right|^2 + {\sigma}^2 } \geq \pi_{u,j} \Gamma_j , \forall u \in \mathcal{U}, j \in \mathcal{J}.  & (\text{\footnotesize{nonconvex}})		 
		\end{cases}
\end{equation}
\hrulefill
% Proposition 4
\begin{equation} \nonumber
	\widebar{\mathrm{F}}_{1}, \widebar{\mathrm{F}}_{2} \Leftrightarrow
		\begin{cases}
				\widebar{\mathrm{G}}_{1}: \left\| \left[ \mathbf{h}_u^\mathrm{H} \widebar{\mathbf{W}}_u, \sigma \right] \right\|_2^2 \leq \frac{1}{\Gamma_j} \left| \mathbf{h}_u^\mathrm{H} {\mathbf{w}_u} \right|^2  + \left( 1 - \alpha_{u,j} \right) L_{\mathrm{max},u}^2, \forall u \in \mathcal{U}, j \in \mathcal{J}, & (\text{\footnotesize{nonconvex}}) \nonumber 
				\\
				\widebar{\mathrm{G}}_{2}: \left\| \left[ \mathbf{h}_u^\mathrm{H} \mathbf{W}, \sigma \right] \right\|_2^2 \leq \frac{1}{\Gamma_j} \left| \mathbf{h}_u^\mathrm{H} \mathbf{m} \right|^2  + \left( 1 - \pi_{u,j} \right) L_{\mathrm{max},u}^2, \forall u \in \mathcal{U}, j \in \mathcal{J}.  & (\text{\footnotesize{nonconvex}}) \nonumber	 
		\end{cases}
\end{equation}
\hrulefill
% Proposition 5
\begin{equation} \nonumber
	\widebar{\mathrm{G}}_{1} \Leftrightarrow
		\begin{cases}
				   	\widebar{\mathrm{H}}_{1}: \mathfrak{Re} \left\lbrace {\mathbf{h}}_u^\mathrm{H} {\mathbf{w}_u} \right\rbrace \geq 0, \forall u \in \mathcal{U}, ~~ \widebar{\mathrm{H}}_{2}: \mathfrak{Im} \left\lbrace {\mathbf{h}}_u^\mathrm{H} {\mathbf{w}_u} \right\rbrace = 0, \forall u \in \mathcal{U}, & (\text{\footnotesize{linear}}) \nonumber
				   	%\\
				   	%\widebar{\mathrm{H}}_{2}: \mathfrak{Im} \left\lbrace {\mathbf{h}}_u^\mathrm{H} {\mathbf{w}_u} \right\rbrace = 0, \forall u \in \mathcal{U}, & (\text{\footnotesize{linear}}) \nonumber	
				   	\\
				   	\widebar{\mathrm{H}}_{3}: \left\| \left[ \mathbf{h}_u^\mathrm{H} \widebar{\mathbf{W}}_u, \sigma \right] \right\|_2 \leq \frac{1}{\sqrt{\Gamma_j}} \mathfrak{Re} \left\lbrace {\mathbf{h}}_u^\mathrm{H} {\mathbf{w}_u} \right\rbrace + \left( 1 - \alpha_{u,j} \right) L_{\mathrm{max},u}, \forall u \in \mathcal{U}, j \in \mathcal{J}. & (\text{\footnotesize{convex}})\nonumber		 
		\end{cases}
\end{equation}
\hrulefill
% Proposition 6
\begin{equation} \nonumber
	\widebar{\mathrm{G}}_{2} \Leftrightarrow
		\begin{cases}
				   	\widebar{\mathrm{I}}_{1}: \mathfrak{Re} \left\lbrace {\mathbf{h}}_u^\mathrm{H} \mathbf{m} \right\rbrace \geq 0, \forall u \in \mathcal{U}, & (\text{\footnotesize{linear}})\nonumber
				   	\\
				   	\widebar{\mathrm{I}}_{2}: \left\| \left[ \mathbf{h}_u^\mathrm{H} \mathbf{W}, \sigma \right] \right\|_2  \leq \frac{1}{\sqrt{\Gamma_j}} \mathfrak{Re} \left\lbrace {\mathbf{h}}_u^\mathrm{H} \mathbf{m} \right\rbrace + \left( 1 - \pi_{u,j} \right) L_{\mathrm{max},u}, \forall u \in \mathcal{U}, j \in \mathcal{J}. & (\text{\footnotesize{convex}}) \nonumber	 
		\end{cases}
\end{equation}
\hrulefill
\vspace*{0pt}
\end{figure*}

% Section: Proposed Algorithms for solving P'
\subsection{Proposed Algorithm} \label{subsection_proposed_algorithm_discrete_rates}

We propose the \texttt{OPT-MISOCP} algorithm to solve the nonconvex MINLPs $ \mathcal{P}_\mathrm{DWSR}' $ and $ \mathcal{P}_\mathrm{DWEE}' $. Instead of treating these problems as general nonconvex MINLPs, we propose a sequence of transformations to overcome the nonconvexities, thereby transforming them into the convex MISOCPs $ \mathcal{P}_\mathrm{DWSR} $ and $ \mathcal{P}_\mathrm{DWEE} $, respectively, whose global optima can be found using branch-and-bound (BnB) and interior-point methods (IPMs). Specifically, BnB is used for decomposing the binary variables of the MISOCP, whereas IPMs are used for solving the underlying SOCPs. In the following, we describe the proposed algorithm for $ \mathcal{P}_\mathrm{DWSR}' $ and then we extend it to $ \mathcal{P}_\mathrm{DWEE}' $.

% Subsection
\subsubsection{Circumventing Integer Multiplicative Couplings} \label{subsubsection_circumventing_integer_multiplicative_couplings}

To cope with the multiplicative coupling between binary variables in constraints $ \widebar{\mathrm{C}}_{12} $, $ \widebar{\mathrm{C}}_{14} $, appearing in the form of $ \chi_u \kappa_j $, we transform such products into the intersection of linear combinations. We introduce new variables $ \pi_{u,j} = \chi_u \kappa_j $, and equivalently rewrite constraints $ \widebar{\mathrm{C}}_{12} $, $ \widebar{\mathrm{C}}_{14} $ as constraints $ \widebar{\mathrm{D}}_{1} - \widebar{\mathrm{D}}_{6} $, shown at the top of this page (cf. \textbf{Appendix~\ref{appendix_proposition_1}}).

% Subsection
\subsubsection{Circumventing Mixed-Integer Multiplicative Couplings} \label{subsubsection_circumventing_mixed_integer_multiplicative_couplings}

To deal with the mixed-integer multiplicative couplings in constraints $ \widebar{\mathrm{C}}_{6} $, $ \widebar{\mathrm{C}}_{9} $, $ \widebar{\mathrm{D}}_{5} $, appearing in the form of $ \mathbf{w}_u \mu_u $ and $ \mathbf{m} \psi $, we reformulate such products as linear relations without altering the nature of the problem. Thus, constraints $ \widebar{\mathrm{C}}_{6}, \widebar{\mathrm{C}}_{9}, \widebar{\mathrm{D}}_{5} $ are equivalently rewritten as constraints $ \widebar{\mathrm{E}}_{1} - \widebar{\mathrm{E}}_{5} $, shown at the top of this page (cf. \textbf{Appendix \ref{appendix_proposition_2}}).

% Subsection
\subsubsection{Circumventing Integer Additive Couplings} \label{subsubsection_circumventing_integer_additive_couplings}

The additive couplings of binary variables  in $ \widebar{\mathrm{E}}_{4} $, $ \widebar{\mathrm{E}}_{5} $, appearing in the form $ \sum_{j \in \mathcal{J}} \alpha_{u,j} \Gamma_j $ and $ \sum_{j \in \mathcal{J}} \pi_{u,j} \Gamma_j $, pose a difficulty for convexification since multiple binary variables and their corresponding target SINRs are combined. However, since the couplings are linear and sum to at most one, we can handle them by expanding them into several constraints (i.e., as multiple choice constraints), such that each of the resulting constraints depends on one binary variable only. Thus, constraints $ \widebar{\mathrm{E}}_{4} $, $ \widebar{\mathrm{E}}_{5} $ are equivalently recast as $ \widebar{\mathrm{F}}_{1} $, $ \widebar{\mathrm{F}}_{2} $, shown at the top of this page (cf. \textbf{Appendix \ref{appendix_proposition_3}}).

% Subsection
\subsubsection{Reformulating the SINR Constraints via the Big-M Method} \label{subsubsection_reformulating_sinr_constraints_big_m_method}

To deal with the disjunctiveness caused by the binary variables, which lead to different SINR requirements for the admitted and non-admitted UEs, in $ \widebar{\mathrm{F}}_{1} $, $ \widebar{\mathrm{F}}_{2} $, we merge these two cases into a single one via the Big-M method. Thus, by defining $ \widebar{\mathbf{W}}_u = \left[ \Delta_\mathrm{SIC} \mathbf{m}, \mathbf{w}_1, \dots, \mathbf{w}_{u-1}, \mathbf{w}_{u+1}, \dots, \mathbf{w}_U \right] $ and $ L_{\mathrm{max},u}^2 = \left\| {\mathbf{h}}_u \right\|_2^2 P_\mathrm{tx}^\mathrm{max} + \sigma^2 $, we recast constraints $ \widebar{\mathrm{F}}_{1} $, $ \widebar{\mathrm{F}}_{2} $ as $ \widebar{\mathrm{G}}_{1}, \widebar{\mathrm{G}}_{2} $, shown at the top of this page (cf. \textbf{Appendix~\ref{appendix_proposition_4}}).

\subsubsection{Convexifying the Private SINR Constraints} \label{subsubsection_convexifying_private_SINR_constraints}

Although constraint $ \widebar{\mathrm{G}}_{1} $ is nonconvex, as it contains a difference of convex functions, it can be convexified without changing its feasible set. Thus, constraint $ \widebar{\mathrm{G}}_{1} $ can be expressed as $ \widebar{\mathrm{H}}_{1} $, $ \widebar{\mathrm{H}}_{2} $, $ \widebar{\mathrm{H}}_{3} $, shown at the top of this page (cf. \textbf{Appendix~\ref{appendix_proposition_5}}).

% Subsection
\subsubsection{Convexifying the Common SINR Constraints} \label{subsubsection_convexifying_common_SINR_constraints}

The nonconvex constraint $ \widebar{\mathrm{G}}_{2} $ can be replaced by the inner convex approximations $ \widebar{\mathrm{I}}_{1} $, $ \widebar{\mathrm{I}}_{2} $, shown at the top of this page, which may shrink the original feasible set (cf. \textbf{Appendix \ref{appendix_proposition_6}}).

% Subsection
\subsubsection{Adding Cutting Planes to Tighten the Feasible Domain} \label{subsubsection_adding_cutting_planes_tighten_feasible_domain}

To reduce the search complexity caused by BnB branching for the binary variables, we add problem-specific cutting planes, which do not impact optimality (cf. \textbf{Appendix \ref{appendix_proposition_10}}). We add cuts $ \widebar{\mathrm{J}}_{1} $ to tighten the feasible set, which can help accelerating the optimization. In addition, we include $ \widebar{\mathrm{J}}_{2} $ as an upper bound of the sum-rate, which facilitates early stopping:
% Proposition 7
\begin{align} 
	& \widebar{\mathrm{J}}_{1}: \mathfrak{Re} \left\lbrace {\mathbf{h}}_u^\mathrm{H} {\mathbf{w}_u} \right\rbrace \geq \sigma \textstyle \sum_{j \in \mathcal{J}} \alpha_{u,j} \sqrt{\Gamma_j}, \forall u \in \mathcal{U}, & (\text{\footnotesize{linear}}) \nonumber
	\\
	& \widebar{\mathrm{J}}_{2}: \textstyle \sum_{u \in \mathcal{U}} \big( \sum_{j \in \mathcal{J}} \alpha_{u,j} R_j + C_u \big) \leq (K+1) R_{J}. & (\text{\footnotesize{linear}}) \nonumber
\end{align}	

% Remark
\vspace{1mm}
\noindent \textit{{\textsc{Remark 2:}} In our simulations, a remarkable improvement in runtime was observed with the addition of $ \widebar{\mathrm{J}}_{1} $ and $ \widebar{\mathrm{J}}_{2} $, which accelerated the optimization $ 3 - 20 $ times compared to the case without them. }

% Subsection
\subsubsection{Outlining the Algorithm and Its Extension to Solve $ \mathcal{P}_\mathrm{DWEE}' $} \label{subsubsection_outlining_opt_misocp_algorithm}

Recapitulating the results above, problem $ \mathcal{P}_\mathrm{DWSR}' $ is recast as
% Problem P_DWSR
\begin{align} \nonumber
	% Objective
	\mathcal{P}_\mathrm{DWSR}: \max_{
			\substack{
						\boldsymbol{\Theta} 
				 }
  	} 
  	f_\mathrm{DWSR} \left( \mathbf{c}, \boldsymbol{\alpha} \right) ~ \mathrm{s.t.} ~ \boldsymbol{\Theta} \in \mathcal{C},
\end{align}
where $ \boldsymbol{\Theta} = \left( \mathbf{W}, \mathbf{m}, \mathbf{c}, \boldsymbol{\chi}, \boldsymbol{\mu}, \boldsymbol{\alpha}, \boldsymbol{\kappa}, \boldsymbol{\pi}, \psi \right)  $ and $ \mathcal{C} $ is the feasible set of $ \boldsymbol{\Theta} $ defined by $ \widebar{\mathrm{C}}_{1} - \widebar{\mathrm{C}}_{5}, \widebar{\mathrm{C}}_{7}, \widebar{\mathrm{C}}_{8}, \widebar{\mathrm{C}}_{10}, \widebar{\mathrm{C}}_{11} $, $ \widebar{\mathrm{C}}_{13}, \widebar{\mathrm{C}}_{15}, \widebar{\mathrm{C}}_{16}, \widebar{\mathrm{D}}_{1} - \widebar{\mathrm{D}}_{4}, \widebar{\mathrm{D}}_{6}, \widebar{\mathrm{E}}_{1} - \widebar{\mathrm{E}}_{3}, \widebar{\mathrm{H}}_{1} - \widebar{\mathrm{H}}_{3}, \widebar{\mathrm{I}}_{1}, \widebar{\mathrm{I}}_{2}, \widebar{\mathrm{J}}_{1}, \widebar{\mathrm{J}}_{2} $. Analogous to $ \mathcal{P}_\mathrm{DWSR}' $, we recast problem $ \mathcal{P}_\mathrm{DWEE}' $ as
% Problem P_DWEE
\begin{align} \nonumber
	% Objective
	\mathcal{P}_\mathrm{DWEE}: \min_{
			\substack{
						\boldsymbol{\Theta}
				 }
  	} 
  	\frac{\frac{1}{\eta_\mathrm{eff}} \left( \sum_{u \in \mathcal{U}} \left\| \mathbf{w}_u \right\|^2_2 + \left\| \mathbf{m} \right\|^2_2 \right) + P_\mathrm{cir}} { \sum_{u \in \mathcal{U}} \omega_u \left( \sum_{j \in \mathcal{J}} \alpha_{u,j} R_j + C_u \right) } ~ \mathrm{s.t.} ~ \boldsymbol{\Theta} \in \mathcal{C},
\end{align}
where we have transformed the maximization of $ f_\mathrm{DWEE} \left( \mathbf{W}, \mathbf{m}, \mathbf{c}, \boldsymbol{\mu}, \boldsymbol{\alpha}, \psi \right) $ into the minimization of its reciprocal $ \frac{1}{ f_\mathrm{DWEE} \left( \mathbf{W}, \mathbf{m}, \mathbf{c}, \boldsymbol{\mu}, \boldsymbol{\alpha}, \psi \right)} $. In addition, we have removed the mixed-integer couplings from the objective function, as described in Section~\ref{subsubsection_circumventing_mixed_integer_multiplicative_couplings}. Hence, the objective function of $ \mathcal{P}_\mathrm{DWEE} $ is convex. Problems $ \mathcal{P}_\mathrm{DWSR} $ and $ \mathcal{P}_\mathrm{DWEE} $ are MISOCPs, which can be solved globally optimal via BnB and IPMs with off-the-shelf solvers, such as \texttt{MOSEK} and \texttt{GUROBI}, as the problems are convex in the continuous variables. %Specifically, after applying the transformations from Section \ref{subsubsection_circumventing_integer_multiplicative_couplings} to Section \ref{subsubsection_adding_cutting_planes_tighten_feasible_domain}, all constraints are convex except for the integrality qualifications of the binary variables.

% Remark
\vspace{1mm}
\noindent \textit{{\textsc{Remark 3:}} Due to the inner convexification of the feasible sets of $ \mathcal{P}_\mathrm{DWSR}' $ ($ \mathcal{P}_\mathrm{DWEE}' $) in Section~\ref{subsubsection_convexifying_common_SINR_constraints}, a globally optimal solution for $ \mathcal{P}_\mathrm{DWSR} $ ($ \mathcal{P}_\mathrm{DWEE} $) is feasible for $ \mathcal{P}_\mathrm{DWSR}' $ ($ \mathcal{P}_\mathrm{DWEE}' $) but not necessarily globally optimal for $ \mathcal{P}_\mathrm{DWSR}' $ ($ \mathcal{P}_\mathrm{DWEE}' $). However, such solution is found to be near-optimal for $ \mathcal{P}_\mathrm{DWSR}' $ ($ \mathcal{P}_\mathrm{DWEE}' $), as shown in Section~\ref{subsection_complexity_optimality_convergence_initialization}, where we compare $ \mathcal{P}_\mathrm{DWSR} $ against an upper bound of $ \mathcal{P}_\mathrm{DWSR}' $, showing a negligible performance loss.}

\vspace{1mm}
\noindent \textit{{\textsc{Remark 4:}} 
Problems $ \mathcal{P}_\mathrm{DWSR} $ and $ \mathcal{P}_\mathrm{DWEE} $ employ discrete rates but allow dynamic rate allocation, i.e., the assigned rates can vary within the designated values of the set of discrete rates $ \mathcal{J} $. Our formulation allows to finds optimal rates for unicast and multicast signals, adapting to the channel characteristics and aiming to maximize the objective function.}

% Subsection
\subsubsection{Computational Complexity} \label{section_computational_complexity_opt_misocp} Problems $ \mathcal{P}_\mathrm{DWSR} $ and $ \mathcal{P}_\mathrm{DWEE} $ involve $ N_v = (U+1) N_\mathrm{tx} + U $ continuous variables and $ N_c = 5UJ + 7U + 4 $ linear and convex constraints. The dimension of the underlying SOCP is $ N_d = 2 J N_\mathrm{tx} U^2 + U^2 + 3UJ + 7U + 2 U N_\mathrm{tx} + 2 N_\mathrm{tx} $ for any fixed values of binary variables. Therefore, the complexities of problems $ \mathcal{P}_\mathrm{DWSR} $ and $ \mathcal{P}_\mathrm{DWEE} $ is $ \mathcal{C}_{\texttt{OPT-MISOCP}} = \mathcal{O} \left( N_p N_c^{0.5} N_v^2 N_d \right) $, where $ N_p $ is the number of solutions evaluated by the BnB solver. The worst-case for $ N_p $ is given by $ N_p^\mathrm{all} = {U \choose K} \left( \sum_{m = 0}^K {K \choose m} J^{K+1-m} \right) $. In practice, $ N_p \ll N_p^\mathrm{all} $ since BnB methods are capable of pruning infeasible and suboptimal branches, thus reducing the search complexity.

% Problem Modeling, Problem Formulation and Proposed Algorithm: Optimization of Beamforming, User Admission and Continuous Rates with Imperfect SIC
% Problem Formulation
\section{Problem Formulation and Proposed Algorithm for Continuous-Rate RSMA} \label{section_problem_modeling_problem_formulation_proposed_algorithm_continuous_rates}

In this section, we formulate and solve the WSR and WEE maximization problems for RSMA to optimize the beamforming, user admission, and continuous rates, while accounting for imperfect SIC.

\subsection{Problem Formulation} \label{subsection_problem_formulation_continuous_rates}

% Problems
\begin{figure*}[!t]
% ensure that we have normalsize text
\normalsize
% Problem Q_prima
\begin{align} 
	% Objective
	\begin{matrix} \mathcal{Q}_\mathrm{CWSR}' \\ \mathcal{Q}_\mathrm{CWEE}' \end{matrix}: & \max_{
			\substack{
						\mathbf{W}, \mathbf{m}, \mathbf{c}, \boldsymbol{\chi}, \psi 
			 }
	} 
	%\\
	& 	 & 
	\begin{cases}
		f_\mathrm{CWSR}' \left( \mathbf{W}, \mathbf{m}, \mathbf{c} \right) \triangleq  \sum_{u \in \mathcal{U}} \omega_u \big( \log_2 \big( 1 + \mathsf{SINR}^{(\mathrm{p})}_u \big) + C_u \big) \nonumber
		\\
		f_\mathrm{CWEE}' \left( \mathbf{W}, \mathbf{m}, \mathbf{c} \right) \triangleq \frac{ \sum_{u \in \mathcal{U}} \omega_u \left( \log_2 \left( 1 + \mathsf{SINR}^{(\mathrm{p})}_u \right) + C_u \right) }{\frac{1}{\eta_\mathrm{eff}} \left(  \sum_{u \in \mathcal{U}} \left\| \mathbf{w}_u \chi_u \right\|^2_2 + \left\| \mathbf{m} \psi \right\|^2_2 \right) + P_\mathrm{cir}} & \nonumber
	\end{cases} & \begin{matrix} (\text{\footnotesize{nonconvex}}) \\ (\text{\footnotesize{nonconvex}}) \end{matrix}
	\\
	% Constraint C1
	& ~~~~~ \mathrm{s.t.} & \widebar{\mathrm{C}}_{1}: ~ & \chi_{u} \in \left\lbrace 0, 1\right\rbrace, \forall u \in \mathcal{U}, & (\text{\footnotesize{binary}}) \nonumber	
	\\
	% Constraint C2
	& & \widebar{\mathrm{C}}_{2}: ~ & \textstyle \sum_{u \in \mathcal{U}} \chi_u = K, & (\text{\footnotesize{linear}}) \nonumber
	\\
	% Constraint C5
	& & \widebar{\mathrm{C}}_{5}: ~ & \psi \in \left\lbrace 0, 1\right\rbrace, & (\text{\footnotesize{binary}}) \nonumber
	\\
	% Constraint C6
	& & \widebar{\mathrm{C}}_{6}: ~ & \textstyle \sum_{u \in \mathcal{U}} \left\| \mathbf{w}_u \chi_u \right\|^2_2 + \left\| \mathbf{m} \psi \right\|^2_2 \leq P_\mathrm{tx}^\mathrm{max}, & (\text{\footnotesize{nonconvex}})\nonumber
	\\
	% Constraint C13
	& & \widebar{\mathrm{C}}_{13}: ~ & C_u \geq 0, \forall u \in \mathcal{U}, & (\text{\footnotesize{linear}}) \nonumber	
	\\ 
	% Constraint C17
	& & \widebar{\mathrm{C}}_{17}: ~ &  C_u \leq \psi \chi_u S_\mathrm{max}, \forall u \in \mathcal{U}, & (\text{\footnotesize{linear}}) \nonumber	
	\\ 
	% Constraint C18
	& & \widebar{\mathrm{C}}_{18}: ~ & \textstyle \sum_{i \in \mathcal{U}} C_i \leq \log_2 \big( 1 + \mathsf{SINR}^{(\mathrm{c})}_u \big) + (1 - \chi_u) S_\mathrm{max}, \forall u \in \mathcal{U}, & (\text{\footnotesize{nonconvex}}) \nonumber	
	\\ 
	% Constraint C19
	& & \widebar{\mathrm{C}}_{19}: ~ & \log_2 \big( 1 + \mathsf{SINR}^{(\mathrm{p})}_u \big) + C_u \geq R_\mathrm{min} \chi_u, \forall u \in \mathcal{U}. & (\text{\footnotesize{nonconvex}}) \nonumber
\end{align}
\hrulefill
% Problem Q_CWSR_n
\begin{align} 
	% Objective
	\mathcal{Q}_{\mathrm{CWSR}_n}: & \max_{
			\substack{
						\mathbf{W}, \mathbf{m}, \mathbf{c}
			 }
	} 
	%\\
	& & f_{\mathrm{CWSR}_n} \left( \mathbf{W}, \mathbf{m}, \mathbf{c} \right) & (\text{\footnotesize{nonconvex}}) \nonumber
  	\\
	% Constraint 6
	& ~~~ \mathrm{s.t.} & \widebar{\mathrm{C}}_{6}: ~ & \textstyle \sum_{u \in \mathcal{U}_n'} \left\| \mathbf{w}_u \right\|_2^2 + \left\| \mathbf{m} \right\|_2^2  \leq P_\mathrm{tx}^\mathrm{max}, & (\text{\footnotesize{convex}}) \nonumber
	\\
	% Constraint C13
	& & \widebar{\mathrm{C}}_{13}: ~ & C_u \geq 0, \forall u \in \mathcal{U}_n', & (\text{\footnotesize{linear}}) \nonumber	
	\\ 	
	% Constraint C18
	& & \widebar{\mathrm{C}}_{18}: ~ &  \textstyle \sum_{i \in \mathcal{U}_n'} C_i \leq \log_2 \bigg( 1 + \frac{\left| \mathbf{h}^\mathrm{H}_u \mathbf{m} \right|^2 } { \sum_{i \in \mathcal{U}_n'} \left| \mathbf{h}_u^\mathrm{H} \mathbf{w}_i \right|^2 + {\sigma}^2} \bigg), \forall u \in \mathcal{U}_n', & (\text{\footnotesize{nonconvex}}) \nonumber	
	\\ 
	% Constraint C19
	& & \widebar{\mathrm{C}}_{19}: ~ & \log_2 \bigg( 1 + \textstyle \frac{\left| \mathbf{h}^\mathrm{H}_u \mathbf{w}_u \right|^2 } {\Delta_\mathrm{SIC}^2 \left| \mathbf{h}^\mathrm{H}_u \mathbf{m} \right|^2 +  \sum_{i \neq u, i \in \mathcal{U}_n'} \left| \mathbf{h}_u^\mathrm{H} \mathbf{w}_i \right|^2 + {\sigma}^2} \bigg) + C_u \geq R_\mathrm{min}, \forall u \in \mathcal{U}_n', & (\text{\footnotesize{nonconvex}}) \nonumber
	\\
  	% Constraint C20
   	& & \widebar{\mathrm{C}}_{20}: ~ & \left\| \mathbf{m}  \right\|_2^2 \leq \psi_0 P_\mathrm{tx}^\mathrm{max}, & (\text{\footnotesize{convex}}) \nonumber
\end{align}
\hrulefill
\vspace*{2pt}
\end{figure*}

We consider again WSR and WEE maximization as objectives, as in Section~\ref{subsection_problem_formulation_discrete_rates}. Thus, we define the corresponding optimization problems $ \mathcal{Q}_\mathrm{CWSR}' $ and $ \mathcal{Q}_\mathrm{CWEE}' $, shown at the top of the next page. To account for continuous rates, we have applied the following changes to problems $ \mathcal{P}_\mathrm{DWSR}' $ and $ \mathcal{P}_\mathrm{DWEE}' $ in Section~\ref{subsection_problem_formulation_discrete_rates}. First, we have eliminated binary variables $ \boldsymbol{\alpha}, \boldsymbol{\kappa} $ used for discrete rate selection. Second, we have reduced the number of binary variables by dropping $ \boldsymbol{\mu} $ and only use $ \boldsymbol{\chi}$ since $ \boldsymbol{\mu} = \boldsymbol{\chi} $, as rates are continuous. Hence, we could remove constraints $ \widebar{\mathrm{C}}_{3}, \widebar{\mathrm{C}}_{4}, \widebar{\mathrm{C}}_{7} - \widebar{\mathrm{C}}_{12} $ and employ Shannon's capacity formula to redefine constraints $ \widebar{\mathrm{C}}_{14}, \widebar{\mathrm{C}}_{15}, \widebar{\mathrm{C}}_{16} $. Specifically, we have replaced constraint $ \widebar{\mathrm{C}}_{14} $ with $ \widebar{\mathrm{C}}_{17}: C_u \leq \psi \chi_u S_\mathrm{max}, \forall u \in \mathcal{U} $, and constraint $ \widebar{\mathrm{C}}_{15} $ with $ \widebar{\mathrm{C}}_{18}: \textstyle \sum_{i \in \mathcal{U}} C_i \leq \log_2 \big( 1 + \mathsf{SINR}^{(\mathrm{c})}_u \big) + (1 - \chi_u) S_\mathrm{max}, \forall u \in \mathcal{U} $, where $ S_\mathrm{max} = \max_{u \in \mathcal{U}} \log_2 \left( 1+ \frac{P_\mathrm{tx}^\mathrm{max}}{\sigma^2} \big\| {\mathbf{h}}_u \big\|_2^2 \right) $ is an upper bound for the common rate. Note that $ \widebar{\mathrm{C}}_{18} $ is tighter when $ \chi_u = 1 $, and therefore the sum of the common rates is bounded by the minimum common rate of all admitted UEs. Finally, we have replaced $ \widebar{\mathrm{C}}_{16} $ with $ \widebar{\mathrm{C}}_{19}: \log_2 \big( 1 + \mathsf{SINR}^{(\mathrm{p})}_u \big) + C_u \geq R_\mathrm{min} \chi_u, \forall u \in \mathcal{U} $, and also redefined the objective functions using Shannon's capacity formula as $ f_\mathrm{CWSR}' \left( \mathbf{W}, \mathbf{m}, \mathbf{c} \right) \triangleq \textstyle \sum_{u \in \mathcal{U}} \omega_u \big( \log_2 \big( 1 + \mathsf{SINR}^{(\mathrm{p})}_u \big) + C_u \big) $ and $ f_\mathrm{CWEE}' \left( \mathbf{W}, \mathbf{m}, \mathbf{c} \right) \triangleq \frac{ \sum_{u \in \mathcal{U}} \omega_u \left( \log_2 \left( 1 + \mathsf{SINR}^{(\mathrm{p})}_u \right) + C_u \right) }{\frac{1}{\eta_\mathrm{eff}} \left( \sum_{u \in \mathcal{U}} \left\| \mathbf{w}_u \chi_u \right\|^2_2 + \left\| \mathbf{m} \psi \right\|^2_2 \right) + P_\mathrm{cir}} $.

% Remark
\vspace{1mm}
\noindent \textit{{\textsc{Remark 5:}} Problems $ \mathcal{Q}_\mathrm{CWSR}' $, $ \mathcal{Q}_\mathrm{CWEE}' $ are nonconvex MINLPs, and compared to  $ \mathcal{P}_\mathrm{CWSR}' $, $ \mathcal{P}_\mathrm{CWEE}' $, they assume continuous rates. In addition, their structure is more complex than that of $ \mathcal{P}_\mathrm{CWSR}' $, $ \mathcal{P}_\mathrm{CWEE}' $, as they involve multiplicative couplings of continuous variables, which are not present in $ \mathcal{P}_\mathrm{CWSR}' $, $ \mathcal{P}_\mathrm{CWEE}' $. }

% Section: Proposed Algorithms for solving Q': RSMA-MISOCP and SDMA-MISOCP
\subsection{Proposed Algorithm} \label{subsection_proposed_algorithm_continuous_rates}

To solve $ \mathcal{Q}_\mathrm{CWSR}' $ and $ \mathcal{Q}_\mathrm{CWEE}' $, we leverage successive convex approximation (SCA), semidefinite relaxation (SDR), and binary enumeration. In particular, we enumerate all combinations of admitted UEs and then solve the underlying nonconvex subproblem for each combination via the proposed \texttt{OPT-SCA-SDR} algorithm, which finds a KKT point by exploiting SCA and SDR. In the following, we describe the proposed algorithm by considering $ \mathcal{Q}_\mathrm{CWSR}' $ and then we extend it to $ \mathcal{Q}_\mathrm{CWEE}' $.

\subsubsection{Enumerating the Binary Variables} \label{subsubsection_enumerating_binary_variables}

Let $ N $ be the total number of combinations of admitted UEs and $\mathcal{N} = \left\lbrace 1, \dots, N \right\rbrace $ the set collecting them. Considering a given combination $ n \in \mathcal{N} $, problem $ \mathcal{Q}_\mathrm{CWSR}' $ reduces to $ \mathcal{Q}_{\mathrm{CWSR}_n} $, shown at the top of this page. In particular, $ f_{\mathrm{CWSR}_n}  \left( \mathbf{W}, \mathbf{m}, \mathbf{c} \right) \triangleq \sum_{u \in \mathcal{U}_n'} \omega_u \Big( \log_2 \Big( 1 + \frac{\left| \mathbf{h}^\mathrm{H}_u \mathbf{w}_u \right|^2 } {\Delta_\mathrm{SIC}^2 \left| \mathbf{h}^\mathrm{H}_u \mathbf{m} \right|^2 + \sum_{i \neq u, i \in \mathcal{U}_n'} \left| \mathbf{h}_u^\mathrm{H} \mathbf{w}_i \right|^2 + {\sigma}^2} \Big) + C_u \Big)  $, and $ \mathcal{U}_n' \subseteq \mathcal{U} $ denotes the set of admitted UEs in combination $ n $, such that $ \mu_u = 1 $, $ \forall u \in \mathcal{U}_n' $ and $ | \mathcal{U}_n' | = K $. For notational simplicity, we reset the UE indices in $ \mathcal{U}_n' $, such that $ \mathcal{U}_n' = \left\lbrace 1, \dots, K \right\rbrace $. Here, constraint $ \widebar{\mathrm{C}}_{20} $ is included to eliminate the coupling $ \mathbf{m} \psi_0 $ in an analogous manner as in Section~\ref{subsubsection_circumventing_mixed_integer_multiplicative_couplings} for constraint $ \mathrm{E_1} $. We have not included $ \mathrm{C_{17}} $ because it is implied by $ \widebar{\mathrm{C}}_{18}, \widebar{\mathrm{C}}_{20} $ when $ \psi_0 $ is given. We adopt $ \psi_0 = 1 $ for RSMA and $ \psi_0 = 0 $ for SDMA.

\subsubsection{Transforming the Problem via Sublevel and Superlevel Sets} \label{subsubsection_transforming_problem_via_sublevel_superlevel_sets}

We introduce nonnegative variables $ \boldsymbol{\gamma} \in \mathbb{R}_{+}^{K}, \boldsymbol{\rho} \in \mathbb{R}_{+}^{K}, \boldsymbol{\lambda} \in \mathbb{R}_{+}^{K}, \boldsymbol{\tau} \in \mathbb{R}_{+}^{K} $, and $ \beta \in \mathbb{R}_{+} $ to define sublevel and superlevel sets, thereby transforming problem $ \mathcal{Q}_{\mathrm{CWSR}_n} $ into $ \widetilde{\mathcal{Q}}_{\mathrm{CWSR}_n} $. In \textbf{\textbf{Appendix \ref{appendix_proposition_7}}}, we show that $ \mathcal{Q}_{\mathrm{CWSR}_n} $ and $ \widetilde{\mathcal{Q}}_{\mathrm{CWSR}_n} $ are equivalent. Specifically, we bound the private SINRs from below via $ \frac{\left| \mathbf{h}^\mathrm{H}_u \mathbf{w}_u \right|^2 } {\Delta_\mathrm{SIC}^2 \left| \mathbf{h}^\mathrm{H}_u \mathbf{m} \right|^2 + \sum_{i \neq u, i \in \mathcal{U}_n'} \left| \mathbf{h}_u^\mathrm{H} \mathbf{w}_i \right|^2 + {\sigma}^2} \geq \gamma_u - 1 $. Also, we bound the interference at each UE from above by including $ \Delta_\mathrm{SIC}^2 \left| \mathbf{h}^\mathrm{H}_u \mathbf{m} \right|^2 + $ $ \sum_{i \neq u, i \in \mathcal{U}_n'} \left| \mathbf{h}_u^\mathrm{H} \mathbf{w}_i \right|^2 + {\sigma}^2 \leq \rho_u $. Following the same idea, we include $ \frac{\left| \mathbf{h}^\mathrm{H}_u \mathbf{m} \right|^2 } { \sum_{i \in \mathcal{U}_n'} \left| \mathbf{h}_u^\mathrm{H} \mathbf{w}_i \right|^2 + {\sigma}^2} \geq \tau_u - 1$ and $ \sum_{i \in \mathcal{U}_n'} \left| \mathbf{h}_u^\mathrm{H} \mathbf{w}_i \right|^2 + {\sigma}^2 \leq \lambda_u $ to bound the common SINRs and the interference. Furthermore, we bound the objective function from below, such that $ f_{\mathrm{CWSR}_n} \left( \mathbf{W}, \mathbf{m}, \mathbf{c} \right) \geq \beta $, thus defining a new objective function $ f_{\mathrm{CWSR}_n} \left( \beta \right) \triangleq \beta $. Upon applying these transformations to $ \mathcal{Q}_{\mathrm{CWSR}_n} $, we obtain problem $ \widetilde{\mathcal{Q}}_{\mathrm{CWSR}_n} $, shown at the top of the next page.

% Problems
\begin{figure*}[!t]
% ensure that we have normalsize text
\normalsize
% Problem Q_tilde_n
\begin{align} 
	% Objective
	\widetilde{\mathcal{Q}}_{\mathrm{CWSR}_n}: & \max_{
			\substack{
						\mathbf{W}, \mathbf{m}, \mathbf{c}, \boldsymbol{\gamma}, \boldsymbol{\rho}, \boldsymbol{\lambda}, \boldsymbol{\tau}, \beta
			 }
	}
	& & f_{\mathrm{CWSR}_n} \left( \beta \right) \triangleq \beta \nonumber
  	\\
	% Constraint K1
	& ~~~~~~~~~ \mathrm{s.t.} & \widebar{\mathrm{K}}_{1}: ~ & \left| \mathbf{h}^\mathrm{H}_u \mathbf{w}_u \right|^2 \geq \left( \gamma_u - 1\right) \rho_u, \forall u \in \mathcal{U}_n', & (\text{\footnotesize{nonconvex}}) \nonumber
	\\
	% Constraint K2
	& & \widebar{\mathrm{K}}_{2}: ~ & \Delta_\mathrm{SIC}^2 \left| \mathbf{h}^\mathrm{H}_u \mathbf{m} \right|^2 + \textstyle \sum_{i \neq u, i \in \mathcal{U}_n'} \left| \mathbf{h}_u^\mathrm{H} \mathbf{w}_i \right|^2 + {\sigma}^2 \leq \rho_u, \forall u \in \mathcal{U}_n', & (\text{\footnotesize{convex}}) \nonumber
	\\
	% Constraint K3
	& & \widebar{\mathrm{K}}_{3}: ~ & \beta - \textstyle \sum_{u \in \mathcal{U}_n'} \omega_u \left( \log_2 \left( \gamma_u \right) + C_u \right) \leq 0, & (\text{\footnotesize{convex}}) \nonumber
	\\
	% Constraint K4
	& & \widebar{\mathrm{K}}_{4}: ~ & \left| \mathbf{h}^\mathrm{H}_u \mathbf{m} \right|^2  \geq \left( \tau_u - 1 \right) \lambda_u, \forall u \in \mathcal{U}_n', & (\text{\footnotesize{nonconvex}}) \nonumber
	\\
	% Constraint K5
	& & \widebar{\mathrm{K}}_{5}: ~ & \textstyle \sum_{i \in \mathcal{U}_n'} \left| \mathbf{h}_u^\mathrm{H} \mathbf{w}_i \right|^2 + {\sigma}^2 \leq \lambda_u, \forall u \in \mathcal{U}_n', & (\text{\footnotesize{convex}}) \nonumber
	\\
	% Constraint K6
	& & \widebar{\mathrm{K}}_{6}: ~ & \textstyle \sum_{i \in \mathcal{U}_n'} C_i - \log_2 \left( \tau_u \right) \leq 0, \forall u \in \mathcal{U}_n', & (\text{\footnotesize{convex}}) \nonumber  
	\\
	% Constraint K7
	& & \widebar{\mathrm{K}}_{7}: ~ & R_\mathrm{min} - \log_2 \left( \gamma_u \right) - C_u \leq 0, \forall u \in \mathcal{U}_n', & (\text{\footnotesize{convex}}) \nonumber 
	\\
	% Constraint K8
	& & \widebar{\mathrm{K}}_{8}: ~ & \beta \geq 0, & (\text{\footnotesize{linear}}) \nonumber
	\\
	% Constraint K9
	& & & \widebar{\mathrm{C}}_{6}, \widebar{\mathrm{C}}_{13}, \widebar{\mathrm{C}}_{20}. \nonumber
\end{align}
\hrulefill
% Constraints L1-L10
\begin{equation} \nonumber
	\widebar{\mathrm{C}}_{6}, \widebar{\mathrm{C}}_{20}, \widebar{\mathrm{K}}_{1}, \widebar{\mathrm{K}}_{2}, \widebar{\mathrm{K}}_{4}, \widebar{\mathrm{K}}_{5} \Leftrightarrow
		\begin{cases}
			   	\widebar{\mathrm{L}}_{1}: \textstyle \sum_{u \in \mathcal{U}_n'} \mathrm{Tr} \left( \mathbf{W}_u \right) + \mathrm{Tr} \left( \mathbf{M} \right) \leq P_\mathrm{tx}^\mathrm{max}, ~~ \widebar{\mathrm{L}}_{2}: \mathrm{Tr} \left( \mathbf{M} \right) \leq \psi_0 P_\mathrm{tx}^\mathrm{max}, & (\text{\footnotesize{linear}}) \nonumber
			   	%\\
			   	%\widebar{\mathrm{L}}_{2}: \mathrm{Tr} \left( \mathbf{M} \right) \leq \psi_0 P_\mathrm{tx}^\mathrm{max}, & (\text{\footnotesize{linear}}) \nonumber
			   	\\
			   	\widebar{\mathrm{L}}_{3}: \left( \gamma_u - 1\right) \rho_u - \mathbf{h}_u^\mathrm{H} {\mathbf{W}_u} \mathbf{h}_u \leq 0, \forall u \in \mathcal{U}_n', & (\text{\footnotesize{nonconvex}})
			   	\\
			   	\widebar{\mathrm{L}}_{4}: \Delta_\mathrm{SIC}^2 \mathbf{h}^\mathrm{H}_u \mathbf{M} \mathbf{h}_u + \textstyle \sum_{ i \neq u, i \in \mathcal{U}_n' } \mathbf{h}_u^\mathrm{H} {\mathbf{W}_i} \mathbf{h}_u + \sigma^2 \leq \rho_u, \forall u \in \mathcal{U}_n', & (\text{\footnotesize{linear}})
			   	\\
			   	\widebar{\mathrm{L}}_{5}: \left( \tau_u - 1 \right) \lambda_u - \mathbf{h}_u^\mathrm{H} {\mathbf{M}} \mathbf{h}_u \leq 0, \forall u \in \mathcal{U}_n', & (\text{\footnotesize{nonconvex}})
			   	\\
			   	\widebar{\mathrm{L}}_{6}: \textstyle \sum_{ i \in \mathcal{U}_n' } \mathbf{h}_u^\mathrm{H} {\mathbf{W}_i} \mathbf{h}_u + \sigma^2 \leq \lambda_u, \forall u \in \mathcal{U}_n', & (\text{\footnotesize{linear}})
			   	\\
			   	\widebar{\mathrm{L}}_{7}: \mathbf{W}_u \succcurlyeq 0, \forall u \in \mathcal{U}_n', & (\text{\footnotesize{linear}})
			    \\
			   	\widebar{\mathrm{L}}_{8}: \mathbf{M} \succcurlyeq 0, & (\text{\footnotesize{linear}})	
			   	\\
			   	\widebar{\mathrm{L}}_{9}: \mathrm{Rank} \left( \mathbf{W}_u \right) \leq 1, \forall u \in \mathcal{U}_n', ~~ \widebar{\mathrm{L}}_{10}: \mathrm{Rank} \left( \mathbf{M} \right) \leq \psi_0. & (\text{\footnotesize{nonconvex}})	
			   	%\\
			   	%\widebar{\mathrm{L}}_{10}: \mathrm{Rank} \left( \mathbf{M} \right) \leq \psi_0, & (\text{\footnotesize{nonconvex}})	
		\end{cases}
\end{equation}
\hrulefill
% Constraints M1-M2
\begin{equation} \nonumber
	\widebar{\mathrm{L}}_{3}, \widebar{\mathrm{L}}_{5} \Leftrightarrow
		\begin{cases}
			   	\widebar{\mathrm{M}}_{1}: \frac{\bar{\Omega}_{1,u}^{(t)}}{2} \gamma_u^2 + \frac{1}{2 \bar{\Omega}_{1,u}^{(t)}} \rho_u^2 - \rho_u - \mathbf{h}_u^\mathrm{H} {\mathbf{W}_u} \mathbf{h}_u \leq 0, \forall u \in \mathcal{U}_n', & (\text{\footnotesize{convex}}) \nonumber
			   	\\
			   	\widebar{\mathrm{M}}_{2}: \frac{\bar{\Omega}_{2,u}^{(t)}}{2} \tau_u^2 + \frac{1}{2 \bar{\Omega}_{2,u}^{(t)}} \lambda_u^2 - \lambda_u - \mathbf{h}_u^\mathrm{H} {\mathbf{M}} \mathbf{h}_u \leq 0, \forall u \in \mathcal{U}_n', & (\text{\footnotesize{convex}}) \nonumber
		\end{cases}
\end{equation}
\hrulefill
% Constraints M3-M4
\begin{equation} \nonumber
	\widebar{\mathrm{L}}_{9}, \widebar{\mathrm{L}}_{10} \Leftrightarrow
		\begin{cases}
			   	\widebar{\mathrm{M}}_{3}: \zeta_0 \mathbf{I} - {\mathbf{T}_0^{(t)}}^\mathrm{H} \mathbf{M} {\mathbf{T}_0^{(t)}} \succcurlyeq \mathbf{0}, ~~ \widebar{\mathrm{M}}_{4}: \zeta_u \mathbf{I} - {\mathbf{T}_u^{(t)}}^\mathrm{H} \mathbf{W}_u {\mathbf{T}_u^{(t)}} \succcurlyeq \mathbf{0}, \forall u \in \mathcal{U}_n', & (\text{\footnotesize{convex}}) \nonumber
			   	%\\
			   	%\widebar{\mathrm{M}}_{4}: \zeta_u \mathbf{I} - {\mathbf{T}_u^{(t)}}^\mathrm{H} \mathbf{W}_u {\mathbf{T}_u^{(t)}} \succcurlyeq \mathbf{0}, \forall u \in \mathcal{U}_n', & (\text{\footnotesize{convex}}) \nonumber
		\end{cases}
\end{equation}
\hrulefill
\vspace*{2pt}
\end{figure*}

\subsubsection{Leveraging Semidefinite Programming} \label{subsubsection_leveraging_semidefinite_programming}

%To circumvent in part the nonconvexity of constraints $ \widebar{\mathrm{K}}_{1}, \widebar{\mathrm{K}}_{4} $, we employ semidefinite programming, which linearizes the quadratic terms on the left-hand side. For this, we introduce positive semidefinite variables $ \mathbf{W}_u \in \mathbb{C}^{N_\mathrm{tx} \times N_\mathrm{tx}} $ and $ \mathbf{M} \in \mathbb{C}^{N_\mathrm{tx} \times N_\mathrm{tx}} $, which replace $ \mathbf{w}_u \mathbf{w}_u^\mathrm{H} $ and $ \mathbf{m} \mathbf{m}^\mathrm{H} $, respectively. The newly introduced variables also affect constraints $ \widebar{\mathrm{K}}_{2}, \widebar{\mathrm{K}}_{5} $, $ \widebar{\mathrm{C}}_{6}, \widebar{\mathrm{C}}_{20} $. Thus, constraints $ \widebar{\mathrm{C}}_{6}, \widebar{\mathrm{C}}_{20}, \widebar{\mathrm{K}}_{1}, \widebar{\mathrm{K}}_{2}, \widebar{\mathrm{K}}_{4}, \widebar{\mathrm{K}}_{5} $ are equivalently reformulated as $ \widebar{\mathrm{L}}_{1} - \widebar{\mathrm{L}}_{10} $, as shown below:

By employing semidefinite programming and introducing positive semidefinite variables $ \mathbf{W}_u \in \mathbb{C}^{N_\mathrm{tx} \times N_\mathrm{tx}} $ and $ \mathbf{M} \in \mathbb{C}^{N_\mathrm{tx} \times N_\mathrm{tx}} $, which replace $ \mathbf{w}_u \mathbf{w}_u^\mathrm{H} $ and $ \mathbf{m} \mathbf{m}^\mathrm{H} $, respectively, constraints $ \widebar{\mathrm{C}}_{6} $, $ \widebar{\mathrm{C}}_{20} $, $ \widebar{\mathrm{K}}_{1} $, $ \widebar{\mathrm{K}}_{2} $, $ \widebar{\mathrm{K}}_{4} $, $ \widebar{\mathrm{K}}_{5} $ can be equivalently reformulated to $ \widebar{\mathrm{L}}_{1} - \widebar{\mathrm{L}}_{10} $, shown at the top of the next page. In doing so, the nonconvexity of constraints $ \widebar{\mathrm{K}}_{1}, \widebar{\mathrm{K}}_{4} $ are circumvented in part as the quadratic terms on the left-hand side are linearized. The newly introduced variables also affect constraints $ \widebar{\mathrm{K}}_{2}, \widebar{\mathrm{K}}_{5} $, $ \widebar{\mathrm{C}}_{6}, \widebar{\mathrm{C}}_{20} $. The positive semidefiniteness of $ \mathbf{W}_u $ and $ \mathbf{M} $ are specified by $ \widebar{\mathrm{L}}_{7} $, $ \widebar{\mathrm{L}}_{8} $, whereas $ \widebar{\mathrm{L}}_{9}, \widebar{\mathrm{L}}_{10} $ allow for the private and common signals to be used or not. Considering the equivalence between $ \widebar{\mathrm{C}}_{6}, \widebar{\mathrm{C}}_{20}, \widebar{\mathrm{K}}_{1}, \widebar{\mathrm{K}}_{2}, \widebar{\mathrm{K}}_{4}, \widebar{\mathrm{K}}_{5} $ and $ \widebar{\mathrm{L}}_{1} - \widebar{\mathrm{L}}_{10} $, we define problem 
%\begin{align} \nonumber
%	\widehat{\mathcal{Q}}_{\mathrm{CWSR}_n}: \max_{ \widehat{\mathbf{W}}, \mathbf{M}, \mathbf{c}, \boldsymbol{\gamma}, \boldsymbol{\rho}, \boldsymbol{\lambda}, \boldsymbol{\tau}, \beta} f_{\mathrm{CWSR}_n} \left( \beta \right) ~ \mathrm{s.t.} ~ \widebar{\mathrm{C}}_{13}, \widebar{\mathrm{K}}_{3}, \widebar{\mathrm{K}}_{6} - \widebar{\mathrm{K}}_{8}, \widebar{\mathrm{L}}_{1} - \widebar{\mathrm{L}}_{10}, 
%\end{align}
\begin{align} 
	% Objective
	\widehat{\mathcal{Q}}_{\mathrm{CWSR}_n}: & \max_{
			\substack{
					\widehat{\mathbf{W}}, \mathbf{M}, \mathbf{c}, \boldsymbol{\gamma}, \boldsymbol{\rho}, \boldsymbol{\lambda}, \boldsymbol{\tau}, \beta
			 }
	} 
	& & f_{\mathrm{CWSR}_n} \left( \beta \right) & \nonumber	
	\\
	% Constraints
	& ~~~~~~~~~~~ \mathrm{s.t.} & & \widebar{\mathrm{C}}_{13}, \widebar{\mathrm{K}}_{3}, \widebar{\mathrm{K}}_{6} - \widebar{\mathrm{K}}_{8}, \widebar{\mathrm{L}}_{1} - \widebar{\mathrm{L}}_{10}, & \nonumber	
\end{align}
where $ \widehat{\mathbf{W}} = \left( \mathbf{W}_1, \dots, \mathbf{W}_K \right) $. We note that $ \widehat{\mathcal{Q}}_{\mathrm{CWSR}_n} $ is equivalent to $ \widetilde{\mathcal{Q}}_{\mathrm{CWSR}_n} $ and $ \mathcal{Q}_{\mathrm{CWSR}_n} $ since the feasible set and objective function are not affected by the applied transformation of the constraints.

\subsubsection{Addressing the Nonconvex Constraints} \label{subsubsection_addressing_nonconvex_constraints}

To cope with the nonconvex constraints $ \widebar{\mathrm{L}}_{3}, \widebar{\mathrm{L}}_{5}, \widebar{\mathrm{L}}_{9}, \widebar{\mathrm{L}}_{10} $, we adopt an iterative approach whereby we sequentially approximate these constraints by convex approximations. 

$ \bullet $ \emph{Quasi-convex constraints:} To circumvent the quasi-convex constraints $ \widebar{\mathrm{L}}_{3}, \widebar{\mathrm{L}}_{5} $, we replace them with the inner convex approximations $ \widebar{\mathrm{M}}_{1}, \widebar{\mathrm{M}}_{3} $, shown at the top of the next page, where $ t $  is the iteration index, and $ \bar{\Omega}_{1,u}^{(t)} $, $ \bar{\Omega}_{2,u}^{(t)} $, $ u \in \mathcal{U}_n' $, are parameters adapted iteratively. In recasting $ \widebar{\mathrm{L}}_{3}, \widebar{\mathrm{L}}_{5} $ as $ \widebar{\mathrm{M}}_{1}, \widebar{\mathrm{M}}_{2} $, we have employed the arithmetic-geometric mean inequality, which states that $ a b \leq \frac{a^2}{2} + \frac{b^2}{2} $ for $ a, b \in \mathbb{R}_{+} $. By introducing a new parameter $ \Phi \in \mathbb{R}_{+} $ and applying transformations $ a \leftarrow \sqrt{\Phi}a $ and $ b \leftarrow \sqrt{\frac{1}{\Phi}}b $, we obtain inequality $ a b \leq \frac{\Phi }{2} a^2 + \frac{1}{2 \Phi } b^2 $, which becomes tight when $ \Phi = \frac{b}{a} $ \cite{beck2010:sequential-parametric-convex-approximation-method-applications-nonconvex-truss-topology-design-problems}. Note that $ \frac{\Phi }{2} a^2 + \frac{1}{2 \Phi } b^2 $ is a convex overestimate of $ a b $, which decouples $ a $ and $ b $, allowing to circumvent the nonconvexity of the product $ ab $. Exploiting this observation, we introduce parameter $ \bar{\Omega}_{1,u}^{(t)} $ and apply the parameterized inequality to the product $ \gamma_u \rho_u $ in $ \widebar{\mathrm{L}}_{3} $, such that $ \gamma_u \rho_u \leq \frac{\bar{\Omega}_{1,u}^{(t)} }{2} \gamma_u^2 + \frac{1}{2 \bar{\Omega}_{1,u}^{(t)} } \rho_u^2 $. We proceed in a similar manner with constraint $ \widebar{\mathrm{L}}_{5} $ by introducing $ \Omega_{2,u}^{(t)} $, which yields $ \tau_u \lambda_u \leq \frac{\bar{\Omega}_{2,u}^{(t)} }{2} \tau_u^2 + \frac{1}{2 \bar{\Omega}_{2,u}^{(t)} } \lambda_u^2 $. Next, we replace products $  \gamma_u \rho_u $ and $ \tau_u \lambda_u $ with their respective convex overestimates, $ \frac{\bar{\Omega}_{1,u}^{(t)} }{2} \gamma_u^2 + \frac{1}{2 \bar{\Omega}_{1,u}^{(t)} } \rho_u^2 $ and $ \frac{\bar{\Omega}_{2,u}^{(t)} }{2} \tau_u^2 + \frac{1}{2 \bar{\Omega}_{2,u}^{(t)} } \lambda_u^2 $, thus yielding $ \widebar{\mathrm{M}}_{1}, \widebar{\mathrm{M}}_{2} $. In each iteration $ t $, we update the parameters according to $ \bar{\Omega}_{1,u}^{(t)} = \frac{\rho_u^{(t-1)} }{\gamma_u^{(t-1)} } $ and $ \bar{\Omega}_{2,u}^{(t)} = \frac{\lambda_u^{(t-1)} }{\tau_u^{(t-1)} } $ making it possible to sequentially adapt the convex approximation. Upon replacing $ \widebar{\mathrm{L}}_{3}, \widebar{\mathrm{L}}_{5} $ with $ \widebar{\mathrm{M}}_{1}, \widebar{\mathrm{M}}_{2} $ in problem $ \widehat{\mathcal{Q}}_{\mathrm{CWSR}_n} $, and then solving it, an optimal solution to this modified problem will be feasible for $ \widehat{\mathcal{Q}}_{\mathrm{CWSR}_n} $, $ \widetilde{\mathcal{Q}}_{\mathrm{CWSR}_n} $, and $ \mathcal{Q}_{\mathrm{CWSR}_n} $ since the feasible set of $ \widebar{\mathrm{M}}_{1}, \widebar{\mathrm{M}}_{2} $ is contained in that of $ \widebar{\mathrm{L}}_{3}, \widebar{\mathrm{L}}_{5} $. However, the solution will not necessarily be globally optimal for $ \widehat{\mathcal{Q}}_{\mathrm{CWSR}_n} $, $ \widetilde{\mathcal{Q}}_{\mathrm{CWSR}_n} $, and $ \mathcal{Q}_{\mathrm{CWSR}_n} $ due to the possible reduction of the feasible set caused by the inner convexification in $ \widebar{\mathrm{M}}_{1}, \widebar{\mathrm{M}}_{2} $.

$ \bullet $ \emph{Rank constraints:} In order to cope with rank constraints $ \widebar{\mathrm{L}}_{9}, \widebar{\mathrm{L}}_{10} $, we adopt the iterative method proposed in \cite{sun2019:iterative-rank-penalty-method-nonconvex-quadratically-constrained-quadratic-programs}, which is described as follows. We first reformulate $ \widebar{\mathrm{L}}_{9} $, $ \widebar{\mathrm{L}}_{10} $ as $ \widebar{\mathrm{M}}_{3} $, $ \widebar{\mathrm{M}}_{4} $, shown at the top of this page. Then, we penalize the objective function by adding cost function $ \sum_{u \in \mathcal{U}_n' \cup \left\lbrace 0 \right\rbrace} p_u^{(t)} \zeta_u $, which promotes rank minimization and enforces $ \mathbf{M} $ and $ \mathbf{W}_u $ to have rank at most one, as shown in \textbf{Appendix \ref{appendix_proposition_8}}. The $ \zeta_u $, $ \forall u \in \mathcal{U}_n' \cup \left\lbrace 0 \right\rbrace $, are slack variables and $ p_u^{(t)} \in \mathbb{R}_{+} $, $ \forall u \in \mathcal{U}_n' \cup \left\lbrace 0 \right\rbrace $, represent the penalty weights in iteration $ t $. Matrices $ \widebar{\mathbf{M}}^{(t-1)} $ and $ \widebar{\mathbf{W}}_u^{(t-1)} $ are the respective solutions for $ \mathbf{M} $ and $ \mathbf{W}_u $, obtained in iteration $ t - 1 $. Also, $ \mathbf{T}_0^{(t)} \in \mathbb{C}^{N_\mathrm{tx} \times (N_\mathrm{tx} - 1)} $ is formed by the eigenvectors of the $ N_\mathrm{tx} - 1 $ smallest eigenvalues of $ \widebar{\mathbf{M}}^{(t-1)} $, whereas $ \mathbf{T}_u^{(t)} \in \mathbb{C}^{N_\mathrm{tx} \times (N_\mathrm{tx} - 1)} $ is formed by the eigenvectors of the $ N_\mathrm{tx} - 1 $ smallest eigenvalues of $ \widebar{\mathbf{W}}_u^{(t-1)} $.

% Subsection
\subsubsection{Outlining the Algorithm and Its Extension to Solve $ \mathcal{Q}_\mathrm{CWEE}' $} \label{subsubsection_outlining_opt_sca_sdr_algorithm}

The transformation of constraints $ \widebar{\mathrm{L}}_{3}, \widebar{\mathrm{L}}_{5}, \widebar{\mathrm{L}}_{9}, \widebar{\mathrm{L}}_{10} $ into $ \widebar{\mathrm{M}}_{1} - \widebar{\mathrm{M}}_{4} $, leads to the following problem
\begin{align} 
	% Objective
	\bar{\mathcal{Q}}_{\mathrm{CWSR}_n}^{(t)}: & \max_{
			\substack{
						\widehat{\mathbf{W}}, \mathbf{M}, \mathbf{c}, \boldsymbol{\gamma}, \boldsymbol{\rho}, \\ \boldsymbol{\lambda}, \boldsymbol{\tau}, \boldsymbol{\zeta}, \beta
			 }
	} 
	& & f_{\mathrm{CWSR}_n} \left( \beta \right) - \textstyle \sum_{u \in \mathcal{U}_n' \cup \left\lbrace 0 \right\rbrace} p_u^{(t)} \zeta_u & \nonumber	
	\\
	% Constraints
	& ~~~~~~~~ \mathrm{s.t.} & & \widebar{\mathrm{C}}_{13}, \widebar{\mathrm{K}}_{3}, \widebar{\mathrm{K}}_{6} - \widebar{\mathrm{K}}_{8}, \widebar{\mathrm{L}}_{1}, \widebar{\mathrm{L}}_{2},   & \nonumber	
	\\
	& & & \widebar{\mathrm{L}}_{4}, \widebar{\mathrm{L}}_{6} - \widebar{\mathrm{L}}_{8}, \widebar{\mathrm{M}}_{1} - \widebar{\mathrm{M}}_{4}. & \nonumber	
\end{align}

On the other hand, to solve $ \mathcal{Q}_\mathrm{CWEE}' $, we introduce the following problem 
\begin{align} 
	% Objective
	\bar{\mathcal{Q}}_{\mathrm{CWEE}_n}^{(t)}: & \max_{
			\substack{
						\widehat{\mathbf{W}}, \mathbf{M}, \mathbf{c}, \boldsymbol{\gamma}, \boldsymbol{\rho},\\ \boldsymbol{\lambda}, \boldsymbol{\tau}, \boldsymbol{\zeta}, \beta, \theta, \delta
			 }
	} 
	& & f_{\mathrm{CWEE}_n}  \left( \theta \right) - \textstyle \sum_{u \in \mathcal{U}_n' \cup \left\lbrace 0 \right\rbrace} p_u^{(t)} \zeta_u & \nonumber	
	\\
	% Constraints
	& ~~~~~~~~ \mathrm{s.t.} & & \widebar{\mathrm{C}}_{13}, \widebar{\mathrm{K}}_{3}, \widebar{\mathrm{K}}_{6} - \widebar{\mathrm{K}}_{8}, \widebar{\mathrm{L}}_{1}, \widebar{\mathrm{L}}_{2}, \widebar{\mathrm{L}}_{4}, & \nonumber	
	\\
	& & & \widebar{\mathrm{L}}_{6} - \widebar{\mathrm{L}}_{8}, \widebar{\mathrm{M}}_{1} - \widebar{\mathrm{M}}_{4}, \widebar{\mathrm{N}}_{1} - \widebar{\mathrm{N}}_{3}, & \nonumber	
\end{align}
where we employed the same procedure as described in Section~\ref{subsubsection_enumerating_binary_variables} to Section~\ref{subsubsection_addressing_nonconvex_constraints}. Compared to $ \bar{\mathcal{Q}}_{\mathrm{CWSR}_n}^{(t)} $, problem $ \bar{\mathcal{Q}}_{\mathrm{CWEE}_n}^{(t)} $ features variables $ \theta $ and $ \delta $, convex constraints $ \widebar{\mathrm{N}}_{1}: \textstyle \sum_{u \in \mathcal{U}_n'} \mathrm{Tr} \left( \mathbf{W}_u \right) + \mathrm{Tr} \left( \mathbf{M} \right) \leq  \eta_\mathrm{eff} \delta $, $ \widebar{\mathrm{N}}_{2}: \frac{\bar{\Omega}_{3}^{(t)}}{2} \theta^2 + \frac{1}{2 \bar{\Omega}_{3}^{(t)}} \delta^2 + \theta P_\mathrm{cir} \leq \beta $ and $ \widebar{\mathrm{N}}_{3}: \theta \geq 0 $, and parameter $ \bar{\Omega}_{3}^{(t)} $. In particular, $ \theta $ is used to bound the objective function $ f_{\mathrm{CWEE}_n} \left( \mathbf{W}, \mathbf{m}, \mathbf{c} \right) $ from below. Variable $ \delta $ is used to bound the transmit power efficiency from above, thereby yielding constraint $ \widebar{\mathrm{N}}_{1} $. The introduction of $ \delta $ and $ \theta $ in the objective function leads to a multiplicative coupling $ \theta \delta $, which is dealt with in the same manner as in Section~\ref{subsubsection_addressing_nonconvex_constraints}, yielding $ \widebar{\mathrm{N}}_{2} $. Also, $ \widebar{\mathrm{N}}_{3} $ is added to ensure the positiveness of the objective function. Parameter $ \bar{\Omega}_{3}^{(t)} $ is updated as $ \bar{\Omega}_3^{(t)} = \frac{\delta^{(t-1)} }{\theta^{(t-1)} } $ and the objective function is penalized by $ \textstyle \sum_{u \in \mathcal{U}_n' \cup \left\lbrace 0 \right\rbrace} p_u^{(t)} \zeta_u $.

% Equations
\begin{figure*}[!t]
% ensure that we have normalsize text
\small
\begin{align} 
	& R_{u,n}^\mathrm{proj} = \left\lbrace R_{j} \mid j = \arg \min_{i \in \mathcal{J}} \widebar{\mathsf{SINR}}^{(\mathrm{p})}_{u,n} - \Gamma_i, \widebar{\mathsf{SINR}}^{(\mathrm{p})}_{u,n} \geq \Gamma_i \right\rbrace \label{equation_projection_private_rate}
\end{align} 
\hrulefill
\begin{align} 
	& \widebar{C}_{u,n}^\mathrm{proj} = \left\lbrace \frac{\widebar{C}_{u,n}}{ \sum_{u \in \mathcal{U}_n'} \widebar{C}_{u,n}} R_{j} \mid j = \arg \min_{i \in \mathcal{J}} \left\lbrace \min_{u \in \mathcal{U}_n'} \widebar{\mathsf{SINR}}^{(\mathrm{c})}_{u,n} \right\rbrace - \Gamma_i, \left\lbrace \min_{u \in \mathcal{U}_n'} \widebar{\mathsf{SINR}}^{(\mathrm{c})}_{u,n} \right\rbrace \geq \Gamma_i \right\rbrace 
	\label{equation_projection_common_rate}
\end{align} 
\hrulefill
\vspace*{2pt}
\end{figure*}

Problems $ \bar{\mathcal{Q}}_{\mathrm{CWSR}_n}^{(t)} $ and $ \bar{\mathcal{Q}}_{\mathrm{CWEE}_n}^{(t)} $ are convex and can be solved optimally via IPMs. Both are solved iteratively, improving the objective function in each iteration until a stop criterion is met, i.e., the difference of the objective function values between successive iterations is less than a threshold $ \epsilon $ or the number of iterations exceeds $ N_\mathrm{iter} $. In \textbf{Appendix \ref{appendix_proposition_9}}, we show that $ \bar{\mathcal{Q}}_{\mathrm{CWSR}_n}^{(t)} $ and $ \bar{\mathcal{Q}}_{\mathrm{CWEE}_n}^{(t)} $ converges to a KKT point. Also, by increasing the penalty weights $ p_u^{(t)} $, variables $ \zeta_u $ decrease in each iteration, leading to $ \zeta_u \rightarrow 0 $ and $ \textstyle \sum_{u \in \mathcal{U}_n' \cup \left\lbrace 0 \right\rbrace} p_u^{(t)} \zeta_u \rightarrow 0 $. This causes $ \mathbf{M} $ and $ \mathbf{W}_u $ to have at most rank one, since the $ N_\mathrm{tx} -1 $ smallest eigenvalues of these matrices are progressively squeezed to zero. Assuming that $ \bar{\mathcal{Q}}_{\mathrm{CWSR}_n}^{(t)} $ converge in iteration $ t^\star $, we have that $ \mathbf{M} \approx \widebar{\mathbf{M}}^{(t^\star-1)} $, where $ \mathbf{M} $ is the solution in iteration $ t^\star $. Via eigendecomposition of $ \mathbf{M} $, we have $ \mathbf{M} = \widetilde{\mathbf{R}}_0 \boldsymbol{\Sigma}_0 \widetilde{\mathbf{R}}_0^\mathrm{H} $, such that $ \widetilde{\mathbf{R}}_0 \widetilde{\mathbf{R}}_0^\mathrm{H} = \mathbf{I} $, $ \boldsymbol{\Sigma}_0 = \mathrm{diag} \left( \sigma_{0,1}, \dots, \sigma_{0,N_\mathrm{tx}} \right)  $, and $ \widetilde{\mathbf{R}}_0 = \left[ \mathbf{r}_0 | \mathbf{R}_0 \right] $. Therefore, $  {\mathbf{T}_0^{(t^\star)}}^\mathrm{H} \mathbf{M} {\mathbf{T}_0^{(t^\star)}} = {\mathbf{T}_0^{(t^\star)}}^\mathrm{H} \left[ \mathbf{r}_0 | \mathbf{R}_0 \right] \boldsymbol{\Sigma}_0 \left[ \mathbf{r}_0 | \mathbf{R}_0 \right]^\mathrm{H} {\mathbf{T}_0^{(t^\star)}} $, which can be further reduced to $ {\mathbf{T}_0^{(t^\star)}}^\mathrm{H} \mathbf{M} {\mathbf{T}_0^{(t^\star)}} = \left[ \mathbf{0} | \mathbf{I} \right] \boldsymbol{\Sigma}_0 \left[ \mathbf{0} | \mathbf{I} \right]^\mathrm{H} = \mathrm{diag} \left( \sigma_{0,2}, \dots, \sigma_{0,N_\mathrm{tx}} \right) $, since $ {\mathbf{T}_0^{(t^\star)}}^\mathrm{H} \mathbf{r}_0 \approx \mathbf{0} $ and $ {\mathbf{T}_0^{(t^\star)}}^\mathrm{H} \mathbf{R}_0 \approx \mathbf{I} $. Considering these outcomes and $ \widebar{\mathrm{M}}_{3} $, we obtain $ \zeta_0 \mathbf{I} \succcurlyeq \mathrm{diag} \left( \sigma_{0,2}, \dots, \sigma_{0,N_\mathrm{tx}} \right) $, which leads to $ \sigma_{0,2}, \dots, \sigma_{0,N_\mathrm{tx}} \rightarrow 0 $ as $ \zeta_0 \rightarrow 0 $. Because $ \sigma_{0,1} $ is not affected by this procedure, $ \sigma_{0,1} $ can be different from zero or even zero, i.e., $ \mathbf{M} $ can be at most rank-one. Following the same reasoning, we can obtain equivalent results for $ {\mathbf{T}_u^{(t^\star)}}^\mathrm{H} \mathbf{W}_u {\mathbf{T}_u^{(t^\star)}} $, $ \forall u \in \mathcal{U}_n' $. The solutions that satisfy $ \widebar{\mathrm{L}}_{9}, \widebar{\mathrm{L}}_{10} $ are recovered via eigendecomposition of $ \mathbf{M} $ and $ \mathbf{W}_u $, i.e., $ \mathbf{m}  = \sqrt{\sigma_{0,1}} \mathbf{r}_0 $ and $ \mathbf{w}_u  = \sqrt{\sigma_{u,1}} \mathbf{r}_u $, where $ \sigma_{u,1} $ and $ \mathbf{r}_u $ are the largest eigenvalue and principal eigenvector of $ \mathbf{W}_u $, respectively. The same analysis applies to $ \bar{\mathcal{Q}}_{\mathrm{CWEE}_n}^{(t)} $ as the constraints are the same.

\begin{table*}[t!]
	\renewcommand{\arraystretch}{1.25}
	\setlength\tabcolsep{2pt} % default value: 6pt
	\scriptsize
	\caption{Rates and target SINRs for various CQIs.}
	\centering
	\begin{tabular}{|c|c|c|c|c|c|c|c|c|c|c|c|c|c|c|c|}
		% Header
		\hline
		{\bf \scriptsize CQI $ \left( j \right) $} & 1 & 2 & 3 & 4 & 5 & 6 & 7 & 8 & 9 & 10 & 11 & 12 & 13 & 14 & 15 \\ 
		\hline
		\hline
		% Rows                                 
		{\bf \scriptsize Modulation} & \multicolumn{6}{c|}{QPSK} & \multicolumn{3}{c|}{16QAM} & \multicolumn{6}{c|}{64QAM} \\ 
		\hline
		{\bf \scriptsize Coding rate} & 0.0762 & 0.1172 & 0.1885 & 0.3008 & 0.4385 & 0.5879 & 0.3691 & 0.4785 & 0.6016 & 0.4551 & 0.5537 & 0.6504 & 0.7539 & 0.8525 & 0.9258 \\ 
		\hline
		{\bf \scriptsize Rate} $ \left( R_j \right) $ [bps/Hz] & 0.1523 & 0.2344 & 0.3770 & 0.6016 & 0.8770 & 1.1758 & 1.4766 & 1.9141 & 2.4063 & 2.7305 & 3.3223 &  3.9023 & 4.5234 & 5.1152 & 5.5547 \\ 
		\hline
		{\bf \scriptsize Target SINR} $ \left( \Gamma_j \right) $ & 0.1128 & 0.2159 & 0.3892 & 0.6610 & 1.0962 & 1.7474 & 2.8113 & 4.3321 & 7.0081 & 10.6316 & 16.6648 & 25.8345 & 38.4503 & 60.0620 & 95.6974 \\ 
		\hline
	\end{tabular}
	\label{table_rates_sinr}
	\vspace{-2mm}
\end{table*}

\subsubsection{Projecting the Continuous Rates}
Due to the use of Shannon's capacity formula, the rates obtained by solving $ \bar{\mathcal{Q}}_{\mathrm{CWSR}_n}^{(t)} $ and $ \bar{\mathcal{Q}}_{\mathrm{CWEE}_n}^{(t)} $ are continuous. To meet the MCS specifications, these rates are projected, i.e., approximated to the closest feasible discrete rates. Thus, the best solution with projected rates is given by $ f_\mathrm{CWSR}^\mathrm{proj} \left( \mathbf{W}, \mathbf{m}, \mathbf{c} \right) \triangleq \max_{n \in \mathcal{N}} \sum_{u \in \mathcal{U}_n'} \omega_u \left( R_{u,n}^\mathrm{proj} + {C}_{u,n}^\mathrm{proj} \right) $ and $ f_\mathrm{CWEE}^\mathrm{proj} \left( \mathbf{W}, \mathbf{m}, \mathbf{c} \right) \triangleq \max_{n \in \mathcal{N}} \frac{ \sum_{u \in \mathcal{U}} \omega_u \left( \log_2 \left( 1 + \widebar{\mathsf{SINR}}^{(\mathrm{p})}_{u,n} \right) + \widebar{C}_{u,n} \right) }{\frac{1}{\eta_\mathrm{eff}} \left(  \sum_{u \in \mathcal{U}} \left\| \widebar{\mathbf{w}}_{u,n} \right\|^2_2 + \left\| \widebar{\mathbf{m}}_n \right\|^2_2 \right) + P_\mathrm{cir}} $, where $ R_{u,n}^\mathrm{proj} $ and $ \widebar{C}_{u,n}^\mathrm{proj} $ are defined in (\ref{equation_projection_private_rate}) and (\ref{equation_projection_common_rate}), shown at the top of this page, whereas $ R_{j} $, $\Gamma_j $ were introduced in Section~\ref{section_system_model}. In particular, $ \widebar{\mathsf{SINR}}^{(\mathrm{p})}_{u,n} $ and $ \widebar{\mathsf{SINR}}^{(\mathrm{c})}_{u,n} $ are respectively the highest discrete private and common SINRs that can be achieved by $ \mathsf{UE}_u $ in $ \mathcal{U}_n' $, which are mapped to their respective discrete rates $ R_{u,n}^\mathrm{proj} $ and $ \widebar{C}_{u,n}^\mathrm{proj} $. Besides, $ \widebar{C}_{u,n} $, $ \widebar{\mathbf{w}}_{u,n} $, and $ \widebar{\mathbf{m}}_n $ are the common rate portion of $ \mathsf{UE}_u $, the private precoder of $ \mathsf{UE}_u $, and the common precoder of the $ n $-th combination $ \mathcal{U}_n' $, respectively. After evaluating all $ N $ combinations of admitted UEs, we pick the combination achieving the highest objective function value.

% Remark
%\noindent \textit{{\textsc{Remark:}} We do not use BnB to branch the binary variables $ \left\lbrace \boldsymbol{\chi}, \psi \right\rbrace $ because of rate projection which is highly nonlinear as shown in (\ref{equation_projection_private_rate}) and (\ref{equation_projection_common_rate}). In particular, the best continuous-rate solution is not necessarily mapped to the best projected rates.  }

% Subsection
\subsubsection{Computational Complexity} \label{section_computational_complexity_opt_sca_sdr} The computational complexities of solving $ {\mathcal{Q}}_{\mathrm{CWSR}} $ and $ {\mathcal{Q}}_{\mathrm{CWEE}} $ are similar, which is given by $ \mathcal{C}_{\texttt{OPT-SCA-SDR}} = \mathcal{O} \left( N_q N_r N_c^{0.5} N_v^2 N_d \right) $, where $ N_q = 2 { U  \choose K } $ is the total number of combinations of admitted UEs, $ N_r $ is the number of iterations needed for convergence, $ N_c = 9K + 6 $ is the total number of constraints, $ N_v = 2 K N_\mathrm{tx} + 9 N_\mathrm{tx} + 2K + 11 $ is the number of decision variables, and $ N_d = N_\mathrm{tx}^2 K^3 + N_\mathrm{tx}^2 K^2 + 5 K N_\mathrm{tx}^2 + K^3 + 3 N_\mathrm{tx}^2 + K^2 + 2K + 1 $ is the dimension of the SDP program.

\section{Simulation Results} \label{section_results}

We evaluate the WSR and WEE for several configurations, varying the number of UEs, number of admitted UEs, and transmit powers. We consider two cases, namely, two-user settings (\textbf{Scenario I} to \textbf{Scenario IV}) and multiuser settings (\textbf{Scenario V} to \textbf{Scenario VIII}). For the first set of scenarios, we adopt deterministic channels and do not include user admission to gain insight regarding the impact of discrete rates, which is done by modifying constraint $ \widebar{\mathrm{C}}_{2} $ as $ \sum_{u \in \mathcal{U}} \chi_u \leq K $. In particular, we consider a system consisting of a BS with $ N_\mathrm{tx} = 4 $ antennas and $ U = 2 $ UEs with channels $ \mathbf{h}_1 = \left[ 1, 1, 1, 1 \right]^\mathrm{H} $, $ \mathbf{h}_2 = \left[ 1, e^{j \phi}, e^{j 2 \phi}, e^{j 3 \phi} \right]^\mathrm{H} $, where $ \phi = \left\lbrace \frac{\pi}{9}, \frac{2\pi}{9}, \frac{3\pi}{9}, \frac{4\pi}{9} \right\rbrace $ controls the similarity of the channels, whereas the noise power is set to $ \sigma^2 = 30 $ dBm, as in \cite{mao2018:rsma-downlink-communication-systems-bridging-generalizing-outperforming-sdma-noma}. For the second set of scenarios, we adopt UMi line-of-sight (LOS)/non-LOS (NLOS) channels \cite{3gpp.38.913} with carrier frequency $ f_\mathrm{c} = 41 $ GHz, $ N_\mathrm{p} = 4 $ paths, bandwidth $ \mathrm{BW} = 100 $ MHz, noise figure $ \mathrm{NF} = 5 $ dB, and noise power $ \sigma^2 = -174 + \mathrm{NF} + 10 \log_{10} (\mathrm{BW} / \text{Hz}) $ dBm. For this case, we consider two types of channels, i.e., correlated and uncorrelated, in order to assess the performance for different channel conditions. The uncorrelated and correlate channels model the cases when the UEs are distributed across the entire sector of $ 120^{\circ} $ and within a narrower sector of $ 10^{\circ} $, respectively. Also, we consider $ J = 15 $ MCSs with target SINRs corresponding to $ 10\% $ BLER \cite{kovalchukov2019:accurate-approximation-resource-request-distributions-mmwave-3gpp},  shown in Table~\ref{table_rates_sinr}. %but the developed algorithms are applicable for any SINR and MCS values. 
% Table: Simulation settings
\begin{table*}[t!]
	\setlength\tabcolsep{5.0pt} % default value: 6pt
	\renewcommand{\arraystretch}{1.2}% Wider
	\tiny
	\centering
	\caption{Simulation parameters.}
	\begin{tabular}{|c||>{\columncolor[gray]{0.9}} c c >{\columncolor[gray]{0.9}} c c >{\columncolor[gray]{0.9}} c c|>{\columncolor[gray]{0.9}} c c >{\columncolor[gray]{0.9}} c c >{\columncolor[gray]{0.9}}c c|}
		% Header
		\hline
		% Rows
		Scenario
		& Objective
		& $ P_\mathrm{tx}^\mathrm{max} [\text{dBm}]$
		& $ \sigma^2 [\text{dBm}]$
		& $ N_\mathrm{tx} $ 
		& $ U $ 
		& $ K $ 
		& $ \Delta_\mathrm{SIC} $ 
		& $ \eta_\mathrm{eff} $ 
		& $ P_\mathrm{dyn} [\text{dBm}]$
		& $ P_\mathrm{sta} [\text{dBm}]$
		& Weights 
		& Channels \\ 
		\hline
		I & WSR & $ 40, 50 $ & $ 30 $ & $ 4 $ & $ 2 $ & $ 2 $ & $ 0 $ & $ - $ & $ - $ & $ - $ & Various \cite{mao2018:rsma-downlink-communication-systems-bridging-generalizing-outperforming-sdma-noma} & Deterministic \cite{mao2018:rsma-downlink-communication-systems-bridging-generalizing-outperforming-sdma-noma} \\
		\hline
		II & WSR & $ 40, 45, 50 $ & $ 30 $ & $ 4 $ & $ 2 $ & $ 2 $ & $ 0 $ & $ - $ & $ - $ & $ - $ &  Various \cite{mao2018:rsma-downlink-communication-systems-bridging-generalizing-outperforming-sdma-noma} & Deterministic \cite{mao2018:rsma-downlink-communication-systems-bridging-generalizing-outperforming-sdma-noma} \\
		\hline
%		III & SE & $ 30, 35, 40, 45 $ & $ 30 $ & $ 4 $ & $ 4 $ & $ 2 $ & $ 4 $ & $ 2 $ & $ 0 $ & $ - $ & $ - $ & $ - $ & Deterministic \cite{mao2018:rsma-downlink-communication-systems-bridging-generalizing-outperforming-sdma-noma} \\
%		\hline
		III & WSR & $ 30, 40, 50 $ & $ 30 $ & $ 4 $ & $ 2 $ & $ 2 $ & $ \left[ 0, 1 \right] $ & $ - $ & $ - $ & $ - $ & Various \cite{mao2018:rsma-downlink-communication-systems-bridging-generalizing-outperforming-sdma-noma} & Deterministic \cite{mao2018:rsma-downlink-communication-systems-bridging-generalizing-outperforming-sdma-noma} \\
		\hline
		IV & WEE & $ 30, 40 $ & $ 30 $ & $ 4 $ & $ 2 $ & $ 2 $ & $ 0 $ & $ 0.35 $ & $ 33 $ & $ 38 $ & Various \cite{mao2018:rsma-downlink-communication-systems-bridging-generalizing-outperforming-sdma-noma} & Deterministic \cite{mao2018:rsma-downlink-communication-systems-bridging-generalizing-outperforming-sdma-noma} \\
		%\hline
%		VI & EE & $ 20, 30, 40 $ & $ 30 $ & $ 2 $ & $ 2 $ & $ 2 $ & $ 2 $ & $ 2 $ & $ 0 $ & $ 0.25, 0.35, 0.45 $ & $ 33 $ & $ 34, 38, 42 $ & Deterministic \cite{mao2018:rsma-downlink-communication-systems-bridging-generalizing-outperforming-sdma-noma} \\
%		\hline
		%V & WEE & $ 20, 30, 40 $ & $ 30 $ & $ 4 $ & $ 2 $ & $ 2 $ & $ 0 $ & $ 0.25, 0.35, 0.45 $ & $ 33 $ & $ 34, 38, 42 $ & Various \cite{mao2018:rsma-downlink-communication-systems-bridging-generalizing-outperforming-sdma-noma} & Deterministic \cite{mao2018:rsma-downlink-communication-systems-bridging-generalizing-outperforming-sdma-noma} \\
		%\hline
		%VI & WEE & $ 30, 40, 50 $ & $ 30 $ & $ 4 $ & $ 2 $ & $ 2 $ & $ 0 $ & $ 0.35 $ & $ 33 $ & $ 38 $ & Various \cite{mao2018:rsma-downlink-communication-systems-bridging-generalizing-outperforming-sdma-noma} & Deterministic \cite{mao2018:rsma-downlink-communication-systems-bridging-generalizing-outperforming-sdma-noma} \\
		\hline
		\hline
		V & WSR & $ 40 $  & 3GPP \cite{3gpp2022:38.886} & $ 16 $ & $ 2, ..., 6 $ & $ U $ & $ 0 $ & $ - $ & $ - $ & $ - $ & Uniform & 3GPP \cite{3gpp.38.913} \\
		\hline
		VI & WSR & $ 10, \dots, 40 $  & 3GPP \cite{3gpp2022:38.886} & $ 16 $ & $ 6 $ & $ 3 $ & $ 0 $ & $ - $ & $ - $ & $ - $ & Uniform & 3GPP \cite{3gpp.38.913} \\
		\hline
		%IX & WSR & $ 40 $  & 3GPP \cite{3gpp2022:38.886} & $ 16 $ & $ 6 $ & $ 3 $ & $ 18 $ & $ 9 $ & $ \left[ 0, 1 \right] $ & $ - $ & $ - $ & $ - $ & Uniform & 3GPP \cite{3gpp.38.913} \\
		%\hline
		%\hline
		VII & WEE & $ 40 $  & 3GPP \cite{3gpp2022:38.886} & $ 16 $ & $ 2, ..., 6 $ & $ U $ & $ 0 $ & $ 0.35 $ & $ 33 $ & $ 38 $ & Uniform & 3GPP \cite{3gpp.38.913} \\
		\hline
		VIII & WEE & $ 10, \dots, 40 $  & 3GPP \cite{3gpp2022:38.886} & $ 16 $ & $ 6 $ & $ 3 $ & $ 0 $ & $ 0.35 $ & $ 33 $ & $ 38 $ & Uniform & 3GPP \cite{3gpp.38.913}\\
		%\hline
		%XII & EE & $ 10, \dots, 40 $  & \cite{} & $ 16 $ & $ 2, ..., 6 $ & $ U $ & $ 3, ..., 18 $ & $ U_\mathrm{total} $ & $ 0 $ & $ - $ & $ - $ & $ - $ & 3GPP \cite{} \\
		\hline
	\end{tabular}
	\label{table_simulation_settings}
	\vspace{-3mm}\\
\end{table*}
% Figure: Scenario 0
\begin{figure*}[!t]
	% Scenario S0a (Row 1)
 	\begin{subfigure}[b]{0.24\textwidth}
		\begin{center}
			\resizebox{\textwidth}{!}
			{%
			\includegraphics[width=1\textwidth]{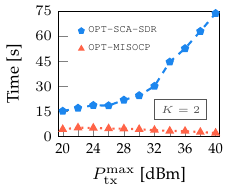}
			}
			\caption{Average runtime of \texttt{OPT-SCA-SDR} and \texttt{OPT-MISOCP} for $ U = 4 $, $ K = 2 $, considering uncorrelated channels.}
			\label{figure_scenario_s0a}
		\end{center}
 	\end{subfigure}
    \hfill 
	% Scenario S0b (Row 1)
 	\begin{subfigure}[b]{0.24\textwidth}
		\begin{center}
			\resizebox{\textwidth}{!}
			{%
			\includegraphics[width=1\textwidth]{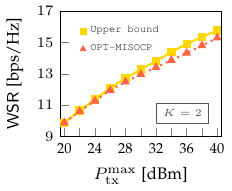}
			}
			\caption{Average WSR of \texttt{OPT-MISOCP} and upper bound for $ U = 4 $, $ K = 2 $, considering uncorrelated channels.}
			\label{figure_scenario_s0b}
		\end{center}
 	\end{subfigure}
    \hfill 
	% Scenario S0d (Row 1)
 	\begin{subfigure}[b]{0.24\textwidth}
		\begin{center}
			\resizebox{\textwidth}{!}
			{%
			\includegraphics[width=1\textwidth]{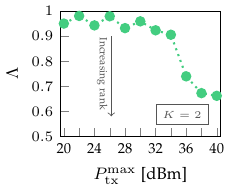}
			}
			\caption{Average `rank-oneness' of the upper bound for $ U = 4 $, $ K = 2 $, considering uncorrelated channels.}
			\label{figure_scenario_s0d}
		\end{center}
 	\end{subfigure}
    \hfill 
	% Scenario S0e (Row 1)
 	\begin{subfigure}[b]{0.24\textwidth}
		\begin{center}
			\resizebox{\textwidth}{!}
			{%
			\includegraphics[width=1\textwidth]{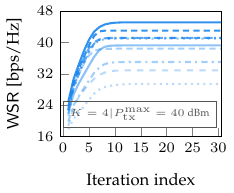}
			}
			\caption{Convergence of \texttt{OPT-SCA-SDR} for $ 10 $ channel realizations, $ U = K = 4 $, considering uncorrelated channels.}
			\label{figure_scenario_s0e}
		\end{center}
 	\end{subfigure}
    \caption{Analysis of time complexity, optimality, and convergence. }
 	\label{figure_scenario_s0}
 	\vspace{-4mm}
\end{figure*}

For the optimization of $ \bar{\mathcal{Q}}_{\mathrm{CWSR}_n}^{(t)} $ and $ \bar{\mathcal{Q}}_{\mathrm{CWEE}_n}^{(t)} $, we initialize the variables $ \gamma_u $, $ \rho_u $, $ \tau_u $, $ \lambda_u $, $ \delta_u $, $ \theta_u $, $ \forall u \in \mathcal{U} $, as $ \gamma_u^{(0)} = 1 $, $ \rho_u^{(0)} = 1 $, $ \tau_u^{(0)} = 1 $, $ \lambda_u^{(0)} = 1 $, $ \delta_u^{(0)} = 1 $, $ \theta_u^{(0)} = 1 $. In addition, we initialize the penalty factor $ p_u $ as $ p_u^{(0)} = 0.01 $, $ \forall u \in \mathcal{U} $, which is updated in each iteration $ t $ as $ p_u^{(t+1)} = \min \left\lbrace p_\mathrm{inc} \cdot p_u^{(t)}, p_\mathrm{max} \right\rbrace $, where $ p_\mathrm{inc} = 4 $ and $ p_\mathrm{max} = 1000 $. As for the stopping criterion, we consider the threshold $ \epsilon = 0.0001 $ and the maximum number of iterations $ N_\mathrm{iter} = 120 $. The simulation results depict the average over $ N_\mathrm{ch} = 100 $ channel realizations assuming $ R_\mathrm{min} = R_1 $ (see Table \ref{table_rates_sinr}), unless specified otherwise. The maximum distance between the BS and UEs is $ D_\mathrm{BS} = 60 $~m. The formulated optimization problems are solved using \texttt{CVX} and \texttt{MOSEK}. The parameter settings employed in the considered scenarios are specified in Table \ref{table_simulation_settings}. Furthermore, we compare the following algorithms\footnote{Our formulation allows us to obtain SDMA as a particular case of RSMA. However, NOMA cannot be obtained from it as NOMA requires multiple SIC stages and optimal decoding order, which our model does not feature. Still, to shed light on the performance of RSMA and NOMA, we have included results for a two-UE case in \textbf{Appendix~\ref{appendix_proposition_12}}.}.  

$ \bullet $ \noindent \textbf{\texttt{OPT-MISOCP}:} As proposed in Section~\ref{subsection_proposed_algorithm_discrete_rates} for discrete rates. By setting $ \psi = 0 $, it reduces to SDMA.

$ \bullet $ \noindent \textbf{\texttt{OPT-SCA-SDR}:} As proposed in Section~\ref{subsection_proposed_algorithm_continuous_rates} for continuous rates. By setting $ \psi = 0 $, it reduces to SDMA.

%$ \bullet $ \noindent \textbf{\texttt{OPT-MISDR}:} It is proposed in Section~\ref{section_proposed_algorithm_opt_misdr}. It is an upper bound for \texttt{OPT-MISOCP}.

$ \bullet $ \noindent \textbf{\texttt{RND-MISOCP}:} Variant of \texttt{OPT-MISOCP}, which assumes random user admission.

$ \bullet $ \noindent \textbf{\texttt{RND-SCA-SDR}:} Variant of \texttt{OPT-SCA-SDR}, which assumes random user admission.

%$ \bullet $ \noindent \textbf{\texttt{RND-MISDR}:} This is a variant of \texttt{OPT-MISDR} that admits UEs chosen randomly.

$ \bullet $ \noindent \textbf{\texttt{PR-OPT-SCA-SDR}:} Obtained from \texttt{OPT-SCA-SDR} upon projecting the rates, as shown in (\ref{equation_projection_private_rate}) and (\ref{equation_projection_common_rate}). 

$ \bullet $ \noindent \textbf{\texttt{PR-RND-SCA-SDR}:} Obtained from \texttt{RND-SCA-SDR} upon projecting the rates, as shown in (\ref{equation_projection_private_rate}) and (\ref{equation_projection_common_rate}).

\subsection{Complexity, Optimality, and Convergence} \label{subsection_complexity_optimality_convergence_initialization}

In this section, we quantify the runtime complexity of \texttt{OPT-MISOCP} and \texttt{OPT-SCA-SDR}, evaluate the optimality of \texttt{OPT-MISOCP} with respect to an upper bound, and analyze the convergence of \texttt{OPT-SCA-SDR}. For the results shown in Fig. \ref{figure_scenario_s0}, we consider the WSR problem with uncorrelated channels for $ U = 4 $, $ K = \left\lbrace 2, 4 \right\rbrace $, weights $ \omega_1 = \dots = \omega_4 = 1 $, and $ N_\mathrm{ch} = 10 $ channel realizations.

\emph{Runtime complexity:} We compare the runtime complexity of \texttt{OPT-MISOCP} and \texttt{OPT-SCA-SDR}. In Fig. \ref{figure_scenario_s0a}, we observe that for the considered parameters, \texttt{OPT-MISOCP} is $ 4 - 36 $ times faster than \texttt{OPT-SCA-SDR} since the former exploits BnB, which circumvents the need of an exhaustive search. In contrast, \texttt{OPT-SCA-SDR} considers all possible combinations of admitted UEs. Furthermore, \texttt{OPT-SCA-SDR} is an iterative scheme, which needs to solve multiple instances of the problem until a stop criterion is met. We notice that \texttt{OPT-SCA-SDR} needs more time to converge as the transmit power increases. In particular, higher transit powers facilitate higher WSRs, and therefore more iterations are needed before the stopping criterion is satisfied. On the other hand, the runtime of \texttt{OPT-MISOCP} remains constant and even slightly decreases for higher transmit powers. This is due to constraint $ \mathrm{J}_2 $, introduced in Section~\ref{subsubsection_adding_cutting_planes_tighten_feasible_domain}, which allows early stopping. Additional results on the worst-case complexity of \texttt{OPT-MISOCP} and \texttt{OPT-SCA-SDR} derived in Section~\ref{section_computational_complexity_opt_misocp} and Section~\ref{section_computational_complexity_opt_sca_sdr}, respectively, are provided in \textbf{Appendix \ref{appendix_proposition_11}}.

% Remark
\vspace{1mm}
\noindent \textit{{\textsc{Remark 6:}} We observed that for small numbers of UEs, e.g., $ U = \left\lbrace 4, 5 \right\rbrace $, \texttt{OPT-MISOCP} has an affordable runtime. However, as $ U $ increases beyond these values, the runtime of \texttt{OPT-MISOCP} grows substantially, as more binary variables are involved. To keep \texttt{OPT-MISOCP} affordable, it can be combined with simple subcarrier allocation policy to avoid co-processing multiple UEs simultaneously and allowing for RRM parallelization.}

%Besides, \texttt{OPT-MISOCP} had a comparable runtime complexity to \texttt{OPT-SCA-SDR}, when random admission was. The reaosn is that the UEs to be served are already decided, ptimisoc had to solve various instances but iteratively.

%$ \mathcal{P}_\mathrm{DWSR} $ and $ \mathcal{P}_\mathrm{DWEE} $ are approximately $ \mathcal{O} \approx N_s ( K^{3.5} J^{0.5} + K^{3.5} N_\mathrm{tx} J^{0.5} + K^{2.5} N_\mathrm{tx}^2 J^{0.5} ) $, where $ N_s $ is the total number of binary variables evaluations needed by the BnB solver. The upper bound of $ N_s $ is equal to $ N_s^\mathrm{all} = (J^{K+1} J^{K}) {U \choose K} $ which is equivalent to an exhaustive search, i.e., over all possible groupings of $ K $ out of $ U $ with all rate allocations. However, in practice, $ N_s \ll N_s^\mathrm{all} $ since BnB procedures are capable of pruning infeasible or suboptimal branches.}

% Figure: Scenario S1
\begin{figure*}[!t]
	% Legend
 	\begin{subfigure}[b]{\textwidth}
		\begin{center}
			\resizebox{!}{0.5cm}
			{%
			\includegraphics[width=1\textwidth]{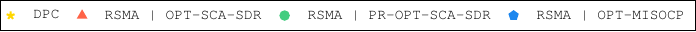}
			}
		\end{center}
 	\end{subfigure}
	\centering
	% Scenario S1a (Row 1)
 	\begin{subfigure}[b]{0.24\textwidth}
		\begin{center}
			\resizebox{\textwidth}{!}
			{%
			\includegraphics[width=1\textwidth]{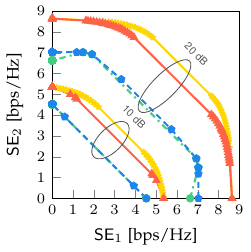}
			}
			\caption{$ \phi = \frac{\pi}{9}$ }
			\label{figure_scenario_s1a}
		\end{center}
 	\end{subfigure}
    \hfill 
	% Scenario S1b (Row 1)
 	\begin{subfigure}[b]{0.24\textwidth}
		\begin{center}
			\resizebox{\textwidth}{!}
			{%
			\includegraphics[width=1\textwidth]{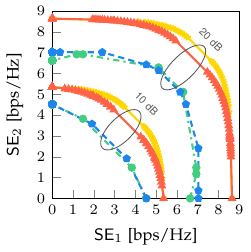}
			}
			\caption{ $ \phi = \frac{2\pi}{9}$ }
			\label{figure_scenario_s1b}
		\end{center}
 	\end{subfigure}
    \hfill 
	% Scenario S1c (Row 1)
 	\begin{subfigure}[b]{0.24\textwidth}
		\begin{center}
			\resizebox{\textwidth}{!}
			{%
			\includegraphics[width=1\textwidth]{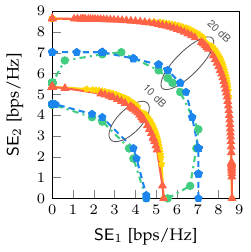}
			}
			\caption{ $ \phi = \frac{3\pi}{9}$ }
			\label{figure_scenario_s1c}
		\end{center}
 	\end{subfigure}
    \hfill 
	% Scenario S1d (Row 1)
 	\begin{subfigure}[b]{0.24\textwidth}
		\begin{center}
			\resizebox{\textwidth}{!}
			{%
			\includegraphics[width=1\textwidth]{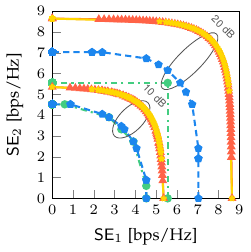}
			}
			\caption{ $ \phi = \frac{4\pi}{9}$ }
			\label{figure_scenario_s1d}
		\end{center}
 	\end{subfigure}
    \caption{(\textit{Scenario I}) Two-user SE region of RSMA with discrete and continuous rates for $ \frac{P_\mathrm{tx}^\mathrm{max}}{\sigma^2} = \left\lbrace 10, 20 \right\rbrace $ dB. \emph{Since \texttt{OPT-SCA-SDR} does not account for rate saturation, it continues upgrading the private rates, not necessarily leading to improved performance upon rate projection. In contrast, \texttt{OPT-MISOCP} considers that the rates are bounded and discrete, promoting more appropriate usage of power. Specifically, \texttt{OPT-MISOCP} uses the surplus of power to upgrade weaker private or common signals, preventing severe rate saturation of other signals.}}
 	\label{figure_scenario_s1}
 	\vspace{-3mm}
\end{figure*}
% Figure: Scenario 2
\begin{figure*}[!t]
	% Legend
 	\begin{subfigure}[b]{\textwidth}
		\begin{center}
			\resizebox{!}{0.5cm}
			{%
			\includegraphics[width=1\textwidth]{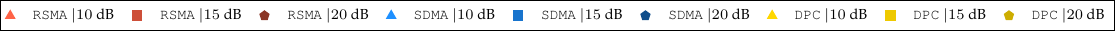}
			}
		\end{center}
 	\end{subfigure}
	% Scenario S2a (Row 1)
 	\begin{subfigure}[b]{0.24\textwidth}
		\begin{center}
			\resizebox{\textwidth}{!}
			{%
			\includegraphics[width=1\textwidth]{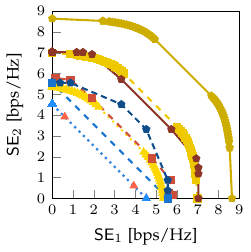}
			}
			\caption{$ \phi = \frac{\pi}{9} $}
			\label{figure_scenario_s2a}
		\end{center}
 	\end{subfigure}
    \hfill 
	% Scenario S2b (Row 1)
 	\begin{subfigure}[b]{0.24\textwidth}
		\begin{center}
			\resizebox{\textwidth}{!}
			{%
			\includegraphics[width=1\textwidth]{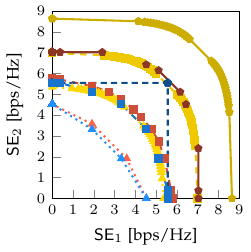}
			}
			\caption{$ \phi = \frac{2\pi}{9} $}
			\label{figure_scenario_s2b}
		\end{center}
 	\end{subfigure}
    \hfill 
	% Scenario S2c (Row 1)
 	\begin{subfigure}[b]{0.24\textwidth}
		\begin{center}
			\resizebox{\textwidth}{!}
			{%
			\includegraphics[width=1\textwidth]{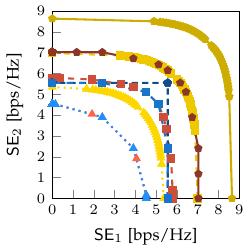}
			}
			\caption{$ \phi = \frac{3\pi}{9} $}
			\label{figure_scenario_s2c}
		\end{center}
 	\end{subfigure}
    \hfill 
	% Scenario S2d (Row 1)
 	\begin{subfigure}[b]{0.24\textwidth}
		\begin{center}
			\resizebox{\textwidth}{!}
			{%
			\includegraphics[width=1\textwidth]{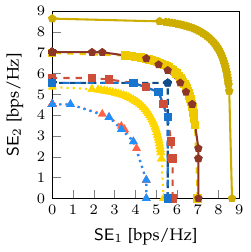}
			}
			\caption{$ \phi = \frac{4\pi}{9} $}
			\label{figure_scenario_s2d}
		\end{center}
 	\end{subfigure}
    \caption{(\textit{Scenario II}) Two-user SE region of RSMA and SDMA with discrete rates using \texttt{OPT-MISOCP} for $ \frac{P_\mathrm{tx}^\mathrm{max}}{\sigma^2} = \left\lbrace 10, 15, 20 \right\rbrace $ dB. \emph{The advantage of RSMA stems from its capability of using the surplus of power to transmit the common signal, even in scenarios with highly uncorrelated channels, which SDMA is unable to do.}}
 	\label{figure_scenario_s2}
 	\vspace{-4mm}
\end{figure*}

\emph{Optimality:} We compare the WSR performance of \texttt{OPT-MISOCP} to an upper bound that we devise using SDR to demonstrate that \texttt{OPT-MISOCP} can yield near-optimal solutions for $ \mathcal{P}_\mathrm{DWSR}' $. This upper bound is used to analyze the impact of the convexification procedure used in \texttt{OPT-MISOCP}. Since the upper bound has a larger feasible set due to the rank-one relaxation, it finds solutions that yield higher objective function values than \texttt{OPT-MISOCP}. However, such solutions are not necessarily feasible for problem $ \mathcal{P}_{\mathrm{DWSR}} $. In Fig. \ref{figure_scenario_s0b}, we observe that the performance gap between \texttt{OPT-MISOCP} and the upper bound is generally small, although it slightly increases to $ 3\% $ for higher transmit powers. To explain this result, we show in Fig. \ref{figure_scenario_s0d}, the ratio of the principal eigenvalue to the sum of all eigenvalues, which we denote by $ \Lambda $, i.e., $ \Lambda $ portrays the `rank-oneness' of the upper bound solutions. Specifically, it is defined as $ \Lambda = \frac{1}{\min \left\lbrace 1, \mathrm{Rank} \left( \mathbf{X}_0 \right) \right\rbrace + \sum_{u \in \mathcal{U}} \min \left\lbrace 1, \mathrm{Rank} \left( \mathbf{X}_u \right) \right\rbrace } \frac{\sum_{u \in \mathcal{U} \cup \left\lbrace 0 \right\rbrace } \hat{\lambda}_{\mathrm{max},u} }{ \sum_m \hat{\lambda}_{m,u} } $, where $ \hat{\lambda}_{m,u} $ is the $ m $-th eigenvalue of $ \mathbf{X}_u \succcurlyeq \mathbf{0} $, $ u \in \mathcal{U} \cup \left\lbrace 0 \right\rbrace $. Here, $ \mathbf{X}_u $ is the private precoder for $ \mathsf{UE}_u $ and $ \mathbf{X}_0 $ is the precoder for the common signal, obtained by the upper bound. $ \Lambda $ reveals that the upper bound solutions have ranks higher than one, and therefore are not feasible for problem $ \mathcal{P}_{\mathrm{DWSR}} $, thus explaining the performance gap. 

\emph{Convergence:} In Fig. \ref{figure_scenario_s0e}, we show the convergence of \texttt{OPT-SCA-SDR} for $ N_\mathrm{ch} = 10 $ channel realizations.

\noindent \textit{\textbf{Scenario I}: Two-User SE Region for Continuous/Discrete RSMA Rates} 

In Fig. \ref{figure_scenario_s1}, we compare the SE of RSMA with discrete and continuous rates to investigate the impact of rate discretization. For all considered cases, \texttt{OPT-MISOCP} and \texttt{PR-OPT-SCA-SDR} exhibit similar performance when $ \frac{P_\mathrm{tx}^\mathrm{max}}{\sigma^2} = 10 $ dB. This occurs because the rates obtained by \texttt{OPT-SCA-SDR} are small due to the low transmit power, and therefore projection does not have a significant impact. However, the performance gap between them can become large when $ \frac{P_\mathrm{tx}^\mathrm{max}}{\sigma^2} = 20 $ dB due to the higher rates achieved, which can lead to more noticeable projection losses. For instance, the difference is negligible in Fig. \ref{figure_scenario_s1a}, whereas it is more evident in Fig. \ref{figure_scenario_s1c}. The reason is that the channels become less correlated as $ \phi $ increases, making the common rate less relevant for \texttt{OPT-SCA-SDR}. This causes \texttt{OPT-SCA-SDR} to be noticeably impacted by rate projection, as the private rates may experience heavy saturation while the common rate remains small. Fig. \ref{figure_scenario_s1d} shows an extreme case with low channel correlation, which causes \texttt{OPT-SCA-SDR} to opt for SDMA. In this case, the loss due to projection is higher than in Fig. \ref{figure_scenario_s1b} and Fig. \ref{figure_scenario_s1c} since the common rate is zero, and the private rates saturate at $ R_J $ (see Table \ref{table_rates_sinr}). On the other hand, \texttt{OPT-MISOCP} can prevent rate saturation losses as it takes the rate discretization into account, and expends any surplus of power to improve the common rate. For reference, we have included dirty paper coding (\texttt{DPC}), which is capacity-achieving for continuous rates \cite{viswanathan2003:downlink-capacity-evaluation-cellular-networks-known-interference-cancellation}. We observe that the proposed \texttt{OPT-SCA-SDR} can approach the performance of \texttt{DPC}, especially in Fig. \ref{figure_scenario_s1d}, thus demonstrating that \texttt{OPT-SCA-SDR} produces high-quality solutions.

In summary, the optimal partitioning of information, transmitted via unicast and multicast signals, can differ significantly across different scenarios since the partitioning depends largely on the channel characteristics. The optimal partitioning also depends on the specific constraints of the RRM problem. Specifically, \texttt{OPT-MISOCP} is subject to discrete rates, while \texttt{OPT-SCA-SDR} assumes continuous rates. Although the resulting partitionings differ in these two cases, the respective values are optimal in each case, given the constraints.

\noindent \textit{\textbf{Scenario II}: Two-User SE Region with Discrete Rates for RSMA and SDMA} 

In Fig. \ref{figure_scenario_s2}, we compare the SE of RSMA and SDMA using discrete rates to elucidate the performance gap between them for different transmit powers. In Fig. \ref{figure_scenario_s2a}, RSMA and SDMA have nearly the same performance when $ \frac{P_\mathrm{tx}^\mathrm{max}}{\sigma^2} = 10 $ dB, however, RSMA outperforms SDMA when $ \frac{P_\mathrm{tx}^\mathrm{max}}{\sigma^2} = \left\lbrace 15, 20 \right\rbrace $ dB. SDMA is unable to cope well with high channel correlation, showing little improvement even as the transmit power increases. In contrast, RSMA can take advantage of high channel correlation to achieve considerable improvement. In Fig. \ref{figure_scenario_s2b} to Fig. \ref{figure_scenario_s2d}, RSMA outperforms SDMA by a small margin when $ \frac{P_\mathrm{tx}^\mathrm{max}}{\sigma^2} = \left\lbrace 10, 15 \right\rbrace $ dB as the channels are less correlated, thus making the transmission of the common signal more expensive. However, RSMA clearly outperforms SDMA when $ \frac{P_\mathrm{tx}^\mathrm{max}}{\sigma^2} = 20 $ dB since SDMA saturates (i.e., UEs are served at rate $ R_{J} = 5.5547 $ bps/Hz), whereas RSMA can still improve as it can use the surplus of power to support a common signal.

\noindent \textit{\textbf{Scenario III}: Two-User SE Region with Imperfect SIC for RSMA} 

% Figure: Scenario S3
% Figure: Scenario 3
\begin{figure*}[!t]
	% Legend 1
 	\begin{subfigure}[b]{0.49\textwidth}
		\begin{center}
			\resizebox{!}{0.5cm}
			{%
			\includegraphics[width=1\textwidth]{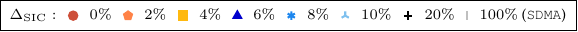}
			}
		\end{center}
 	\end{subfigure}
 	\hfill
 	% Legend 2
 	\begin{subfigure}[b]{0.49\textwidth}
		\begin{center}
			\resizebox{!}{0.5cm}
			{%
			\includegraphics[width=1\textwidth]{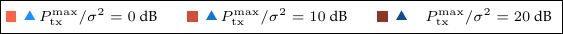}
			}
		\end{center}
 	\end{subfigure}
 	\hfill
	% Scenario S3a (Row 1)
 	\begin{subfigure}[b]{0.24\textwidth}
		\begin{center}
			\resizebox{\textwidth}{!}
			{%
			\includegraphics[width=1\textwidth]{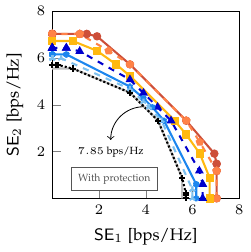}
			}
			\caption{$ \frac{P_\mathrm{tx}^\mathrm{max}}{\sigma^2} = 20 $ dB and $ \phi = \frac{\pi}{9} $}
			\label{figure_scenario_s3a}
		\end{center}
 	\end{subfigure}
    \hfill 
	% Scenario S3b (Row 1)
 	\begin{subfigure}[b]{0.24\textwidth}
		\begin{center}
			\resizebox{\textwidth}{!}
			{%
			\includegraphics[width=1\textwidth]{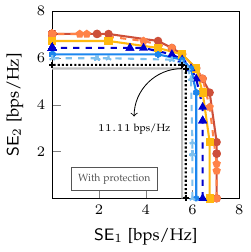}
			}
			\caption{$ \frac{P_\mathrm{tx}^\mathrm{max}}{\sigma^2} = 20 $ dB and $ \phi = \frac{4\pi}{9} $}
			\label{figure_scenario_s3b}
		\end{center}
 	\end{subfigure}
    \hfill 
	% Scenario S3c (Row 1)
 	\begin{subfigure}[b]{0.24\textwidth}
		\begin{center}
			\resizebox{\textwidth}{!}
			{%
			\includegraphics[width=1\textwidth]{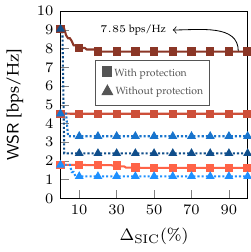}
			}
			\caption{$ \omega_1 = \omega_2 = 1 $ and $ \phi = \frac{\pi}{9} $}
			\label{figure_scenario_s3c}
		\end{center}
 	\end{subfigure}
    \hfill 
	% Scenario S3d (Row 1)
 	\begin{subfigure}[b]{0.24\textwidth}
		\begin{center}
			\resizebox{\textwidth}{!}
			{%
			\includegraphics[width=1\textwidth]{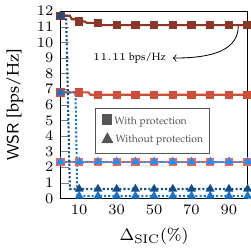}
			}
			\caption{$ \omega_1 = \omega_2 = 1 $ and $ \phi = \frac{4\pi}{9} $}
			\label{figure_scenario_s3d}
		\end{center}
 	\end{subfigure}
    \caption{(\textit{Scenario III}) Two-user SE region of RSMA with discrete rates and imperfect SIC using \texttt{OPT-MISOCP} for $ \frac{P_\mathrm{tx}^\mathrm{max}}{\sigma^2} = \left\lbrace 0, 10, 20 \right\rbrace $ dB and various $ \Delta_\mathrm{SIC} $ values. \emph{Accounting for potentially imperfect SIC has an enormous performance benefit. In the worst case, RSMA collapses to SDMA, still providing outstanding performance compared to the case without protection. In the presence of large unmanaged residuals of the common signal, due to an imperfect SIC, the private rates cannot be guaranteed, thus collapsing to zero due to the inability to fulfill the target SINRs required for successful decoding.}}
 	\label{figure_scenario_s3}
 	\vspace{-3mm}
\end{figure*}
% Figure: Scenario S4
\begin{figure*}[!t]
	% Legend
 	\begin{subfigure}[b]{\textwidth}
		\begin{center}
			\resizebox{!}{0.5cm}
			{%
			\includegraphics[width=1\textwidth]{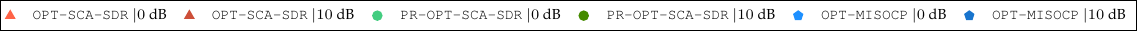}
			}
		\end{center}
 	\end{subfigure}
	\centering
	% Scenario S4a (Row 1)
 	\begin{subfigure}[b]{0.24\textwidth}
		\begin{center}
			\resizebox{\textwidth}{!}
			{%
			\includegraphics[width=1\textwidth]{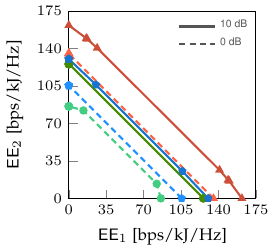}
			}
			\caption{ $ \phi = \frac{\pi}{9}$ }
			\label{figure_scenario_s4a}
		\end{center}
 	\end{subfigure}
    \hfill 
	% Scenario S4b (Row 1)
 	\begin{subfigure}[b]{0.24\textwidth}
		\begin{center}
			\resizebox{\textwidth}{!}
			{%
			\includegraphics[width=1\textwidth]{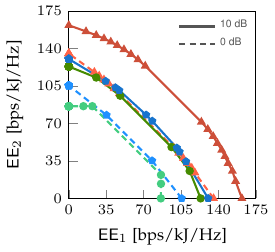}
			}
			\caption{ $ \phi = \frac{2\pi}{9}$ }
			\label{figure_scenario_s4b}
		\end{center}
 	\end{subfigure}
    \hfill 
	% Scenario S4c (Row 1)
 	\begin{subfigure}[b]{0.24\textwidth}
		\begin{center}
			\resizebox{\textwidth}{!}
			{%
			\includegraphics[width=1\textwidth]{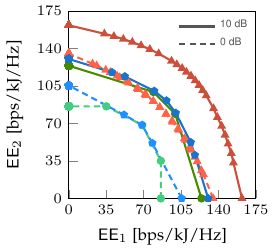}
			}
			\caption{ $ \phi = \frac{3\pi}{9}$ }
			\label{figure_scenario_s4c}
		\end{center}
 	\end{subfigure}
    \hfill 
	% Scenario S4d (Row 1)
 	\begin{subfigure}[b]{0.24\textwidth}
		\begin{center}
			\resizebox{\textwidth}{!}
			{%
			\includegraphics[width=1\textwidth]{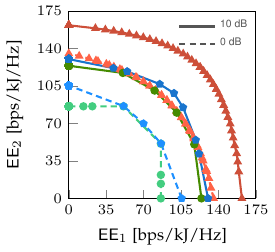}
			}
			\caption{ $ \phi = \frac{4\pi}{9}$ }
			\label{figure_scenario_s4d}
		\end{center}
 	\end{subfigure}
    \caption{(\textit{Scenario IV}) Two-user EE region of RSMA with discrete and continuous rates for $ \frac{P_\mathrm{tx}^\mathrm{max}}{\sigma^2} = \left\lbrace 0, 10 \right\rbrace $ dB, $ \eta_\mathrm{eff} = 0.35 $, $ P_\mathrm{dyn} = 33 $ dBm, and $ P_\mathrm{sta} = 38 $ dBm. \emph{ As it uses the transmit power more judiciously, \texttt{OPT-MISOCP} has a notable advantage over \texttt{PR-OPT-SCA-SDR}, ensuring high discrete rates with minimal power consumption, leading to improved EE. In contrast, \texttt{PR-OPT-SCA-SDR} is not aware of rate discretization, and therefore the precoders have larger powers than necessary, which impacts the EE upon rate projection}.}
 	\label{figure_scenario_s4}
 	\vspace{-3mm}
\end{figure*}

In Fig. \ref{figure_scenario_s3}, we evaluate the SE of RSMA for various levels of protection against imperfect SIC as well as without protection. When we consider protection, we assume a given $ \Delta_\mathrm{SIC} \neq 0\% $, which is taken into account for the optimization. Therefore, the BS guarantees the allocated rates for the UEs up to the selected value of $ \Delta_\mathrm{SIC} $. When we neglect protection, we assume $ \Delta_\mathrm{SIC} = 0\% $ for the optimization even though the UEs may suffer from imperfect SIC. Therefore, the allocated rates may not be guaranteed. In Fig. \ref{figure_scenario_s3a} and Fig. \ref{figure_scenario_s3b}, protection against imperfect SIC is considered. We observe that endowing RSMA with a higher robustness against imperfect SIC, i.e., larger $ \Delta_\mathrm{SIC} $, produces a more noticeable decrease in the SE because the private SINRs are optimized to deal with additional interference due to $ \Delta_\mathrm{SIC} \neq 0\% $ (see Section~\ref{subsubsection_imperfect_SIC}). Also, we observe that values up to $ \Delta_\mathrm{SIC} = 4\% $ do not affect the SE performance substantially while providing adequate protection. However, RSMA almost collapses to SDMA when $ \Delta_\mathrm{SIC} = 20\% $, as the common rates become very small. In fact, RSMA smartly switches to SDMA for values larger than $ \Delta_\mathrm{SIC} = 20\% $ since the high protection against imperfect SIC prevents enhancement of the private SINRs. The results for SDMA are identical to those for RSMA with $ \Delta_\mathrm{SIC} = 100\% $. In Fig. \ref{figure_scenario_s3c} and Fig. \ref{figure_scenario_s3d}, we evaluate the impact of not accounting for imperfect SIC on the WSR performance, where we consider the same scenarios in Fig. \ref{figure_scenario_s3a} and Fig. \ref{figure_scenario_s3b}, and equal weights, i.e., $ \omega_1 = \omega_2 = 1 $. In Fig. \ref{figure_scenario_s3c}, the impact of imperfect SIC is small when $ \frac{P_\mathrm{tx}^\mathrm{max}}{\sigma^2} = 0 $ dB because information is predominantly transmitted via the common signal which is not affected by imperfect SIC. When $ \frac{P_\mathrm{tx}^\mathrm{max}}{\sigma^2} = \left\lbrace 10, 20 \right\rbrace $ dB, the common and private rates increase since higher MCSs can be selected. This also implies that potential unmanaged residuals of the common signal may cause the private rates to collapse more noticeably, e.g., the SE drops from $ 7.85 $ bps/Hz to $ 2.41 $ bps/Hz (when $ \frac{P_\mathrm{tx}^\mathrm{max}}{\sigma^2} = 20 $ dB) and from $ 4.52 $ bps/Hz to $ 3.32 $ bps/Hz (when $ \frac{P_\mathrm{tx}^\mathrm{max}}{\sigma^2} = 10 $ dB). However, the system performs well when protection against imperfect SIC is considered. In particular, for high $ \Delta_\mathrm{SIC} $, RSMA transitions to SDMA thereby avoiding further private SINRs degradation. In Fig. \ref{figure_scenario_s3d}, we observe the same trend as in Fig. \ref{figure_scenario_s3c}, although the degradation due to imperfect SIC is more conspicuous when protection against imperfect SIC is neglected. This occurs because the channels are highly uncorrelated, making the private rates even more prominent than in Fig. \ref{figure_scenario_s3c}, with the consequent potential risk of much larger degradation in case of SIC failure.

\noindent \textit{\textbf{Scenario IV}: Two-User EE Region with Continuous/Discrete Rates for RSMA} 

% Figure: Scenario 9
\begin{figure*}[!t]
	\centering
	% Legend
% 	\begin{subfigure}[b]{\textwidth}
%		\begin{center}
%			\resizebox{!}{0.5cm}
%			{%
%			\begin{tikzpicture}
%			    \begin{axis}[%
%			    hide axis,
%			    xmin = 10,
%			    xmax = 50,
%			    ymin = 0,
%			    ymax=0.4,
%				legend style = {row sep = 0.01cm},
%			 	legend style = {column sep = 0.25cm},
%		    	legend cell align = {left},
%		    	legend columns = 4,
%		    	legend pos = north east,
%		    	legend style = {at = {(0.0, 0)}, anchor = south west, font = \fontsize{7}{6}\selectfont, text depth = .ex, fill = none},
%			    ]
%
%				% RSMA (Continuous): OPT-MISOCP
%			 	\addlegendimage{color = Tomato1, mark = square*, line width = 1pt, solid, only marks, mark options = {fill = Tomato1, solid, scale = 1}} \addlegendentry{\texttt{RSMA|OPT-MISOCP}}
%
%				% RSMA (Continuous): PR-OPT-SCA-SDR
%			 	\addlegendimage{color = DodgerBlue2, mark = square*, line width = 1pt, solid, only marks, mark options = {fill = DodgerBlue2, solid, scale = 1}} \addlegendentry{\texttt{RSMA|PR-OPT-SCA-SDR}}
%
%				% SDMA (Continuous): OPT-MISOCP
%			 	\addlegendimage{color = Gold1, mark = triangle*, line width = 1pt, solid, only marks, mark options = {fill = Gold1, solid, scale = 1}} \addlegendentry{\texttt{SDMA|OPT-MISOCP}}
%			 	
%				% SDMA (Continuous): PR-OPT-SCA-SDR
%			 	\addlegendimage{color = SeaGreen3, mark = triangle*, line width = 1pt, solid, only marks, mark options = {fill = SeaGreen3, solid, scale = 1}} \addlegendentry{\texttt{SDMA|PR-OPT-SCA-SDR}}
%
%			    \end{axis}
%			\end{tikzpicture}
%			}
%		\end{center}
% 	\end{subfigure}
	% Scenario S9a
 	\begin{subfigure}[b]{0.38\textwidth}
		\begin{center}
			\resizebox{\textwidth}{!}
			{%
			\includegraphics[width=1\textwidth]{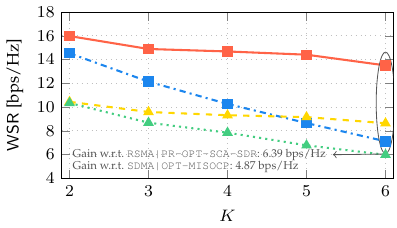}
			}
			\caption{Correlated channels.}
			\label{figure_scenario_s7a}
		\end{center}
 	\end{subfigure}
 	\hfill
	% Legend
 	\begin{subfigure}[b]{0.2\textwidth}
		\begin{center}
			\resizebox{\textwidth}{!}
			{%
			\includegraphics[width=1\textwidth]{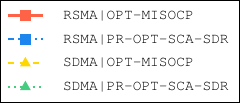}
			}
			\vspace{1.6cm}
		\end{center}
 	\end{subfigure}
 	\hfill	
	% Scenario S9b
 	\begin{subfigure}[b]{0.38\textwidth}
		\begin{center}
			\resizebox{\textwidth}{!}
			{%
			\includegraphics[width=1\textwidth]{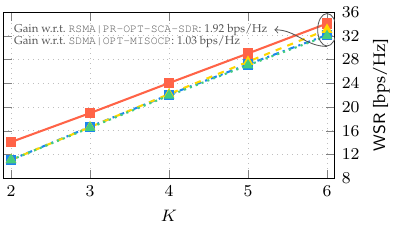}
			}
			\caption{Uncorrelated channels.}
			\label{figure_scenario_s7b}
		\end{center}
 	\end{subfigure}
    \caption{(\textit{Scenario V}) WSR of RSMA and SDMA as a function of the number of admitted UEs. \emph{In Fig. \ref{figure_scenario_s7a}, \texttt{RSMA|OPT-MISOCP} has an advantage of $ 6.39 $ bps/Hz ($ \uparrow 89.7 \% $ gain) and $ 4.87 $ bps/Hz ($ \uparrow 56.3 \% $ gain) with respect to \texttt{RSMA|PR-OPT-SCA-SDR} and \texttt{SDMA|OPT-MISOCP}, respectively, when $ U = 6 $. In Fig. \ref{figure_scenario_s7b}, \texttt{RSMA|OPT-MISOCP} has an advantage of $ 1.92 $ bps/Hz ($ \uparrow 5.9 \% $ gain) and $ 1.03 $ bps/Hz ($ \uparrow 3.1 \% $ gain) compared to \texttt{RSMA|PR-OPT-SCA-SDR} and \texttt{SDMA|OPT-MISOCP}, respectively, when $ K = 6 $.}}
 	\label{figure_scenario_s7}
 	\vspace{-3mm}
\end{figure*}
% Figure: Scenario 8
\begin{figure*}[!t]
	\centering
	% Scenario S8a
 	\begin{subfigure}[b]{0.38\textwidth}
		\begin{center}
			\resizebox{\textwidth}{!}
			{%
			\includegraphics[width=1\textwidth]{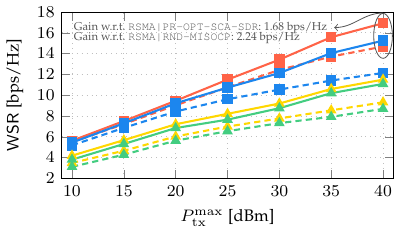}
			}
			\caption{Correlated channels.}
			\label{figure_scenario_s8a}
		\end{center}
 	\end{subfigure}
 	\hfill
	% Legend
 	\begin{subfigure}[b]{0.2\textwidth}
		\begin{center}
			\resizebox{\textwidth}{!}
			{%
			\includegraphics[width=1\textwidth]{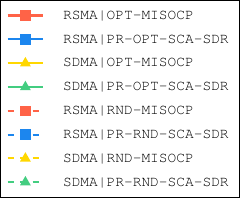}
			}
			\vspace{0.8cm}
		\end{center}
 	\end{subfigure}
 	\hfill	
	% Scenario S8b
 	\begin{subfigure}[b]{0.38\textwidth}
		\begin{center}
			\resizebox{\textwidth}{!}
			{%
			\includegraphics[width=1\textwidth]{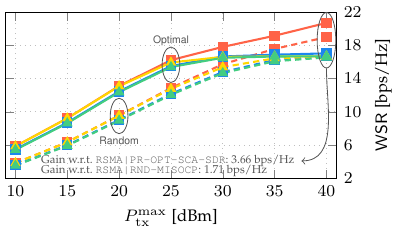}
			}
			\caption{Uncorrelated channels.}
			\label{figure_scenario_s8b}
		\end{center}
 	\end{subfigure}
    \caption{(\textit{Scenario VI}) WSR of RSMA and SDMA with optimal and random UE admission as a function of the transmit power. \emph{In Fig. \ref{figure_scenario_s8a}, \texttt{RSMA|OPT-MISOCP} has an advantage of $ 1.68 $ bps/Hz ($ \uparrow 10.9 \% $ gain) and $ 2.24 $ bps/Hz ($ \uparrow 15.3 \% $ gain) with respect to \texttt{RSMA|PR-OPT-SCA-SDR} and \texttt{RSMA|RND-MISOCP}, respectively, when $ P_\mathrm{tx}^\mathrm{max} = 40 $ dBm. In Fig. \ref{figure_scenario_s8b}, \texttt{RSMA|OPT-MISOCP} has an advantage of $ 3.66 $ bps/Hz ($ \uparrow 21.6 \% $ gain) with respect to \texttt{RSMA|PR-OPT-SCA-SDR} and $ 1.71 $ bps/Hz ($ \uparrow 8.9 \% $ gain) with respect to \texttt{RSMA|RND-MISOCP}, when $ P_\mathrm{tx}^\mathrm{max} = 40 $ dBm.}}
 	\label{figure_scenario_s8}
 	\vspace{-3mm}
\end{figure*}

%  with a better trade-off between the achieved rates and the expended power
In Fig. \ref{figure_scenario_s4}, we compare the EE of RSMA with continuous and discrete rates to investigate the impact of rate discretization. In Fig. \ref{figure_scenario_s4a} to Fig. \ref{figure_scenario_s4d}, the EE of both \texttt{OPT-MISOCP} and \texttt{PR-OPT-SCA-SDR} improve when $ P_\mathrm{tx}^\mathrm{max} $ increases from $ 0 $ dB to $ 10 $ dB, as a higher transmit power allows to find an improved EE operating point with a better tradeoff between the achieved rates and the expended power. When $ \frac{P_\mathrm{tx}^\mathrm{max}}{\sigma^2} = 0 $ dB (dashed lines), \texttt{OPT-MISOCP} surpasses \texttt{PR-OPT-SCA-SDR} showing gains as large as $ 18 $ bps/kJ/Hz, particularly when the UE weights are not equal. When $ \frac{P_\mathrm{tx}^\mathrm{max}}{\sigma^2} = 10 $ dB (solid lines), \texttt{OPT-MISOCP} also outperforms \texttt{PR-OPT-SCA-SDR} although the gap is smaller. The reason for this effect is that \texttt{OPT-MISOCP} can better exploit the limited transmit power when $ \frac{P_\mathrm{tx}^\mathrm{max}}{\sigma^2} = 0 $ dB as it is able to handle discrete rates, whereas \texttt{PR-OPT-SCA-SDR} wastes power yielding rates higher than necessary, thus incurring a loss after projection. However, when $ \frac{P_\mathrm{tx}^\mathrm{max}}{\sigma^2} = 10 $ dB, the power limitation is alleviated, and therefore \texttt{PR-OPT-SCA-SDR} can reduce the performance gap with respect to \texttt{OPT-MISOCP}. We oserve that as channels become less correlated, the EE of \texttt{OPT-MISOCP} and \texttt{PR-OPT-SCA-SDR} improve because interference can be handled more effectively and with less transmit power.

\noindent \textit{\textbf{Scenario V}: Impact of the Number of Admitted UEs on WSR Performance}

In Fig. \ref{figure_scenario_s7}, we compare the WSR of RSMA and SDMA when the number admitted UEs varies. In Fig. \ref{figure_scenario_s7a}, we consider correlated channels, for which we observe that an increasing number of UEs leads to WSR degradation. This occurs because the UEs are located in close proximity of each other, exacerbating interference for every additional UE admitted. We observe that \texttt{RSMA|OPT-MISOCP} has a noticeable advantage over \texttt{SDMA|OPT-MISOCP} since it can exploit the channel similarity via the common signal. We observe a similar behavior for \texttt{RSMA|PR-OPT-SCA-SDR} and \texttt{SDMA|PR-OPT-SCA-SDR} although the difference between them decreases as $ U $ increases. Also, not considering rate discretization can severely affect \texttt{RSMA|PR-OPT-SCA-SDR}, reducing its performance to the extent of being outperformed by \texttt{SDMA|OPT-MISOCP} when $ U = \left\lbrace 5, 6 \right\rbrace $. In Fig. \ref{figure_scenario_s7b}, we consider uncorrelated channels, for which we observe that increasing the number of UEs leads to an improved WSR. This is expected as interference is more easily dealt with in this case. Also, \texttt{RSMA|PR-OPT-SCA-SDR} and \texttt{SDMA|PR-OPT-SCA-SDR} achieve the same performance because RSMA does not devise a common signal. However, \texttt{RSMA|OPT-MISOCP} surpasses \texttt{SDMA|OPT-MISOCP} as it is able to exploit the surplus of power to devise the common signal. Besides, \texttt{SDMA|OPT-MISOCP} performs slightly better than \texttt{RSMA|PR-OPT-SCA-SDR} as it avoids projection losses.

\noindent \textit{\textbf{Scenario VI}: Impact of the Transmit Power on the WSR Performance} 

In Fig. \ref{figure_scenario_s8}, we evaluate the WSR as a function of the transmit power. In Fig. \ref{figure_scenario_s8a}, we consider correlated channels, for which optimal admission leads to a consistently higher WSR compared to random admission of UEs. Besides, RSMA outperforms SDMA in all cases due to the high channel similarity, which allows capitalizing on the multicast signal. Specifically, the performance gap widens as the transmit power increases since higher rates can be allocated to the common signal, whereas SDMA is hampered by high interference. We also observe that \texttt{RSMA|OPT-MISOCP} outperforms \texttt{RSMA|PR-OPT-SCA-SDR} for all considered cases, whereas \texttt{RSMA|RND-MISOCP} performs similarly to \texttt{RSMA|PR-OPT-SCA-SDR} even though \texttt{RSMA|RND-MISOCP} does not control which UEs are admitted. In Fig. \ref{figure_scenario_s8b}, we consider uncorrelated channels, where optimal admission also facilitates additional gains for both RSMA and SDMA compared to random UE admission, particularly when the transmit power is more constrained. Besides, \texttt{SDMA|OPT-MISOCP} performs marginally better than \texttt{RSMA|PR-OPT-SCA-SDR} because the latter collapses to SDMA due to the low channel correlation, thereby experiencing severe saturation upon rate projection. On the other hand, the gains due to optimal UE admission tend to diminish for higher transmit powers. For high transmit powers, \texttt{RSMA|RND-MISOCP} surpasses \texttt{RSMA|PR-OPT-SCA-SDR} as the former accounts for rate discretization, thus avoiding losses due to rate projection.

\noindent \textit{\textbf{Scenario VII}: Impact of the Number of Admitted UEs on WEE Performance}

% Figure: Scenario 10
\begin{figure*}[!t]
	\centering
	% Scenario S10a
 	\begin{subfigure}[b]{0.38\textwidth}
		\begin{center}
			\resizebox{\textwidth}{!}
			{%
			\includegraphics[width=1\textwidth]{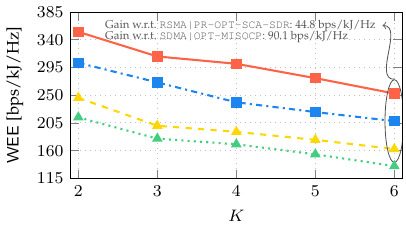}
			}
			\caption{Correlated channels.}
			\label{figure_scenario_s10a}
		\end{center}
 	\end{subfigure}
 	\hfill
	% Legend
 	\begin{subfigure}[b]{0.2\textwidth}
		\begin{center}
			\resizebox{\textwidth}{!}
			{%
			\includegraphics[width=1\textwidth]{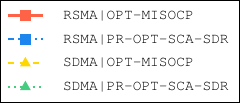}
			}
			\vspace{1.6cm}
		\end{center}
 	\end{subfigure}
 	\hfill	
	% Scenario S10b
 	\begin{subfigure}[b]{0.38\textwidth}
		\begin{center}
			\resizebox{\textwidth}{!}
			{%
			\includegraphics[width=1\textwidth]{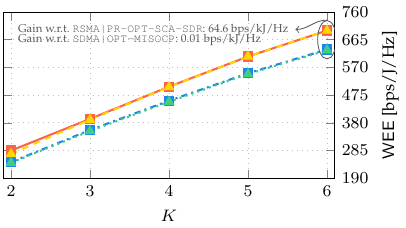}
			}
			\caption{Uncorrelated channels.}
			\label{figure_scenario_s10b}
		\end{center}
 	\end{subfigure}
    \caption{(\textit{Scenario VII}) WEE of RSMA and SDMA as a function of the number of admitted UEs. \emph{In Fig. \ref{figure_scenario_s10a}, \texttt{RSMA|OPT-MISOCP} outperforms \texttt{RSMA|OPT-PR-SCA-SDR} and \texttt{SDMA|OPT-MISOCP} by $ 44.8 $ bps/kJ/Hz ($ \uparrow 21.5 \% $ gain) and $ 90.1 $ bps/kJ/Hz ($ \uparrow 55.4 \% $ gain), respectively, when $ K = 6 $. In Fig. \ref{figure_scenario_s10b}, \texttt{RSMA|OPT-MISOCP} outperforms \texttt{RSMA|OPT-PR-SCA-SDR} by $ 64.6 $ bps/kJ/Hz ($ \uparrow 10.2 \% $ gain), when $ K = 6 $.}}
 	\label{figure_scenario_s10}
 	\vspace{-4mm}
\end{figure*}
% Figure: Scenario 11
\begin{figure*}[!t]
	\centering
	% Scenario S11a
 	\begin{subfigure}[b]{0.38\textwidth}
		\begin{center}
			\resizebox{\textwidth}{!}
			{%
			\includegraphics[width=1\textwidth]{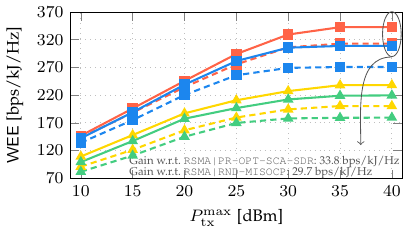}
			}
			\caption{Correlated channels.}
			\label{figure_scenario_s11a}
		\end{center}
 	\end{subfigure}
 	\hfill
	% Legend
 	\begin{subfigure}[b]{0.2\textwidth}
		\begin{center}
			\resizebox{\textwidth}{!}
			{%
			\includegraphics[width=1\textwidth]{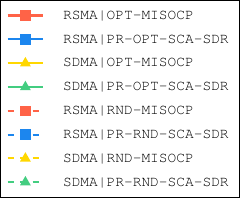}
			}
			\vspace{0.8cm}
		\end{center}
 	\end{subfigure}
 	\hfill	
	% Scenario S11b
 	\begin{subfigure}[b]{0.38\textwidth}
		\begin{center}
			\resizebox{\textwidth}{!}
			{%
			\includegraphics[width=1\textwidth]{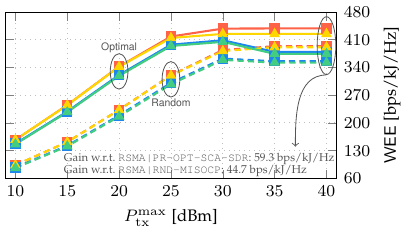}
			}
			\caption{Uncorrelated channels.}
			\label{figure_scenario_s11b}
		\end{center}
 	\end{subfigure}
    \caption{(\textit{Scenario VIII}) WEE of RSMA and SDMA with optimal and random UE admission as a function of the transmit power. \emph{In Fig. \ref{figure_scenario_s11a}, \texttt{RSMA|OPT-MISOCP} outperforms \texttt{RSMA|OPT-PR-SCA-SDR} and \texttt{RSMA|RND-MISOCP} by $ 33.8 $ bps/kJ/Hz ($ \uparrow 10.9 \% $ gain) and $ 29.7 $ bps/kJ/Hz ($ \uparrow 9.5 \% $ gain), respectively, when $ P_\mathrm{tx}^\mathrm{max} = 40 $ dBm. In Fig. \ref{figure_scenario_s11b}, \texttt{RSMA|OPT-MISOCP} outperforms \texttt{RSMA|OPT-PR-SCA-SDR} and \texttt{RSMA|RND-MISOCP} by $ 59.3 $ bps/kJ/Hz ($ \uparrow 15.6 \% $ gain) and $ 44.7 $ bps/kJ/Hz ($ \uparrow 11.4 \% $ gain), respectively, when $ P_\mathrm{tx}^\mathrm{max} = 40 $ dBm.}}
 	\label{figure_scenario_s11}
 	\vspace{-4mm}
\end{figure*}
In Fig. \ref{figure_scenario_s10}, we compare the WEE of RSMA and SDMA as a function of the number of admitted UEs. In Fig. \ref{figure_scenario_s10a}, we consider correlated channels, for which RSMA and SDMA experience a WEE degradation as the number of admitted UEs increases. This occurs because the transmit power needs to be distributed among more UEs, thus affecting the SINRs and the allocated rates. However, RSMA attains a higher performance than SDMA since RSMA is capable of harnessing the high channel similarity. Further, we observe that \texttt{RSMA|OPT-MISOCP} and \texttt{SDMA|OPT-MISOCP} respectively outperform \texttt{RSMA|PR-OPT-SCA-SDR} and \texttt{SDMA|PR-OPT-SCA-SDR} by at least $ 20\%$. In Fig. \ref{figure_scenario_s10b}, we consider uncorrelated channels, for which RSMA collapses to SDMA in most cases, since the common rate is very small or zero due to a low channel correlation. Furthermore, the common rate improves marginally when the number of UEs increases, as it requires a substantially larger transmit power. Specifically, as the number of UEs increases, utilizing the common signal becomes less energy-efficient. We observe that \texttt{RSMA|OPT-MISOCP} outperforms \texttt{RSMA|PR-OPT-SCA-SDR} for all considered values of $ K $.

\noindent \textit{\textbf{Scenario VIII}: Impact of the Transmit Power on the WEE Performance} 

In Fig. \ref{figure_scenario_s11}, we evaluate the WEE as a function of the transmit power. In Fig. \ref{figure_scenario_s11a}, we consider correlated channels, for which RSMA outperforms SDMA as it can exploit the high channel correlation. We also observe that optimal admission performs significantly better than random UE admission, as it allows to select UEs with mutually beneficial channel characteristics that promote EE gains. In addition, increasing the transmit power boosts the WEE as improved operating points can be found. However, this increment saturates after a certain point, as the power required to reach higher rates becomes too costly for a marginal gain in WEE. Besides, \texttt{RSMA|RND-MISOCP} performs similarly to \texttt{RSMA|PR-OPT-SCA-SDR} since its ability to handle discrete rates compensates indirectly for the random selection of UEs. In Fig. \ref{figure_scenario_s11b}, we consider uncorrelated channels, for which we observe that optimal UE admission can lead to substantial gains. Also, \texttt{RSMA|OPT-MISOCP} and \texttt{SDMA|OPT-MISOCP} outperform \texttt{RSMA|OPT-PR-SCA-SDR} and \texttt{RSMA|RND-PR-SCA-SDR}, respectively. Moreover, \texttt{RSMA|OPT-PR-SCA-SDR} and \texttt{SDMA|OPT-PR-SCA-SDR} experience a WEE degradation for larger values of the transmit power because of rate saturation.

\section{Future Research Directions} \label{section_future_research_directions}

In the following, we discuss future avenues of research that can be pursued to extend our current work.

\noindent \textbf{Practical RRM algorithms:} 
With an increasing pool of radio resources, the execution times of RRM algorithms are expected to rise. Thus, having fast, efficient, real-time algorithms capable of operating within short time frames is crucial for practical deployments. This is especially important for reducing end-to-end communication latency and optimizing radio resource utilization efficiency. Achieving fast RRM is plausible through the use of heuristics, which can simplify RRM's complexity by, for example, pre-selecting a subset of admitted UEs to minimize the number of binary variables. However, choosing such a subset of UEs is not trivial in RSMA. Greedy strategies based, for instance, on channel correlation, widely used in SDMA and NOMA, may not be effective in RSMA. Unlike SDMA, which favors low channel correlation, and NOMA, which favors high channel correlation, RSMA presents unique challenges. In particular, the multicast signal of RSMA benefits from correlated channels, whereas the unicast signals benefit from uncorrelated channels. Hence, heuristics for determining a subset of admitted UEs that ensures high performance of RSMA are yet to be developed and require further research. Considering that a significant portion of the computational complexity is linked to the binary variables needed for UE admission and discrete rate selection, one strategy to cope with it is to relax the binary variables and penalize their integrality violation in the objective function, as in \cite{khalili2020:antenna-selection-strategy-energy-efficiency-maximization-uplink-ofdma-networks-multi-objective-approach}. However, fine-tuning the penalty factors associated with these binary variables may become tedious, with no guarantee of obtaining integer solutions in all cases. Overall, further research is needed to develop real-time and efficient algorithms for the RRM of RSMA, particularly when an extended pool of radio resources is considered, which is expected to increase computational complexity. Additionally, integrating mobile data traffic patterns into RRM strategies, as discussed in \cite{fernandez2023:characterizing-modeling-mobile-networks-user-traffic-millisecond-level}, may offer further opportunities for optimization.

\noindent \textbf{Profiting from caching:}
Leveraging the caching capabilities of UEs holds great promise for further improving RSMA performance, as demonstrated by its efficacy in improving the SE of NOMA \cite{xiang2019:cache-aided-non-orthogonal-multiple-access-two-user-case}. Specifically, RSMA can benefit from exploiting content cached at UEs in several ways. Cached content can significantly reduce latency by allowing the BS to transmit only the missing fragments of the requested content. Moreover, cached content can bolster signal quality by aiding in interference cancellation, thus increasing SINR and data rates. However, integrating caching into the RRM design of RSMA poses challenges, including the cache capacity, the size of the requested content, and the specific content cached. In addition, determining which content fragments to cache is non-trivial due to limited cache capacities. Also, exchanging information between the BS and UEs regarding the cached content incurs additional overhead. Furthermore, accounting for cache power consumption is critical, as maintaining cache freshness demands energy. Despite these challenges, optimizing caching jointly with other radio resources holds significant potential for enhancing RSMA's performance.
%To further improve the performance of RSMA, the caching capabilities of the UEs can be leveraged, as this has proven advantageous for enhancing the SE of NOMA \cite{xiang2019:cache-aided-non-orthogonal-multiple-access-two-user-case}. Particularly, RSMA can exploit cached content in UEs for additional gains. Cached content can contribute to minimizing latency. Specifically, if the requested content is partially or fully available at the UE, the BS must only transmit the missing fragments instead of sending the entire content. This also improves the EE, as less energy is needed to make the content available at the UE. Cached content can also contribute to improving signal quality. Specifically, it can be used for canceling part of the interference, resulting in a higher SINR and rate. However, incorporating caching in the RRM design of RSMA is challenging. In particular, finding which parts of the content to cache is not trivial since caches have limited capacities. Also, coordination between the BS and UEs for exchanging information on the cached data can lead to additional overhead. Furthermore, the power consumption of the cache must also be accounted for since maintaining the cache refreshed and updated consumes energy. Finally, optimizing the caching with the rest of the radio resources becomes more complex but promises excellent prospects for improving RSMA's performance.

\noindent \textbf{Integration with OFDM:} 
Given the widespread adoption of orthogonal frequency-division multiplexing (OFDM) in modern wireless communications systems, recent research has focused on integrating OFDM with next-generation multiple-access candidates, such as RSMA. Notably, OFDM-RSMA integration has demonstrated significant advantages in mitigating inter-carrier interference (ICI) \cite{sahin2023:multicarrier-rate-splitting-multiple-access-superiority-ofdm-rsma} caused by Doppler spread and inter-numerology interference (INI) \cite{sahin2023:ofdma-rsma-robust-transmission-intercarrier-interference} caused by different subcarrier spacing (SS) numerologies. These findings have opened new avenues for future research. In particular, the integration of beamforming design in OFDM-RSMA systems remains to be thoroughly investigated, as most previous works focused on SISO scenarios. It is also interesting to explore flexible SS numerology selection, an aspect overlooked in prior works, which only assessed the impact of different numerologies on performance. Furthermore, with discussions on more advanced OFDM variants like orthogonal time-frequency space (OTFS) modulation underway, it is timely to investigate the combination of RSMA with these emerging waveforms, which promise to support ultra-high mobility.

%Given the widespread adoption of OFDM in modern wireless communications systems, recent research has focused on the importance of integrating OFDM with next-generation multiple-access candidates, such as RSMA. It has been shown that OFDM-RSMA can significantly contribute to mitigating ICI \cite{sahin2023:multicarrier-rate-splitting-multiple-access-superiority-ofdm-rsma}, originating from Doppler spread, and INI \cite{sahin2023:ofdma-rsma-robust-transmission-intercarrier-interference}, originating from using different subcarrier spacing (SS) numerologies. These recent results have opened new avenues for future research at the intersection of OFDM and RSMA. In particular, the integration of beamforming design in OFDM-RSMA systems has yet to be fully investigated, as most previous works considered SISO scenarios. Allowing for a flexible selection of SS numerologies is also interesting, as previous works did not focus on optimal SS selection and were limited to evaluating only the impact of different numerologies on performance. Furthermore, since more advanced forms of OFDM, such as OTFS, are being explored in 3GPP discussions, it is timely to investigate the combination of RSMA with these emerging waveforms, which promise to support ultra-high mobility.

\noindent \textbf{CSI estimation:} As RSMA utilizes the same CSI as SDMA for RRM, acquiring and estimating CSI in RSMA systems is expected to remain consistent with the established standardized methods. Specifically, standardized CSI estimation procedures are already in place for unicast services \cite{3gpp.211}. Additionally, 3GPP has implemented CSI estimation procedures tailored for multicast-broadcast services, designed primarily to transmit multimedia content to multiple UEs efficiently \cite{3gpp.23.246}. As RSMA is based on unicast and multicast transmissions, it can seamlessly leverage existing CSI estimation procedures without any changes. However, future RSMA systems may benefit from the joint design of, e.g., reference signals for CSI estimation. This joint design strategy could reduce the overhead compared to employing individual and independent reference signals for unicast and multicast transmissions, opening up novel research directions on CSI estimation for RSMA systems.

\section{Conclusions} \label{section_conclusions}

%In this paper, we proposed two novel RRM problems to investigate the SE and EE of RSMA, taking into account the characteristics of practical wireless systems, namely user admission, discrete rates, and imperfect SIC. 

%In particular, we investigated the maximization of the WSR and WEE of RSMA by jointly optimizing the beamforming, user admission, and discrete rates, while accounting for imperfect SIC. 

%We considered the case of continuous rates, given the widespread adoption of Shannon capacity for SINR-rate modeling. 

%The considered RRM problems resulted in a nonconvex MINLPs, which are generally difficulty to solve. Nevertheless, we developed two algorithms capable of finding high-quality solutions. 

%We revealed that ignoring the practical characteristics of wireless systems for RRM design can have serious repercussions on performance. Specifically, we found that it is important to account for discrete rates in the RRM model to avoid possibly severe rate projection losses. In addition, we recognized the importance of optimal UE admission, which compared to random UE admission, yields greater gains, as it allows to select UEs with channel characteristics that promote mutual benefits. Finally, our results confirmed the benefits of accounting for imperfect SIC to guarantee the allocated RSMA. 

In this paper, two new RRM problems were proposed to investigate the SE and EE of RSMA, taking into account characteristics of practical wireless systems, namely the use of discrete rates in contrast to the widely embraced continuous rates, the need for selective UE admission instead of ubiquitously serving all UEs, and imperfect SIC in lieu of ideal SIC. In particular, we investigated the maximization of the WSR and WEE of RSMA as optimization problems and jointly optimized the beamforming, the UE admission, and the allocation of discrete rates, while accounting for an imperfect SIC. Furthermore, given the widespread adoption of Shannon's capacity formula for SINR-rate modeling in RRM designs, we also considered the case of continuous rates. The considered RRM problems resulted in nonconvex MINLPs, which are generally difficult to solve. Nevertheless, we developed two algorithms capable of finding high-quality solutions. The first algorithm addresses the RRM with discrete rates and transforms the nonconvex MINLP into a MISOCP, which can be solved globally optimally via BnB and IPMs. This algorithm features custom cutting planes that reduce the runtime. The second algorithm addresses the RRM with continuous rates, and solves the nonconvex MINLP using binary enumeration, SDR, and SCA, converging to a KKT point. We revealed that ignoring the practical characteristics of wireless systems in RRM design can have serious repercussions on performance. Specifically, we demonstrated the importance of accounting for discrete rates in the RRM model to avoid potentially severe rate projection losses. In addition, we recognized the importance of selectivity for UE admission, which yields greater gains, as it allows to serve UEs with mutually beneficial channel characteristics that can improve the WSR or WEE. Finally, our results confirmed the benefits of accounting for imperfect SIC to guarantee the allocated rate. Our simulations show that RSMA designed for discrete rates achieves gains of up to $ 89.7\% $ (WSR) and $ 21.5\% $ (WEE) compared to projecting continuous rates onto the admissible set of discrete rates since projection losses are avoided. Furthermore, user admission proves crucial for RSMA as it yields additional gains of up to $ 15.3\% $ (WSR) and $ 11.4\% $ (WEE) compared to random user admission when discrete rates are considered.

\section*{Acknowledgment}
The research is in part funded by the Deutsche Forschungsgemeinschaft (DFG) within the B5G-Cell project (210487104) in SFB 1053 MAKI and the HyRIS project (455077022), by the LOEWE initiative (Hesse, Germany) within the emergenCITY center, and by the European Commission within the DAEMON project (101017109).

%\begin{spacing}{1.18}
\bibliographystyle{IEEEtran}
\bibliography{IEEEabrv,ref}

\clearpage

\begin{appendices}

% Appendices format
\renewcommand{\thesectiondis}[2]{\Alph{section}:}

% Proof of Proposition 1
\setcounter{equation}{0}
\renewcommand{\theequation}{A.\arabic{equation}}
\section{Circumventing Integer Multiplicative Couplings} \label{appendix_proposition_1}

We introduce new variables $ \pi_{u,j} = \chi_u  \kappa_j $, which are binary because of constraints  $ \widebar{\mathrm{C}}_{1}, \widebar{\mathrm{C}}_{10} $. Therefore, we define $ \widebar{\mathrm{D}}_{1}: \pi_{u,j} \in \left\lbrace 0, 1 \right\rbrace, \forall u \in \mathcal{U}, j \in \mathcal{J} $, and employ the McCormick envelopes to linearize the product of binary variables $ \chi_u $, $ \kappa_j $ \cite{mccormick1976:computability-global-solutions-factorable-nonconvex-programs-part-i-convex-underestimating-problems}. Specifically, the product $ \chi_u  \kappa_j $ can be removed if constraints $ \widebar{\mathrm{D}}_{2}: \pi_{u,j} \leq \chi_u, \forall u \in \mathcal{U}, j \in \mathcal{J} $, $ \widebar{\mathrm{D}}_{3}: \pi_{u,j} \leq \kappa_j, \forall u \in \mathcal{U}, j \in \mathcal{J} $, and $ \widebar{\mathrm{D}}_{4}: \pi_{u,j} \geq \chi_u + \kappa_j - 1, \forall u \in \mathcal{U}, j \in \mathcal{J} $, are added. 

In addition, we obtain $ \widebar{\mathrm{D}}_{5}: \frac{\left| \mathbf{h}^\mathrm{H}_u \mathbf{m} \psi \right|^2 } { \sum_{i \in \mathcal{U}} \left| \mathbf{h}_u^\mathrm{H} \mathbf{w}_i \mu_i \right|^2 + {\sigma}^2} \geq  \sum_{j \in \mathcal{J}} \pi_{u,j} \Gamma_j, \forall u \in \mathcal{U} $, and $ \widebar{\mathrm{D}}_{6}: C_u \leq  \sum_{j \in \mathcal{J}} \pi_{u,j} R_j, \forall u \in \mathcal{U} $, upon replacing $ \widebar{\mathrm{D}}_{1} $ in $ \widebar{\mathrm{C}}_{12} $, $ \widebar{\mathrm{C}}_{14} $. Thus, constraints $ \widebar{\mathrm{C}}_{12} $, $ \widebar{\mathrm{C}}_{14} $ are equivalently rewritten as constraints $ \widebar{\mathrm{D}}_{1} - \widebar{\mathrm{D}}_{6} $.

% Proof of Proposition 2
\setcounter{equation}{0}
\renewcommand{\theequation}{B.\arabic{equation}}
\section{Circumventing Mixed-Integer Multiplicative Couplings} \label{appendix_proposition_2}

Let $ \tilde{\mathbf{w}}_u  = \mathbf{w}_u \mu_u $ represent the effective precoder for $ \mathsf{UE}_u $. When $ \mu_u = 0 $, then $ \mathsf{UE}_u $ is not served by a private signal since $ \tilde{\mathbf{w}}_u = \mathbf{0} $. When $ \mu_u = 1 $, then $ \mathsf{UE}_u $ is served by a private signal via precoder $ \tilde{\mathbf{w}}_u = \mathbf{w}_u \neq \mathbf{0} $. We can decouple $ \mathbf{w}_u $ and $ \mu_u $, while obtaining the same effect, by including constraint $ \widebar{\mathrm{E}}_{1}: \left\| \mathbf{w}_u \right\|_2^2 \leq \mu_u P_\mathrm{tx}^\mathrm{max} $. In a similar manner, we can decouple $ \mathbf{m} $ and $ \psi $ by including $ \widebar{\mathrm{E}}_{2}: \left\| \mathbf{m} \right\|_2^2 \leq \psi P_\mathrm{tx}^\mathrm{max} $. 

With the above changes, constraints $ \widebar{\mathrm{C}}_{6} $, $ \widebar{\mathrm{C}}_{9} $, $ \widebar{\mathrm{D}}_{5} $ can be respectively rewritten as $ \widebar{\mathrm{E}}_{3}: \sum_{u \in \mathcal{U}} \left\| \mathbf{w}_u \right\|_2^2 + \left\| \mathbf{m} \right\|_2^2  \leq P_\mathrm{tx}^\mathrm{max} $, $ \widebar{\mathrm{E}}_{4}: \frac{\left| \mathbf{h}_u^\mathrm{H} {\mathbf{w}_u} \right|^2 }{ \Delta_\mathrm{SIC}^2 \left| \mathbf{h}^\mathrm{H}_u \mathbf{m} \right|^2 + \sum_{ i \neq u, i \in \mathcal{U} } \left| \mathbf{h}_u^\mathrm{H} {\mathbf{w}_i} \right|^2 + \sigma^2 } \geq \sum_{j \in \mathcal{J}} \alpha_{u,j} \Gamma_j, \forall u \in \mathcal{U} $, and $ \widebar{\mathrm{E}}_{5}: \frac{\left| \mathbf{h}_u^\mathrm{H} \mathbf{m} \right|^2 }{ \sum_{i \in \mathcal{U}} \left| \mathbf{h}_u^\mathrm{H} {\mathbf{w}_i} \right|^2 + {\sigma}^2 } \geq \sum_{j \in \mathcal{J}} \pi_{u,j} \Gamma_j , \forall u \in \mathcal{U} $. As a result, constraints $ \widebar{\mathrm{C}}_{6}, \widebar{\mathrm{C}}_{9}, \widebar{\mathrm{D}}_{5} $ are equivalently rewritten as constraints $ \widebar{\mathrm{E}}_{1} - \widebar{\mathrm{E}}_{5} $,

\setcounter{equation}{0}
\renewcommand{\theequation}{C.\arabic{equation}}
\section{Circumventing Integer Additive Couplings} \label{appendix_proposition_3}

In the following, we prove the equivalence between $ \widebar{\mathrm{E}}_{4} $ and $ \widebar{\mathrm{F}}_{1} $. Assuming a given $ \mathsf{UE}_u $, we distinguish the following two cases: 

\noindent \circled{\footnotesize{1}} $ \sum_{j \in \mathcal{J}} \alpha_{u,j} = 0 $ ($ \mathsf{UE}_u $ is not served by a private signal) 

\noindent \circled{\footnotesize{2}} $ \sum_{j \in \mathcal{J}} \alpha_{u,j} = 1 $ ($ \mathsf{UE}_u $ is served by a private signal).

\noindent \textbf{Case \circled{\footnotesize{1}} $\Rightarrow$ } When \circled{\footnotesize{1}} is true, constraint $ \widebar{\mathrm{E}}_{4} $ collapses to $ \mathsf{SINR}^{(\mathrm{p})}_u \geq 0 $ since $ \alpha_{u,j} = 0, \forall j \in \mathcal{J} $. Also, when \circled{\footnotesize{1}} is true, constraint $ \widebar{\mathrm{F}}_{1} $ collapses to the intersection of $ J $ constraints, i.e., $ \mathsf{SINR}^{(\mathrm{p})}_u \geq 0 $, $ \forall j \in \mathcal{J} $, which yields $ \mathsf{SINR}^{(\mathrm{p})}_u \geq 0 $, and is consequently equivalent to $ \widebar{\mathrm{E}}_{4} $. 

\noindent \textbf{Case \circled{\footnotesize{2}} $\Rightarrow$} When \circled{\footnotesize{2}} is true, constraint $ \widebar{\mathrm{E}}_{4} $ collapses to $ \mathsf{SINR}^{(\mathrm{p})}_u \geq \Gamma_{j'} $ for some $ j'$ since $ \alpha_{u,j'} = 1 $. Also, when \circled{\footnotesize{2}} is true, constraint $ \widebar{\mathrm{F}}_{1} $ collapses $ \mathsf{SINR}^{(\mathrm{p})}_u \geq \Gamma_{j'} $ and $ \mathsf{SINR}^{(\mathrm{p})}_u \geq 0, \forall j \in \mathcal{J} \backslash j' $, which intersected yield $ \mathsf{SINR}^{(\mathrm{p})}_u \geq \Gamma_{j'} $. This is equivalent to $ \widebar{\mathrm{E}}_{4} $

Thus, the equivalence between $ \widebar{\mathrm{E}}_{4} $ and $ \widebar{\mathrm{F}}_{1}$ was shown. The equivalence between $ \widebar{\mathrm{E}}_{5} $ and $ \widebar{\mathrm{F}}_{2} $ can also be proven using the same procedure above, which we also omit. As a result, constraints $ \widebar{\mathrm{E}}_{4} $, $ \widebar{\mathrm{E}}_{5} $ are equivalently expressed as $ \widebar{\mathrm{F}}_{1} $, $ \widebar{\mathrm{F}}_{2} $,

\setcounter{equation}{0}
\renewcommand{\theequation}{D.\arabic{equation}}
\section{Reformulating the SINR Constraints via the Big-M Method} \label{appendix_proposition_4}

By defining $ \widebar{\mathbf{W}}_u = \left[ \Delta_\mathrm{SIC} \mathbf{m}, \mathbf{w}_1, \dots, \mathbf{w}_{u-1}, \mathbf{w}_{u+1}, \dots, \mathbf{w}_U \right] $, constraint $ \widebar{\mathrm{F}}_{1} $ can be expressed as $ \left\| \left[ \mathbf{h}_u^\mathrm{H} \widebar{\mathbf{W}}_u, \sigma \right] \right\|_2^2 \leq \frac{1}{\alpha_{u,j} \Gamma_j} \left| \mathbf{h}_u^\mathrm{H} {\mathbf{w}_u} \right|^2, \forall u \in \mathcal{U}, j \in \mathcal{J} $. From $ \widebar{\mathrm{F}}_{1} $, two cases are obtained: 

\noindent $ \noindent \textbf{Case \circled{\footnotesize{1}}} ~~~ \alpha_{u,j} = 1 \Rightarrow \left\| \left[ \mathbf{h}_u^\mathrm{H} \widebar{\mathbf{W}}_u, \sigma \right] \right\|_2^2 \leq \frac{1}{\Gamma_j} \left| \mathbf{h}_u^\mathrm{H} {\mathbf{w}_u} \right|^2 $ 
 
\noindent $ \noindent \textbf{Case \circled{\footnotesize{2}}} ~~~ \alpha_{u,j} = 0 \Rightarrow \left\| \left[ \mathbf{h}_u^\mathrm{H} \widebar{\mathbf{W}}_u, \sigma \right] \right\|_2^2 \leq \infty $. 

Notice that using $ \infty $ is not necessary as it would suffice to find an upper bound $ L_{\mathrm{max},u} ^2 $ such that $ \left\| \left[ \mathbf{h}_u^\mathrm{H} \widebar{\mathbf{W}}_u, \sigma \right] \right\|_2^2 \leq L_{\mathrm{max},u}^2 $. Therefore, the two cases can be integrated into a single inequality, thus redefining $ \widebar{\mathrm{F}}_{1} $ as $ \widebar{\mathrm{G}}_{1}: \left\| \left[ \mathbf{h}_u^\mathrm{H} \widebar{\mathbf{W}}_u, \sigma \right] \right\|_2^2 \leq \frac{\left| \mathbf{h}_u^\mathrm{H} {\mathbf{w}_u} \right|^2}{\Gamma_j} + \left( 1 - \alpha_{u,j} \right)^2 L_{\mathrm{max},u}^2, \forall u \in \mathcal{U}, j \in \mathcal{J} $, where $ L_{\mathrm{max},u} = \sqrt{ \big\| {\mathbf{h}}_u \big\|_2^2 P_\mathrm{tx}^\mathrm{max} + \sigma^2} $. We follow a similar procedure to transform $ \widebar{\mathrm{F}}_{2} $ into $ \widebar{\mathrm{G}}_{2} $, which we omit here. As a result, we equivalently recast constraints $ \widebar{\mathrm{F}}_{1} $, $ \widebar{\mathrm{F}}_{2} $ as $ \widebar{\mathrm{G}}_{1}, \widebar{\mathrm{G}}_{2} $.

%\begin{align} 
%	& \mathrm{G_3}: \epsilon_u \left\| \mathbf{w}_u \right\|_2 + d_u \leq \left| \hat{\mathbf{h}}_u^H {\mathbf{w}_u} \right|, \forall u \in \mathcal{U}, \nonumber
%	\\
%	& \mathrm{G_4}: \epsilon_u \left\| \mathbf{m} \right\|_2 + r_u \leq \left| \hat{\mathbf{h}}_u^H {\mathbf{m}} \right|, \forall u \in \mathcal{U}, \nonumber
%	\\
%	& \mathrm{G_5}: \left| \hat{\mathbf{h}}_u^H {\mathbf{w}_i} \right| + \epsilon_u \left\| \mathbf{w}_i \right\|_2 \leq q_{u,i}, \forall u \in \mathcal{U}, i \in \mathcal{U}, \nonumber
%	\\
%	& \mathrm{G_6}: \left| \hat{\mathbf{h}}_u^H {\mathbf{m}} \right| + \epsilon_u \left\| \mathbf{m} \right\|_2 \leq t_u, \forall u \in \mathcal{U}, \nonumber
%	\\
%	& \mathrm{G_7}: d_u \geq 0, \forall u \in \mathcal{U}, \nonumber
%	\\
%	& \mathrm{G_8}: r_u \geq 0, \forall u \in \mathcal{U}, \nonumber
%\end{align}	

% Proof of Proposition 5
\setcounter{equation}{0}
\renewcommand{\theequation}{E.\arabic{equation}}
\section{Convexifying the Private SINR Constraints} \label{appendix_proposition_5}

While $ \widebar{\mathrm{G}}_{1} $ is nonconvex in its current form, it can be transformed into a SOC constraint using Jensen's inequality, as $ \left\| \left[ \mathbf{h}_u^\mathrm{H} \widebar{\mathbf{W}}_u, \sigma \right] \right\|_2 \leq \frac{\left| \mathbf{h}_u^\mathrm{H} {\mathbf{w}_u} \right|}{\sqrt{\Gamma_j}}  + \left( 1 - \alpha_{u,j} \right) L_{\mathrm{max},u} $, $ \forall u \in \mathcal{U} $, $ j \in \mathcal{J} $. Note that the above inequality and $ \widebar{\mathrm{G}}_{1} $ are not equivalent but both delimit the same feasible set when $ \alpha_{u,j} = 1 $.  When $ \alpha_{u,j} = 0 $, the inequality still holds without changing the feasible set because of the valid upper bound. 

Besides, since the precoders are invariant to phase shifting, $ \mathbf{w}_u $ and $ \mathbf{w}_u e^{ j \theta_u } $ yield the same SINR. As a result, it is possible to choose a phase $ e^{ j \theta_u } $ such that $ {\mathbf{h}}_u^\mathrm{H} \mathbf{w}_u  $ becomes purely real and nonnegative. Based on this observation, $ \widebar{\mathrm{G}}_{1} $ can be equivalently expressed as constraints $ \widebar{\mathrm{H}}_{1}: \mathfrak{Re} \left\lbrace {\mathbf{h}}_u^\mathrm{H} {\mathbf{w}_u} \right\rbrace \geq 0, \forall u \in \mathcal{U} $, $ \widebar{\mathrm{H}}_{2}: \mathfrak{Im} \left\lbrace {\mathbf{h}}_u^\mathrm{H} {\mathbf{w}_u} \right\rbrace = 0, \forall u \in \mathcal{U} $, $ \widebar{\mathrm{H}}_{3}: \left\| \left[ \mathbf{h}_u^\mathrm{H} \widebar{\mathbf{W}}_u, \sigma \right] \right\|_2 \leq \frac{1}{\sqrt{\Gamma_j}} \mathfrak{Re} \left\lbrace {\mathbf{h}}_u^\mathrm{H} {\mathbf{w}_u} \right\rbrace + \left( 1 - \alpha_{u,j} \right) L_{\mathrm{max},u} $, $ \forall u \in \mathcal{U}, j \in \mathcal{J} $.

% Proof of Proposition 6
\setcounter{equation}{0}
\renewcommand{\theequation}{F.\arabic{equation}}
\section{Convexifying the Common SINR Constraints} \label{appendix_proposition_6}

Note that $ \left| {\mathbf{h}}_u^\mathrm{H} \mathbf{m} \right| \geq \mathfrak{Re} \left\lbrace {\mathbf{h}}_u^\mathrm{H} \mathbf{m} \right\rbrace $ always holds true. Using this relation, we replace $ \widebar{\mathrm{G}}_{2} $ with the convex constraints $ \widebar{\mathrm{I}}_{1}: \mathfrak{Re} \left\lbrace {\mathbf{h}}_u^\mathrm{H} \mathbf{m} \right\rbrace \geq 0, \forall u \in \mathcal{U} $, and $ \widebar{\mathrm{I}}_{2}: \left\| \left[ \mathbf{h}_u^\mathrm{H} \mathbf{W}, \sigma \right] \right\|_2  \leq \frac{1}{\sqrt{\Gamma_j}} \mathfrak{Re} \left\lbrace {\mathbf{h}}_u^\mathrm{H} \mathbf{m} \right\rbrace + \left( 1 - \pi_{u,j} \right) L_{\mathrm{max},u}, \forall u \in \mathcal{U}, j \in \mathcal{J} $, where $ \mathbf{W} = \left[ \mathbf{w}_1, \dots, \mathbf{w}_U \right]^\mathrm{T} $. To obtain $ \widebar{\mathrm{I}}_{2} $, we follow the same procedure as in \textbf{Appendix \ref{appendix_proposition_5}}.

% Proof of Proposition 10
\setcounter{equation}{0}
\renewcommand{\theequation}{F.\arabic{equation}}
\section{Adding Cutting Planes to Tighten the Feasible Domain} \label{appendix_proposition_10}

From constraints $ \widebar{\mathrm{E}}_{4} $, $ \widebar{\mathrm{H}}_{1} $, $ \widebar{\mathrm{H}}_{2} $, we obtain $ \hat{\mathrm{J}}_1: \mathfrak{Re} \left\lbrace {\mathbf{h}}_u^\mathrm{H} {\mathbf{w}_u} \right\rbrace^2 \geq \sum_{j \in \mathcal{J}} \alpha_{u,j} \Gamma_j \big( \Delta_\mathrm{SIC}^2 \left| \mathbf{h}^\mathrm{H}_u \mathbf{m} \right|^2 + \sum_{ i \neq u, i \in \mathcal{U} } \left| \mathbf{h}_u^\mathrm{H} {\mathbf{w}_i} \right|^2 + \sigma^2 \big), \forall u \in \mathcal{U}, j \in \mathcal{J} $. Assuming zero interference and perfect SIC (i.e., $ \Delta_\mathrm{SIC} = 0 $), we obtain a lower bound for $ \mathfrak{Re} \left\lbrace {\mathbf{h}}_u^\mathrm{H} {\mathbf{w}_u} \right\rbrace $ defined as $ \breve{\mathrm{J}}_1: \mathfrak{Re} \left\lbrace {\mathbf{h}}_u^\mathrm{H} {\mathbf{w}_u} \right\rbrace \geq \sigma \sqrt{ \sum_{j \in \mathcal{J}} \alpha_{u,j} \Gamma_j }, \forall u \in \mathcal{U} $. However, since the sum of all $ \alpha_{u,j} $ is at most one for a given $ \mathsf{UE}_u $ (see constraints $ \widebar{\mathrm{C}}_{3}, \widebar{\mathrm{C}}_{7}, \widebar{\mathrm{C}}_{8} $), then $ \breve{\mathrm{J}}_1 $ can be equivalently recast as $ \widebar{\mathrm{J}}_{1}: \mathfrak{Re} \left\lbrace {\mathbf{h}}_u^\mathrm{H} {\mathbf{w}_u} \right\rbrace \geq \sigma \sum_{j \in \mathcal{J}} \alpha_{u,j} \sqrt{\Gamma_j} $, $ \forall u \in \mathcal{U} $, thus defining a new set of cuts. 

In addition, $ \widebar{\mathrm{J}}_{2} $ is included as it allows to terminate early the binary variable branching. In particular, if the upper bound is achieved for some combination of discrete rates for a valid subset of admitted UEs, the algorithm has found an optimal solution, and therefore the process is stopped. Although $ \widebar{\mathrm{J}}_{1} $, $ \widebar{\mathrm{J}}_{2} $ are optional, they tighten the feasible set of the binary variables and contribute to accelerate the search.

%% Equation
%\begin{align}
%	& \mathrm{I_1}: \epsilon_u \left\| \mathbf{m} \right\|_2 + r_u \leq \mathfrak{Re} \left\lbrace \hat{\mathbf{h}}_u^H \mathbf{m} \right\rbrace, \forall u \in \mathcal{U}, \nonumber
%\end{align}

% Proof of Proposition 7
\setcounter{equation}{0}
\renewcommand{\theequation}{G.\arabic{equation}}
\section{Transforming the Problem via Sublevel and Superlevel Sets} \label{appendix_proposition_7}

%We prove the equivalence of the two problems by contradiction. Let us assume that we have obtained an optimal solution for $ \widetilde{\mathcal{Q}}_{\mathrm{CWSR}_n} $ with objective value $ \beta^\star $, and denote with $ \gamma^\star_u $, $ \rho^\star_u $, $ \tau^\star_u $, $ \lambda^\star_u $ the optimal values of variables $ \gamma_u $, $ \rho_u $, $ \tau_u $, $ \lambda_u $, corresponding to $ \mathsf{UE}_u $. We further assume that at the optimum, constraint $ \widebar{\mathrm{K}}_{2} $ for $ \mathsf{UE}_u $ is inactive, i.e., not satisfied with equality, which implies that there exists a strictly smaller $ \rho'_u < \rho^\star_u $, for which $ \widebar{\mathrm{K}}_{2} $ is satisfied. However, a smaller $ \rho'_u $ also implies that there exists a larger $ \gamma'_u > \gamma^\star_u $ that satisfies $ \widebar{\mathrm{K}}_{1} $. At the same time, a larger $ \gamma'_u $ allows the existence of a larger objective value $ \beta' > \beta^\star $ due to $ \widebar{\mathrm{K}}_{3} $, which contradicts the assumption that we have found an optimum. As a result, this outcome refutes the possibility that some inequalities at the optimum could be inactive, and shows that the reformulation with sublevel and superlevel sets is tight as they are attained with equality at the optimum. Similar relations can be derived for $ \widebar{\mathrm{K}}_{4} - \widebar{\mathrm{K}}_{6} $, thus corroborating the equivalence between $ \mathcal{Q}_{\mathrm{CWSR}_n} $ and $ \widetilde{\mathcal{Q}}_{\mathrm{CWSR}_n} $.

In the following, we show by contradiction that constraints $ \mathrm{K}_1 - \mathrm{K}_6 $ are satisfied with equality at the optimum, thus corroborating that $ \mathcal{Q}_{\mathrm{CWSR}_n} $ and $ \widetilde{\mathcal{Q}}_{\mathrm{CWSR}_n} $ are equivalent. Note that $ \mathrm{K}_7 $, $ \mathrm{K}_8 $ are satisfied automatically, if $ \mathrm{K}_1 - \mathrm{K}_6 $ are tight. 

We assume that we have an optimal solution for $ \widetilde{\mathcal{Q}}_{\mathrm{CWSR}_n} $ with objective function value $ \beta^\star $, and denote with $ \gamma^\star_u $, $ \rho^\star_u $, $ \tau^\star_u $, $ \lambda^\star_u $, $ C^\star_u $ the optimal values of $ \gamma_u $, $ \rho_u $, $ \tau_u $, $ \lambda_u $, $ C_u $ corresponding to $ \mathsf{UE}_u $. We further assume that at the optimum, $ \widebar{\mathrm{K}}_{2} $ for $ \mathsf{UE}_u $ is inactive, i.e., not tight, allowing for the existence of a strictly smaller $ \rho'_u < \rho^\star_u $, for which $ \widebar{\mathrm{K}}_{2} $ is also satisfied. However, a smaller $ \rho'_u $ implies that there exists a larger $ \gamma'_u > \gamma^\star_u $ that satisfies $ \widebar{\mathrm{K}}_{1} $ thus allowing for the existence of a larger objective function value $ \beta' > \beta^\star $ due to $ \widebar{\mathrm{K}}_{3} $, hence contradicting the assumption that we have found an optimum. Similarly, we can assume that $ \widebar{\mathrm{K}}_{5} $ for $ \mathsf{UE}_u $ is inactive, allowing for the existence of $ \lambda'_u < \lambda^\star_u $, for which $ \widebar{\mathrm{K}}_{5} $ is also satisfied. At the same time, this allows for the existence of $ \tau'_u > \tau^\star_u $ that satisfies $ \widebar{\mathrm{K}}_{4} $, and the existence of a larger $ C_u' > C_u^\star $ due to $ \widebar{\mathrm{K}}_{6} $. This observation implies that a higher objective function value can be obtained thus contradicting the assumption that we have found an optimum. For further reading, we refer the reader to \cite{tran2012:fast-converging-algorithm-weighted-sum-rate-maximization-multicell-miso-downlink}, where similar deductions were drawn for a different problem. Besides, the above analysis can be employed to show the equivalence between $ \mathcal{Q}_{\mathrm{CWEE}_n} $ and $ \widetilde{\mathcal{Q}}_{\mathrm{CWEE}_n} $, which we omit.

% Proof of Proposition 8
\setcounter{equation}{0}
\renewcommand{\theequation}{H.\arabic{equation}}
\section{Solutions with at Most Rank One} \label{appendix_proposition_8}

We define the Lagrangian of $ \bar{\mathcal{Q}}_{\mathrm{CWSR}_n}^{(t)} $ with respect to $ \mathbf{M} $ (when $ \psi = 1 $, otherwise $ \mathbf{M} = \mathbf{0} $) as 
\begin{align}
\mathcal{L} = & \phi_{\widebar{\mathrm{L}}_{1}} \mathrm{Tr} \left( \mathbf{M} \right) + \phi_{\widebar{\mathrm{L}}_{2}} \mathrm{Tr} \left( \mathbf{M} \right) + \sum_u \phi_{\widebar{\mathrm{L}}_{4}}^u \Delta_\mathrm{SIC}^2 \mathbf{h}_u^\mathrm{H} \mathbf{M} \mathbf{h}_u - \nonumber
\\ 
& \mathrm{Tr} \left( \boldsymbol{\Phi}_{\widebar{\mathrm{L}}_{8}} \mathbf{M} \right) - \sum_u \phi_{\widebar{\mathrm{M}}_{2}}^u \mathbf{h}_u^\mathrm{H} \mathbf{M} \mathbf{h}_u - \nonumber
\\ 
& \mathrm{Tr} \left( \boldsymbol{\Phi}_{\widebar{\mathrm{M}}_{3}} \left( \zeta_0 \mathbf{I} - {\mathbf{T}_0^{(t)}}^\mathrm{H} \mathbf{M} {\mathbf{T}_0^{(t)}} \right) \right) + h(\widehat{\mathbf{W}}), \nonumber
\end{align}
where $ h(\widehat{\mathbf{W}}) $ represents the terms that depend on $ \widehat{\mathbf{W}} $.

Based on the KKT dual feasibility condition, $ \phi_{\widebar{\mathrm{L}}_{1}} \geq 0 $, $ \phi_{\widebar{\mathrm{L}}_{2}} \geq 0 $, $ \boldsymbol{\Phi}_{\widebar{\mathrm{L}}_{8}} \succcurlyeq \mathbf{0} $, $ \boldsymbol{\Phi}_{\widebar{\mathrm{M}}_{3}} \succcurlyeq \mathbf{0} $ are the KKT multipliers associated with constraints $ \widebar{\mathrm{L}}_{4} $, $ \widebar{\mathrm{L}}_{2} $, $ \widebar{\mathrm{L}}_{8} $, $ \widebar{\mathrm{M}}_{3} $, whereas $ \phi_{\widebar{\mathrm{L}}_{4}}^u $, $ \phi_{\widebar{\mathrm{M}}_{2}}^u $ are the KKT multipliers associated with constraints $ \widebar{\mathrm{L}}_{1} $, $ \widebar{\mathrm{M}}_{2} $ for $ \mathsf{UE}_u $. 

By invoking the KKT stationarity condition, we take the derivative of $ \mathcal{L} $ with respect to $ \mathbf{M} $ and equate it to zero, yielding 
\begin{align}
	\boldsymbol{\Phi}_{\widebar{\mathrm{L}}_{8}} = \mathbf{A} + c \mathbf{h}_u \mathbf{h}_u^\mathrm{H}, \nonumber
\end{align}
where $ \mathbf{A} = \left( \phi_{\widebar{\mathrm{L}}_{1}} + \phi_{\widebar{\mathrm{L}}_{2}} \right) \mathbf{I} + {\mathbf{T}_0^{(t)}} \boldsymbol{\Phi}_{\widebar{\mathrm{M}}_{3}} {\mathbf{T}_0^{(t)}}^\mathrm{H}  $ and $ c = \sum_u \phi_{\widebar{\mathrm{L}}_{4}}^u \Delta_\mathrm{SIC}^2 - \phi_{\widebar{\mathrm{M}}_{2}}^u $. 

From the KKT complementary slackness condition, it must hold that $ \boldsymbol{\Phi}_{\widebar{\mathrm{L}}_{8}} \mathbf{M} = \mathbf{0} $ and $ \boldsymbol{\Phi}_{\widebar{\mathrm{M}}_{3}} \left( \zeta_0 \mathbf{I} - {\mathbf{T}_0^{(t)}}^\mathrm{H} \mathbf{M} {\mathbf{T}_0^{(t)}} \right) = \mathbf{0} $. Applying Sylvester's rank inequality to $ \boldsymbol{\Phi}_{\widebar{\mathrm{L}}_{8}} \mathbf{M} = \mathbf{0} $, we obtain that $ \mathrm{Rank} \left( \boldsymbol{\Phi}_{\widebar{\mathrm{L}}_{8}} \right) + \mathrm{Rank} \left( \mathbf{M} \right) \leq N_\mathrm{tx} $. In addition, we note that  $ \boldsymbol{\Phi}_{\widebar{\mathrm{L}}_{8}} $ is Hermitian with $ \mathrm{Rank} \left( \boldsymbol{\Phi}_{\widebar{\mathrm{L}}_{8}} \right) \geq N_\mathrm{tx} - 1 $ since $ \mathbf{A} $ is positive definite. This results in the following two possible cases, i.e., 

$ \bullet ~ \mathrm{Rank} \left( \boldsymbol{\Phi}_{\widebar{\mathrm{L}}_{8}} \right) = N_\mathrm{tx} $

$ \bullet ~ \mathrm{Rank} \left( \boldsymbol{\Phi}_{\widebar{\mathrm{L}}_{8}} \right) = N_\mathrm{tx} - 1 $ 

When $ \mathrm{Rank} \left( \boldsymbol{\Phi}_{\widebar{\mathrm{L}}_{8}} \right) = N_\mathrm{tx} $ then $ \mathbf{M} = \mathbf{0} $, implying that the common signal is not transmitted. When $ \mathrm{Rank} \left( \boldsymbol{\Phi}_{\widebar{\mathrm{L}}_{8}} \right) = N_\mathrm{tx} - 1 $, for some $ c < 0 $, leads to $ \mathrm{Rank} \left( \mathbf{M} \right) \leq 1 $, implying that the common signal could be transmitted. The same conclusions can be obtained for $ \mathbf{W}_u $, $ \forall u \in \mathcal{U}_n' $. The above procedure can also be applied to $ \bar{\mathcal{Q}}_{\mathrm{CWEE}_n}^{(t)} $ as the two problems are similar, which leads to the same conclusion, i.e., the ranks of $ \mathbf{M} $ and $ \mathbf{W}_u $, $ \forall u \in \mathcal{U}_n' $ in both problems are at most one. For further reading, we refer to \cite{alavi2017:robust-beamforming-techniques-nonorthogonal-multiple-access-systems-bounded-channel-uncertainties, fu2020:robust-secure-beamforming-design-two-user-downlink-miso-rate-splitting-systems, xu2020:resource-allocation-irs-assisted-full-duplex-cognitive-radio-systems, sun2019:iterative-rank-penalty-method-nonconvex-quadratically-constrained-quadratic-programs}, where similar problems were considered.

% In particular, we know that $ \boldsymbol{\Phi}_{\mathrm{M_{3}}} \mathbf{C} = \mathbf{0} $ where $ \mathbf{C} = \zeta_0 \mathbf{I} - \mathbf{T}_0^H \mathbf{M} \mathbf{T}_0 $ with $ \mathbf{C} \succcurlyeq\mathbf{0} $, $ \boldsymbol{\Phi}_{\mathrm{L_{8}}} \succcurlyeq\mathbf{0} $. As more iterations elapse, we eventually obtain that $ \mathbf{C} = \mathbf{0} $, and therefore it must be satisfied that upon convergence $ \boldsymbol{\Phi}_{\mathrm{M_{3}}} \succ \mathbf{0} $ is full-rank, thus making $ \mathbf{A} $ positive definite.

% Proof of Proposition 9
\setcounter{equation}{0}
\renewcommand{\theequation}{I.\arabic{equation}}
\section{Convergence Proof} \label{appendix_proposition_9}

Since $ \widehat{\mathcal{Q}}_{\mathrm{CWSR}_n} $ and $ \mathcal{Q}_{\mathrm{CWSR}_n} $ are equivalent, in this proof we employ $ \widehat{\mathcal{Q}}_{\mathrm{CWSR}_n} $, which can be expressed as
\begin{align} 
	% Objective
	\widehat{\mathcal{Q}}_{\mathrm{CWSR}_n} : & \max_{
			\substack{
						\boldsymbol{\nu} \in \mathcal{X}
			 }
	}
	~~~ f \left( \boldsymbol{\nu} \right)  ~~~ \mathrm{s.t.} & &  g_i \left( \boldsymbol{\nu} \right) \leq 0, ~ i \in \mathcal{V}, & \nonumber 
	\\
	& & & h_j \left( \boldsymbol{\nu} \right) \leq 0, ~ j \in \mathcal{W}, & \nonumber
	\\
	& & & \ell_k \left( \boldsymbol{\nu} \right) \leq 0, ~ k \in \mathcal{R}, & \nonumber
\end{align}
where $ \boldsymbol{\nu} $ collects all the decision variables of $ \widehat{\mathcal{Q}}_{\mathrm{CWSR}_n} $; $ {\mathcal{X}} $ denotes the feasible set; $ f \left( \boldsymbol{\nu} \right) $ is the objective function; $ g_i \left( \boldsymbol{\nu} \right) \leq 0, i \in \mathcal{V} $, represent constraints $ \widebar{\mathrm{L}}_{3}, \widebar{\mathrm{L}}_{5} $; $ h_i \left( \boldsymbol{\nu} \right) \leq 0, i \in \mathcal{W} $, represent constraints $ \widebar{\mathrm{L}}_{9}, \widebar{\mathrm{L}}_{10} $; $ \ell_k \left( \boldsymbol{\nu} \right) \leq 0, k \in \mathcal{R} $, represent the rest of constraints; and $ \mathcal{V}, \mathcal{W}, \mathcal{R} $ are index sets. Similarly, we express $ \bar{\mathcal{Q}}_{\mathrm{CWSR}_n}^{(t)} $ as
\begin{align} 
	% Objective
	\bar{\mathcal{Q}}_{\mathrm{CWSR}_n}^{(t)}: & \max_{
			\substack{
						\boldsymbol{\omega} \in \bar{\mathcal{X}}^{(t)} 
			 }
	}
	~~~ \bar{f} \left( \boldsymbol{\omega} \right) ~~~ \mathrm{s.t.} & & G_i \left( \boldsymbol{\omega}, \Omega_i^{(t)} \right) \leq 0, ~ i \in \mathcal{V}, & \nonumber
	\\
	& & & H_j \left( \boldsymbol{\omega} \right) \leq 0, ~ j \in \mathcal{W}, & \nonumber
	\\
	& & & \ell_k \left( \boldsymbol{\omega} \right) \leq 0, ~ k \in \mathcal{R}, & \nonumber
\end{align}
where $ \boldsymbol{\omega} $ collects all the decision variables of $ \bar{\mathcal{Q}}_{\mathrm{CWSR}_n}^{(t)} $; $ \bar{\mathcal{X}}^{(t)} $ denotes the feasible set; $ \bar{f} \left( \boldsymbol{\omega} \right) $ is the objective function; $ G_i \left( \boldsymbol{\omega}, \Omega_i^{(t)} \right) \leq 0, i \in \mathcal{V} $, represent constraints $ \widebar{\mathrm{M}}_{1} $, $ \widebar{\mathrm{M}}_{2} $; and $ H_j \left( \boldsymbol{\omega} \right) \leq 0, ~ j \in \mathcal{W} $, represent constraints $ \widebar{\mathrm{M}}_{3} $, $ \widebar{\mathrm{M}}_{4} $. 

Let $ {\boldsymbol{\omega}}_t $ denote the solution of $ \bar{\mathcal{Q}}_{\mathrm{CWSR}_n}^{(t)} $ and let $ \Omega_i^{(t)} = \Pi \left( \boldsymbol{\omega}_{t-1} \right), i \in \mathcal{V}, $ be the adaptable parameters in $ \widebar{\mathrm{M}}_{1} $, $ \widebar{\mathrm{M}}_{2} $, computed as a function $ \Pi \left( \cdot \right) $ of the previous solution $ \boldsymbol{\omega}_{t-1} $ (the adaptable parameters are $ \bar{\Omega}_{1,u}^{(t)} $, $ \bar{\Omega}_{2,u}^{(t)} $ in problem $ \bar{\mathcal{Q}}_{\mathrm{CWSR}_n}^{(t)} $ and $ \bar{\Omega}_{3,u}^{(t)} $ in problem $ \bar{\mathcal{Q}}_{\mathrm{CWEE}_n}^{(t)} $). 

Since $ \widebar{\mathrm{M}}_{1} $, $ \widebar{\mathrm{M}}_{2} $ are inner approximations for $ \widebar{\mathrm{L}}_{3} $, $ \widebar{\mathrm{L}}_{5} $, i.e., $ G_i \left( \boldsymbol{\omega}, \Omega_i^{(t)} \right) \geq  g_i \left( \boldsymbol{\omega} \right) $, then $ \bar{\mathcal{X}}^{(t)} \subseteq {\mathcal{X}} $. Also, for sufficiently large penalty weights $ p_u^{(t)} $, constraints $ \widebar{\mathrm{M}}_{3} $, $ \widebar{\mathrm{M}}_{4} $ ensure $ \bar{\mathcal{X}}^{(t)} \subseteq {\mathcal{X}} $. Therefore, $ {\boldsymbol{\omega}}_t $ satisfies $ \widehat{\mathcal{Q}}_{\mathrm{CWSR}_n} $. Note that $ {\boldsymbol{\omega}}_t $ also satisfies $ \bar{\mathcal{Q}}^{(t+1)}_{\mathrm{CWSR}_n} $ since $ \Omega_i^{(t)} $ is updated such that $ G_i \left( {\boldsymbol{\omega}}_{t - 1}, \Omega_i^{(t)} \right) = g_i \left( {\boldsymbol{\omega}}_{t - 1} \right) $. As a result, $ {\boldsymbol{\omega}}_t \in \bar{\mathcal{X}}^{(t)} \cap \bar{\mathcal{X}}^{(t+1)} $, implying that $ \bar{f} \left( {\boldsymbol{\omega}}_{t+1} \right) \geq \bar{f} \left( {\boldsymbol{\omega}}_t \right) $ thereby leading to a monotonically non-decreasing sequence $ \left\lbrace \bar{f} \left( {\boldsymbol{\omega}}_t \right) \right\rbrace $. Since $ \bar{\mathcal{X}}^{(t)} $ is compact and $ \bar{\mathcal{Q}}_{\mathrm{CWSR}_n}^{(t)} $ is limited by a power constraint, sequence $ \left\lbrace \bar{f} \left( {\boldsymbol{\omega}}_t \right) \right\rbrace $ is bounded and converges. In particular, the collection of solutions for $ \bar{\mathcal{Q}}_{\mathrm{CWSR}_n}^{(t)} $ define a sequence $ \left\lbrace {\boldsymbol{\omega}}_t \right\rbrace $ that converges to an accumulation point $ {\boldsymbol{\omega}}^\star $, i.e., $ {\boldsymbol{\omega}}_t \rightarrow {\boldsymbol{\omega}}^\star $, which is a KKT point.

Note that $ \boldsymbol{\nu} $ is included in $ \boldsymbol{\omega} $ such that $ \boldsymbol{\omega} = \left( \boldsymbol{\nu}, \boldsymbol{\zeta} \right) $, where $ \boldsymbol{\zeta} $ are the slack variables in $ \widebar{\mathrm{M}}_{3} $, $ \widebar{\mathrm{M}}_{4} $. Since $ \boldsymbol{\omega}^\star $ is an accumulation point of $ \left\lbrace {\boldsymbol{\omega}}_t \right\rbrace $, there exists a subsequence $ \left\lbrace {\boldsymbol{\omega}}_{m_t} \right\rbrace $ such that $ {\boldsymbol{\omega}}_{m_t} \rightarrow {\boldsymbol{\omega}}^\star $. Hence, it also follows that $ {\boldsymbol{\omega}}_{m_t - 1} \rightarrow {\boldsymbol{\omega}}^\star $. Upon convergence at $ {\boldsymbol{\omega}}^\star = \left( {\boldsymbol{\nu}}^\star, {\boldsymbol{\zeta}}^\star \right) $, variables $ \boldsymbol{\zeta}^\star = \mathbf{0} $ and therefore $ \boldsymbol{\nu}^\star $ is an accumulation point. In addition, $ H_j \left( {\boldsymbol{\omega}}^\star \right) = h_j \left( {\boldsymbol{\nu}}^\star \right), j \in \mathcal{W} $, since both enforce feasible sets with ranks of at most one, which is shown in \textbf{Appendix \ref{appendix_proposition_8}}. 

Let $ \mathcal{L} = \mathcal{V} \cup \mathcal{W} \cup \mathcal{R} $ such that $ \mathcal{V} \cap \mathcal{W} = \left\lbrace \emptyset \right\rbrace $, $ \mathcal{W} \cap \mathcal{R} = \left\lbrace \emptyset \right\rbrace $, $ \mathcal{R} \cap \mathcal{V} = \left\lbrace \emptyset \right\rbrace $. Also, let $ \mathcal{I} \supseteq \mathcal{L} $ be the set of active constraints of $ \widehat{\mathcal{Q}}_{\mathrm{CWSR}_n} $ with respect to $ {\boldsymbol{\nu}}^\star $ and let $ \mathcal{I}_{m_t} \supseteq \mathcal{L} $ be the set of active constraints of $ \bar{\mathcal{Q}}_{\mathrm{CWSR}_n}^{(t)} $ with respect to $ {\boldsymbol{\omega}}_{m_t} $.
Now, letting $ t \rightarrow \infty $ for $ \bar{\mathcal{Q}}_{\mathrm{CWSR}_n}^{(t)} $, we obtain that 
\begin{align}
G_i \left( {\boldsymbol{\omega}}_{m_t}, \Pi \left( {\boldsymbol{\omega}}_{m_t - 1} \right) \right) & \rightarrow G_i \left( {\boldsymbol{\omega}}^\star, \Pi \left( {\boldsymbol{\omega}}^\star \right) \right), i \in \mathcal{V}, \nonumber
\\
& = g_i \left( {\boldsymbol{\omega}}^\star \right), i \in \mathcal{V}, \nonumber
\\
& = g_i \left( {\boldsymbol{\nu}}^\star \right), i \in \mathcal{V}, \nonumber
\\[0.2cm]
H_j \left( {\boldsymbol{\omega}}_{m_t} \right) & \rightarrow H_j \left( {\boldsymbol{\omega}}^\star \right), j \in \mathcal{W}, \nonumber
\\
& = h_j \left( {\boldsymbol{\omega}}^\star \right), j \in \mathcal{W}, \nonumber
\\
& = h_j \left( {\boldsymbol{\nu}}^\star \right), j \in \mathcal{W}, \nonumber
\\[0.2cm]
\ell_k \left( {\boldsymbol{\omega}}_{m_t} \right) & \rightarrow \ell_k \left( {\boldsymbol{\omega}}^\star \right), k \in \mathcal{R}, \nonumber
\\
& = \ell_k \left( {\boldsymbol{\nu}}^\star \right), k \in \mathcal{R}. \nonumber 
\end{align}

These above limits suggest that there exists an integer $ T_1 $ for which $ \mathcal{I}_{m_t} \subseteq \mathcal{I}, \forall t > T_1 $. Similarly, we have the following:
\begin{align}
	\nabla_{\boldsymbol{\omega}} \bar{f} \left( {\boldsymbol{\omega}}_{m_t} \right) & \rightarrow \nabla_{\boldsymbol{\omega}} \bar{f} \left( {\boldsymbol{\omega}}^\star \right), \nonumber
	\\
	& = [ \nabla_{\boldsymbol{\nu}} f \left( {\boldsymbol{\nu}}^\star \right) ~ \mathbf{0} ], \nonumber
	\\[0.2cm]
	\nabla_{\boldsymbol{\omega}} G_i \left( {\boldsymbol{\omega}}_{m_t}, \Pi \left( {\boldsymbol{\omega}}_{m_t - 1} \right) \right) & \rightarrow \nabla_{\boldsymbol{\omega}} G_i \left( {\boldsymbol{\omega}}^\star, \Pi \left( {\boldsymbol{\omega}}^\star \right) \right), i \in \mathcal{V}, \nonumber 
	\\
	& = \nabla_{\boldsymbol{\omega}} g_i \left( {\boldsymbol{\omega}}^\star \right), i \in \mathcal{V}, \nonumber
	\\
	& = [\nabla_{\boldsymbol{\nu}} g_i \left( {\boldsymbol{\nu}}^\star \right) ~ \mathbf{0}], i \in \mathcal{V}, \nonumber
	\\[0.2cm]
	\nabla_{\boldsymbol{\omega}} H_j \left( {\boldsymbol{\omega}}_{m_t} \right) & \rightarrow \nabla_{\boldsymbol{\omega}} H_j \left( {\boldsymbol{\omega}}^\star \right), j \in \mathcal{W}, \nonumber 
	\\
	& = [ \nabla_{\boldsymbol{\nu}} h_j \left( {\boldsymbol{\nu}}^\star \right) ~ \mathbf{0}], j \in \mathcal{W}, \nonumber
	\\[0.2cm]
	\nabla_{\boldsymbol{\omega}} \ell_k \left( {\boldsymbol{\omega}}_{m_t} \right) & \rightarrow \nabla_{\boldsymbol{\omega}} \ell_k \left( {\boldsymbol{\omega}}^\star \right), l \in \mathcal{R}, \nonumber
	\\ 
	& = [ \nabla_{\boldsymbol{\nu}} \ell_k \left( {\boldsymbol{\nu}}^\star \right) ~ \mathbf{0}], l \in \mathcal{R}, \nonumber
\end{align}
showing that all constraint gradients of $ \bar{\mathcal{Q}}_{\mathrm{CWSR}_n}^{(t)} $ converge to their corresponding ones in $ \widehat{\mathcal{Q}}_{\mathrm{CWSR}_n} $. These results together with the fact that $ \mathcal{I}_{m_t} \subseteq \mathcal{I}, \forall t \geq T_1 $, imply that there exists an integer $ T_2 > T_1 $ such that $ {\boldsymbol{\omega}}_{m_t} $ is a regular point of $ \bar{\mathcal{Q}}_{\mathrm{CWSR}_n}^{(t)} $ when $ t > T_2 $, for which the KKT conditions are satisfied, i.e., 
\begin{align}
	\bullet ~ & - \nabla_{\boldsymbol{\omega}} \bar{f} \left( {\boldsymbol{\omega}}_{m_t} \right) + \sum_{i \in \mathcal{V}} \mu_i^{m_t} \nabla_{\boldsymbol{\omega}} G_i \left( {\boldsymbol{\omega}}_{m_t}, \Pi \left( {\boldsymbol{\omega}}_{m_t - 1} \right) \right) + \nonumber
	\\ 
	& \sum_{j \in \mathcal{W}} \nabla_{\boldsymbol{\omega}} \mu_j^{m_t} H_j \left( {\boldsymbol{\omega}}_{m_t} \right) + \sum_{k \in \mathcal{R}} \nabla_{\boldsymbol{\omega}} \mu_k^{m_t} \ell_k \left( {\boldsymbol{\omega}}_{m_t} \right) = 0, \nonumber
	\\[0.2cm]
	\bullet ~ & \mu_i^{m_t} G_i \left( {\boldsymbol{\omega}}_{m_t}, \Pi \left( {\boldsymbol{\omega}}_{m_t - 1} \right) \right) = 0, i \in \mathcal{V}, \nonumber
	\\[0.2cm]
	\bullet ~ & \mu_j^{m_t} H_j \left( {\boldsymbol{\omega}}_{m_t} \right) = 0, j \in \mathcal{W},  \nonumber
	\\[0.2cm]
	\bullet ~ & \mu_k^{m_t} \ell_k \left( {\boldsymbol{\omega}}_{m_t} \right) = 0, k \in \mathcal{R}, \nonumber
\end{align}
where $ \mu_l^{m_t} \geq 0, l \in \mathcal{L} $, are KKT multipliers. 

Considering $ t > T_2 $, let $ \mathbf{r}_t = \nabla_{\boldsymbol{\omega}} \bar{f} \left( {\boldsymbol{\omega}}_{m_t} \right) $ and let $ \mathbf{D}_t $ be the matrix whose columns are the gradients of the active constraints, indexed by $ \mathcal{I} $. By complementary slackness, it follows that $ \mu_l^{m_t} = 0, l \notin \mathcal{I} $ for $ t > T_2 $. Thus, the stationarity condition can be expressed in matrix form as $ \mathbf{D}_t \mathbf{b}_t = \mathbf{r}_t $, where $ \mathbf{b}_t  $ is formed by the elements in $ \left\lbrace \mu_l^{m_t} \mid l \in \mathcal{I} \right\rbrace $, which are positive.
Similarly, we define $ \mathbf{r} = \nabla_{\boldsymbol{\omega}} \bar{f} \left( {\boldsymbol{\omega}}^\star \right) $ and $ \mathbf{D} $ as the matrix whose columns are the gradients of the constraints indexed by $ \mathcal{I} $. Thus we have that $ \mathbf{D}_t \rightarrow \mathbf{D} $ and $ \mathbf{r}_t \rightarrow \mathbf{r} $, where $ \mathbf{D}_t $ and $ \mathbf{D} $ are full rank for $ t > T_2 $ due to $ {\boldsymbol{\omega}}_{m_t} $ being a regular point at which the set of active gradients are linearly independent, leading to $ \mathbf{b}_t = \left( \mathbf{D}_t^\mathrm{T} \mathbf{D}_t \right)^{-1} \mathbf{D}_t^\mathrm{T} \mathbf{r}_t $ and $ \mathbf{b}_t \rightarrow \left( \mathbf{D}^\mathrm{T} \mathbf{D} \right)^{-1} \mathbf{D}^\mathrm{T} \mathbf{r} $. Since $ \mu_l^{m_t}, l \in \mathcal{L} $, are either elements from $ \mathbf{r}_t $ or zero, they have a limit which we denote by $ \mu_l^\star, l \in \mathcal{L} $. Thus, letting $ t \rightarrow \infty $, the KKT conditions of $ \bar{\mathcal{Q}}_{\mathrm{CWSR}_n}^{(t)} $ are as follows:
\begin{align}
	\bullet ~ & - \nabla_{\boldsymbol{\omega}} \bar{f} \left( {\boldsymbol{\omega}}^\star \right) + \sum_{i \in \mathcal{I} \cap \mathcal{V}} \nabla_{\boldsymbol{\omega}} \mu_i^\star G_i \left( {\boldsymbol{\omega}}^\star, \Pi \left( {\boldsymbol{\omega}}^\star \right) \right) + \nonumber
	\\  
	& \sum_{j \in \mathcal{I} \cap \mathcal{W}} \nabla_{\boldsymbol{\omega}} \mu_j^\star H_j \left( {\boldsymbol{\omega}}^\star \right) + \sum_{k \in \mathcal{I} \cap \mathcal{R}} \nabla_{\boldsymbol{\omega}} \mu_k^\star \ell_k \left( {\boldsymbol{\omega}}^\star \right) = 0, \nonumber  
	\\[0.2cm]
	\bullet ~ & \mu_i^\star G_i \left( {\boldsymbol{\omega}}^\star, \Pi \left( {\boldsymbol{\omega}}^\star \right) \right)= 0, i \in \mathcal{I} \cap \mathcal{V}, \nonumber   
	\\[0.2cm]
	\bullet ~ & \mu_j^\star H_j \left( {\boldsymbol{\omega}}^\star \right) = 0, j \in \mathcal{I} \cap \mathcal{W}, \nonumber   
	\\[0.2cm]
	\bullet ~ & \mu_k^\star \ell_k \left( {\boldsymbol{\omega}}^\star \right) = 0, k \in \mathcal{I} \cap \mathcal{R}. \nonumber 
\end{align}

Since the following holds true
$ \bar{f} \left( {\boldsymbol{\omega}}^\star \right) = f \left( {\boldsymbol{\omega}}^\star \right) $;
$ G_i \left( {\boldsymbol{\omega}}^\star, \Pi \left( {\boldsymbol{\omega}}^\star \right) \right) = g_i \left( {\boldsymbol{\omega}}^\star \right), i \in \mathcal{V} $; 
$ H_j \left( {\boldsymbol{\omega}}^\star \right) = h_j \left( {\boldsymbol{\omega}}^\star \right), j \in \mathcal{W} $; 
and $ \mu_l^\star = 0, l \notin \mathcal{I} $, then we obtain: 
\begin{align}
	\bullet ~ & - \nabla_{\boldsymbol{\omega}} f \left( {\boldsymbol{\omega}}^\star \right) + \sum_{i \in \mathcal{V}} \nabla_{\boldsymbol{\omega}} \mu_i^\star g_i \left( {\boldsymbol{\omega}}^\star \right) +  & \nonumber 
	\\
	& \sum_{j \in \mathcal{W}} \nabla_{\boldsymbol{\omega}} \mu_j^\star H_j \left( {\boldsymbol{\omega}}^\star \right) + \sum_{k \in \mathcal{R}} \nabla_{\boldsymbol{\omega}} \mu_k^\star \ell_k \left( {\boldsymbol{\omega}}^\star \right) = 0, & \nonumber 
	\\[0.2cm]
	\bullet ~ & \mu_i^\star g_i \left( {\boldsymbol{\omega}}^\star \right) = 0, i \in \mathcal{V}, & \nonumber 
	\\[0.2cm]
	\bullet ~ & \mu_j^\star h_j \left( {\boldsymbol{\omega}}^\star \right) = 0, j \in \mathcal{W}, & \nonumber 
	\\[0.2cm]
	\bullet ~ & \mu_k^\star \ell_k \left( {\boldsymbol{\omega}}^\star \right) = 0, k \in \mathcal{R}, & \nonumber 
\end{align}
proving that $ {\boldsymbol{\omega}}^\star $ is a KKT point as the above KKT conditions are the same for $ \bar{\mathcal{Q}}_{\mathrm{CWSR}_n}^{(t)} $. We can arrive to the same conclusion for problems $ \widehat{\mathcal{Q}}_{\mathrm{CWEE}_n} $ and $ \bar{\mathcal{Q}}_{\mathrm{CWEE}_n}^{(t)} $. For further reading, we refer the reader to \cite{beck2010:sequential-parametric-convex-approximation-method-applications-nonconvex-truss-topology-design-problems}.\

% Proof of Proposition 12
\setcounter{equation}{0}
\renewcommand{\theequation}{I.\arabic{equation}}
\section{Comparing RSMA against NOMA} \label{appendix_proposition_12}

\begin{figure*}[!t]
	% Legend
 	\begin{subfigure}[b]{\textwidth}
		\begin{center}
			\resizebox{!}{0.5cm}
			{%
			\includegraphics[width=1\textwidth]{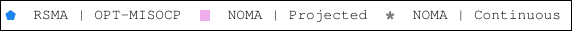}
			}
		\end{center}
 	\end{subfigure}
	\centering
	% Scenario S1a (Row 1)
 	\begin{subfigure}[b]{0.24\textwidth}
		\begin{center}
			\resizebox{\textwidth}{!}
			{%
			\includegraphics[width=1\textwidth]{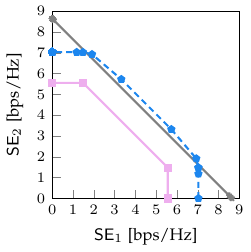}
			}
			\caption{$ \phi = \frac{\pi}{9}$ }
			\label{figure_scenario_s15a}
		\end{center}
 	\end{subfigure}
    \hfill 
	% Scenario S1b (Row 1)
 	\begin{subfigure}[b]{0.24\textwidth}
		\begin{center}
			\resizebox{\textwidth}{!}
			{%
			\includegraphics[width=1\textwidth]{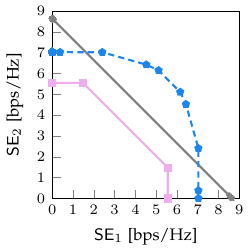}
			}
			\caption{ $ \phi = \frac{2\pi}{9}$ }
			\label{figure_scenario_s15b}
		\end{center}
 	\end{subfigure}
    \hfill 
	% Scenario S1c (Row 1)
 	\begin{subfigure}[b]{0.24\textwidth}
		\begin{center}
			\resizebox{\textwidth}{!}
			{%
			\includegraphics[width=1\textwidth]{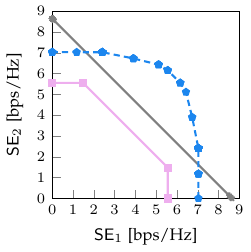}
			}
			\caption{ $ \phi = \frac{3\pi}{9}$ }
			\label{figure_scenario_s15c}
		\end{center}
 	\end{subfigure}
    \hfill 
	% Scenario S1d (Row 1)
 	\begin{subfigure}[b]{0.24\textwidth}
		\begin{center}
			\resizebox{\textwidth}{!}
			{%
			\includegraphics[width=1\textwidth]{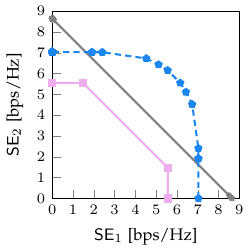}
			}
			\caption{ $ \phi = \frac{4\pi}{9}$ }
			\label{figure_scenario_s15d}
		\end{center}
 	\end{subfigure}
    \caption{Two-user SE region of RSMA with discrete and NOMA with continuous rates assuming $ \frac{P_\mathrm{tx}^\mathrm{max}}{\sigma^2} = 20 $ dB and equal channel strengths.}
    \label{figure_scenario_s15}
 	\vspace{-4mm}
\end{figure*}

In Fig. \ref{figure_scenario_s15}, we show the SE of RSMA and NOMA for a two-UE case, assuming the same parameter settings as in \emph{Scenario I}, which we reproduce in the following. The transmit SNR is $ \frac{P_\mathrm{tx}^\mathrm{max}}{\sigma^2} = 20 $ dB, the channels for $ \mathsf{UE}_1 $ and $ \mathsf{UE}_2 $ are respectively given by $ \mathbf{h}_1 = \left[ 1, 1, 1, 1, 1 \right]^\mathrm{H} $ and $ \mathbf{h}_2 = \left[ 1, e^{j \phi}, e^{j 2 \phi}, e^{j 3 \phi} \right]^\mathrm{H} $, and $ \phi = \left\lbrace \frac{\pi}{9}, \frac{2\pi}{9}, \frac{3\pi}{9}, \frac{4\pi}{9} \right\rbrace $ controls the channel correlation. For RSMA, we employed our proposed approach \texttt{RSMA|OPT-MISOCP}, which inherently handles discrete rates. For NOMA, we implemented the continuous-rate approach in \cite{mao2018:rsma-downlink-communication-systems-bridging-generalizing-outperforming-sdma-noma}, which we named \texttt{NOMA|Continuous}. In addition, we include \texttt{NOMA|Projected}, which represents the rates of \texttt{NOMA|Continuous} after projecting the rates onto the feasible discrete rate set. \emph{In the following, \texttt{NOMA|Continuous} is referential and the comparison is made between \texttt{RSMA|OPT-MISOCP} and \texttt{NOMA|Projected}, as both employ discrete rates.} 

In the considered scenario, there is no difference between the channel strengths, i.e., $ \left\| \mathbf{h}_1 \right\|_2^2 = \left\| \mathbf{h}_2 \right\|_2^2 $. As a result, NOMA cannot successfully exploit SIC to remove interference, leading NOMA to have the same performance as orthogonal multiple access (OMA), e.g., TDMA or FDMA. In addition, we note that rate projection may severely affect NOMA's performance. Conversely, RSMA can adapt to different channel characteristics as shown in Fig. \ref{figure_scenario_s15a} to Fig. \ref{figure_scenario_s15d}. Particularly, as the channel correlation approaches $ \frac{\pi}{2} $, RSMA rates increase since multiuser interference becomes less significant due to the channels turning more orthogonal.

\begin{figure*}[!t]
	% Legend
 	\begin{subfigure}[b]{\textwidth}
		\begin{center}
			\resizebox{!}{0.5cm}
			{%
			\includegraphics[width=1\textwidth]{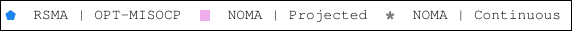}
			}
		\end{center}
 	\end{subfigure}
	\centering
	% Scenario S2e (Row 2)
 	\begin{subfigure}[b]{0.24\textwidth}
		\begin{center}
			\resizebox{\textwidth}{!}
			{%
			\includegraphics[width=1\textwidth]{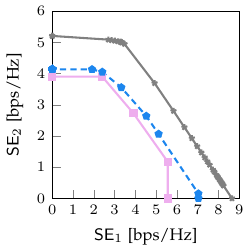}
			}
			\caption{ $ \gamma_1 = 1 $,  $ \gamma_2 = 0.3 $. }
			\label{figure_scenario_s16a}
		\end{center}
 	\end{subfigure}
    \hfill 
	% Scenario S2f (Row 2)
 	\begin{subfigure}[b]{0.24\textwidth}
		\begin{center}
			\resizebox{\textwidth}{!}
			{%
			\includegraphics[width=1\textwidth]{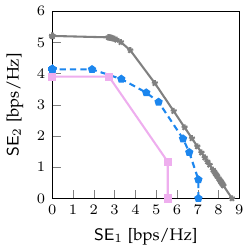}
			}
			\caption{ $ \gamma_1 = 1 $,  $ \gamma_2 = 0.3 $. }
			\label{figure_scenario_s16b}
		\end{center}
 	\end{subfigure}
    \hfill 
	% Scenario S2g (Row 2)
 	\begin{subfigure}[b]{0.24\textwidth}
		\begin{center}
			\resizebox{\textwidth}{!}
			{%
			\includegraphics[width=1\textwidth]{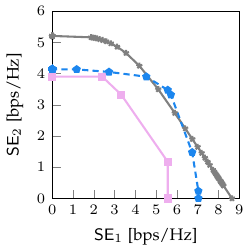}
			}
			\caption{ $ \gamma_1 = 1 $,  $ \gamma_2 = 0.3 $. }
			\label{figure_scenario_s16c}
		\end{center}
 	\end{subfigure}
    \hfill 
	% Scenario S2h (Row 2)
 	\begin{subfigure}[b]{0.24\textwidth}
		\begin{center}
			\resizebox{\textwidth}{!}
			{%
			\includegraphics[width=1\textwidth]{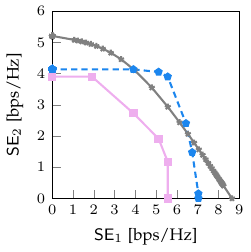}
			}
			\caption{ $ \gamma_1 = 1 $,  $ \gamma_2 = 0.3 $. }
			\label{figure_scenario_s16d}
		\end{center}
 	\end{subfigure}
    \caption{Two-user SE region of RSMA with discrete and NOMA with continuous rates assuming $ \frac{P_\mathrm{tx}^\mathrm{max}}{\sigma^2} = 20 $ dB and unequal channel strengths.}
 	\label{figure_scenario_s16}
 	\vspace{-4mm}
\end{figure*}

Since channels of equal strength do not benefit NOMA, we evaluate a more favorable setup for NOMA in Fig. \ref{figure_scenario_s16}. Specifically, we kept all other parameters unchanged but assumed channels $ \mathbf{h}_1 = \left[ 1, 1, 1, 1, 1 \right]^\mathrm{H} $ and $ \mathbf{h}_2 = 0.3 \left[ 1, e^{j \phi}, e^{j 2 \phi}, e^{j 3 \phi} \right]^\mathrm{H} $ with a pronounced difference in strength, i.e., $ \left\| \mathbf{h}_2 \right\|_2^2 = 0.09 \left\| \mathbf{h}_1 \right\|_2^2 $, which NOMA can exploit to its advantage. Despite the favorable channel conditions for NOMA, we observe in Fig. \ref{figure_scenario_s16a} to Fig. \ref{figure_scenario_s16d} that \texttt{NOMA|Projected} is vastly outperformed by \texttt{RSMA|OPT-MISOCP}. 

Besides, in Fig. \ref{figure_scenario_s15} and Fig. \ref{figure_scenario_s16}, we note that although \texttt{RSMA|OPT-MISOCP} is restricted to the use of discrete rates and \texttt{NOMA|Continuous} is more flexible with continuous rates, \texttt{RSMA|OPT-MISOCP} can outperform \texttt{NOMA|Continuous} in scenarios where the channels have equal channel strength and/or low channel correlation.

% Figure: Scenario S1

% Figure: Scenario S1

% Proof of Proposition 11
\setcounter{equation}{0}
\renewcommand{\theequation}{I.\arabic{equation}}
\section{Numerical Evaluation of Worst-case Complexity} \label{appendix_proposition_11}
% Figure: Scenario 
\begin{figure*}[!h]
	% Legend
 	\begin{subfigure}[b]{\textwidth}
		\begin{center}
			\resizebox{!}{0.5cm}
			{%
			\includegraphics[width=1\textwidth]{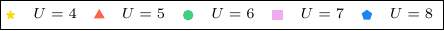}
			}
		\end{center}
 	\end{subfigure}
	\centering
	% Scenario A
 	\begin{subfigure}[b]{0.32\textwidth}
		\begin{center}
			\resizebox{\textwidth}{!}
			{%
			\includegraphics[width=1\textwidth]{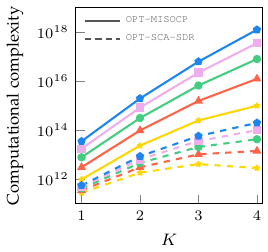}
			}
			\caption{Computational complexity for different values of $ U $ and $ K $, assuming $ J = 15 $.}
			\label{figure_scenario_s12a}
		\end{center}
 	\end{subfigure}
    \hfill 
	% Scenario B
 	\begin{subfigure}[b]{0.32\textwidth}
		\begin{center}
			\resizebox{\textwidth}{!}
			{%
			\includegraphics[width=1\textwidth]{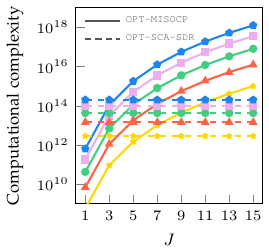}
			}
			\caption{Computational complexity for different values of $ U $ and $ J $, assuming $ K = 4 $.}
			\label{figure_scenario_s12b}
		\end{center}
 	\end{subfigure}
    \hfill 
	% Scenario C
 	\begin{subfigure}[b]{0.32\textwidth}
		\begin{center}
			\resizebox{\textwidth}{!}
			{%
			\includegraphics[width=1\textwidth]{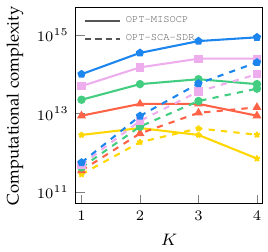}
			}
			\caption{Computational complexity for different values of $ U $ and $ K $, assuming $ J = 7 $, and $ N_p = 2000 $.}
			\label{figure_scenario_s12c}
		\end{center}
 	\end{subfigure}
    \caption{Computation complexity of \texttt{OPT-MISOCP} and \texttt{OPT-SCA-SDR} as function of the number of UEs, number of admitted UEs, and number of discrete rates.}
 	\label{figure_scenario_s12}
\end{figure*}

In Fig. \ref{figure_scenario_s12}, we show the impact of different parameter settings on the computational complexity of \texttt{OPT-MISOCP} and \texttt{OPT-SCA-SDR}. 

In Fig. \ref{figure_scenario_s12a}, we consider $ N_\mathrm{tx} = 16 $, $ U = \left\lbrace 4, 5, 6, 7, 8\right\rbrace $, $ K = \left\lbrace 1, 2, 3, 4\right\rbrace $, and $ J = 15 $. We observe that the computational complexity of \texttt{OPT-MISOCP} is significantly higher than that of \texttt{OPT-SCA-SDR}, especially for larger values of $ U $ and $ K $. This difference exists because \texttt{OPT-MISOCP} contains additional binary variables for discrete rate selection, which are not accounted for in \texttt{OPT-SCA-SDR}. Note that when $ K = U = 4 $, the computational complexity due to UE admission is absent for both approaches since all UEs are admitted. Therefore, for $ K = U = 4 $, the computational complexity of \texttt{OPT-MISOCP} is only due to the beamforming design and discrete rate selection, whereas the computational complexity of \texttt{OPT-SCA-SDR} is only due to beamforming. For \texttt{OPT-SCA-SDR}, we observe that the computational complexity slightly decreases when $ K = U = 4 $ compared to the case when $ K = 3 $ and $ U = 4 $. This reduction in computational complexity occurs because only one UE admission combination is evaluated in the former case, whereas four different UE admission combinations are considered in the latter case. For \texttt{OPT-MISOCP}, we do not observe that the computational complexity decreases when $ K = U = 4 $. The reason is the presence of binary variables for discrete rate selection, whose search complexity dominates over that of UE admission, especially since $ J = 15 $.

In Fig. \ref{figure_scenario_s12b}, we consider $ N_\mathrm{tx} = 16 $, $ U = \left\lbrace 4, 5, 6, 7, 8 \right\rbrace $, $ K = 4 $, and $ J = \left\lbrace 1, 3, 5, 7, 9, 11, 13, 15 \right\rbrace $. As expected, we find that \texttt{OPT-SCA-SDR} is not influenced by $ J $ since it does not account for discrete rates in the RRM design. However, the computational complexity of \texttt{OPT-MISOCP} grows with an increasing value of $ J $ because more rate values become available, increasing the allocation possibilities. Interestingly, for $ J = 3 $, the complexities of \texttt{OPT-MISOCP} and \texttt{OPT-SCA-SDR} are comparable. This outcome is explainable because $ J $ is so small that the computational complexity of both approaches is mainly due to the beamforming design.

Note, however, that the computational complexities used for obtaining Fig. \ref{figure_scenario_s12} correspond to the worst case, i.e., an upper limit. This means that we assumed the worst-case complexity for the continuous variables, i.e., IPM's upper estimates, and the worst-case complexity for the binary variables search, i.e., exhaustive search. In practice, the computational complexity associated with the optimization of continuous and discrete variables is much smaller. Specifically, IPMs run much faster than their upper complexity estimates and BnB methods can explore the binary variable space very efficiently, at a small fraction of the complexity of an exhaustive search. 

In Fig. \ref{figure_scenario_s12c}, we show the computational complexities for the same settings as in Fig. \ref{figure_scenario_s12a}, but assuming practical search complexity values of BnB. In our simulations, we realized that BnB needed to explore a few thousand solutions before finding an optimal solution. Thus, we consider $ N_p = 2000 $ in this case. We also assume $ J = 7 $, which is about half of the number of discrete rates considered in Fig. \ref{figure_scenario_s12a} and Fig. \ref{figure_scenario_s12b}, helping to reduce the number of binary variables. We observe that for the considered parameter configuration, the complexities of both approaches are comparable. Still, these results are theoretical, and therefore, more emphasis should be given to the experimental runtime complexities, such as those presented in Section~\ref{subsection_complexity_optimality_convergence_initialization}. \\

\end{appendices}

\end{document}